\pdfoutput=1
\documentclass[usenatbib]{mn2e}
\usepackage{apjfonts}
\usepackage{amssymb}
\usepackage{amsmath}
\usepackage{ctable}
\usepackage{url}
\usepackage{fixltx2e} 

\newcommand{\be}{\begin{equation}}
\newcommand{\ee}{\end{equation}}

\newcommand{\mbh}{M_{\rm BH}}

\newcommand{\msun}{M_{\sun}}

\newcommand{\qeos}{q_{\rm eos}}

\newcommand{\movieurltwo}{\url{http://www.cfa.harvard.edu/~phopkins/Site/Movies_zoom.html}}

\newcommand\plotone[1]
 {\centering \leavevmode \includegraphics[width={0.99\columnwidth}]{#1}}

\newcommand{\plotside}[1]
 {\centering \leavevmode \includegraphics[width={0.95\textwidth}]{#1}}

\newcommand{\acknowledgments}{\begin{small}\section*{Acknowledgments}\end{small}}
\newcommand\altaffilmark[1]{$^{#1}$}
\newcommand\altaffiltext[1]{$^{#1}$}
\title[The Dynamical AGN Torus]{The Origins of AGN Obscuration: The `Torus' as a 
Dynamical, Unstable Driver of Accretion}

\author[Hopkins et al.]{
\parbox[t]{\textwidth}{ 
Philip F. Hopkins\altaffilmark{1}\thanks{E-mail:phopkins@astro.berkeley.edu},
Christopher C.~Hayward\altaffilmark{2}, 
Desika Narayanan\altaffilmark{3}, 
\&\ Lars Hernquist\altaffilmark{2}
} 
\vspace*{6pt} \\
\altaffiltext{1}{Department of Astronomy, University of California
  Berkeley, Berkeley, CA 94720} \\
\altaffiltext{2}{Harvard-Smithsonian Center for Astrophysics, 60 
Garden Street, Cambridge, MA 02138, USA} \\
\altaffiltext{3}{Steward Observatory, University of Arizona, 933 
N Cherry Ave, Tucson, Az, 85721}
}

\date{Submitted to MNRAS, August, 2011}
\begin{document}
\maketitle
\label{firstpage}

\begin{abstract}

Recent multi-scale simulations have made it possible to follow gas
inflows responsible for high-Eddington ratio accretion onto massive
black holes (BHs) from galactic scales to the BH accretion disk.  When
sufficient gas is driven towards a BH, gravitational instabilities
generically form lopsided, eccentric disks that propagate inwards from
larger radii. The lopsided stellar disk exerts a strong torque on the
gas, driving inflows that fuel the growth of the BH. Here, we
investigate the possibility that the same disk, in its gas-rich phase,
is the putative ``torus'' invoked to explain obscured active galactic
nuclei and the cosmic X-ray background.  The disk is generically thick
and has characteristic $\sim1-10\,$pc sizes and masses resembling
those required of the torus.  Interestingly, the scale heights and
obscured fractions of the predicted torii are substantial even in the
absence of strong stellar feedback providing the vertical support.
Rather, they can be maintained by strong bending modes and
warps/twists excited by the inflow-generating instabilities.  A number
of other observed properties commonly attributed to ``feedback''
processes may in fact be explained entirely by dynamical,
gravitational effects: the lack of alignment between torus and host
galaxy, correlations between local SFR and turbulent gas velocities,
and the dependence of obscured fractions on AGN luminosity or SFR.  We
compare the predicted torus properties with observations of gas
surface density profiles, kinematics, scale heights, and SFR densities
in AGN nuclei, and find that they are consistent in all cases.  We
argue that it is not possible to reproduce these observations and the
observed column density distribution without a clumpy gas
distribution, but allowing for simple clumping on small scales the
predicted column density distribution is in good agreement with
observations from $N_{\rm H}\sim10^{20}-10^{27}\,{\rm cm^{-2}}$.  We
examine how the $N_{\rm H}$ distribution scales with galaxy and AGN
properties.  The dependence is generally simple, but AGN feedback may
be necessary to explain certain trends in obscured fraction with
luminosity and/or redshift.  In our paradigm, the torus is not merely
a bystander or passive fuel source for accretion, but is itself the
mechanism driving accretion. Its generic properties are not
coincidence, but requirements for efficient accretion.

\end{abstract}


\section{Introduction}
\label{sec:intro}

It has long been realized that bright, high-Eddington ratio accretion
(i.e., a quasar) dominates the accumulation of mass in the
supermassive BH population \citep{soltan82,salucci:bhmf,shankar:bhmf,
hopkins:old.age}.  The discovery, in the past decade, of tight
correlations between black hole mass and host spheroid properties
including mass \citep{KormendyRichstone95,magorrian}, velocity dispersion
\citep{FM00,Gebhardt00}, and binding energy or potential well depth
\citep{hopkins:bhfp.obs,hopkins:bhfp.theory,aller:mbh.esph,feoli:bhfp.1} implies that 
black hole (BH) growth is tightly coupled to the process of galaxy and
bulge formation. Increasingly, models invoke feedback processes from
active galactic nuclei (AGN) to explain a host of phenomena, from the
origin of the $M_{\rm BH}-\sigma$ relation, to rapid quenching of star
formation in bulges, to the buildup of the color-magnitude relation
and resolution of the cooling flow problem
\citep[see e.g.][and references therein]{silkrees:msigma,dimatteo:msigma,
springel:red.galaxies,hopkins:groups.ell,hopkins:twostage.feedback,croton:sam,
cattaneo:2009.bhfb.ell.quenching}. 

Observations have demonstrated that most of the accretion luminosity
in the Universe is obscured by large columns of gas and dust
\citep[e.g.][and references therein]{lawrence:receding.torus,
risaliti:seyfert.2.nh.distrib, hill96:sf.abs.in.radio.gal,
simpson99:thermal.imaging.of.radio.gal,
willott00:optical.qso.frac.vs.l,simpson00:ir.photometry.radio.qsos,
hao:local.lf,ueda03:qlf}.  This obscured AGN population dominates the
population of X-ray sources
\citep{miyaji01:sx.qlf,ueda03:qlf,nandra05:z3.faint.qlf,hasinger05:qlf,
steffen03:agn.obsc.vs.l.z,grimes04:obsc.frac.vs.l,hasinger04:obsc.frac.and.xrb,
sazonov.revnivtsev04:local.hx.qlf,barger:qlf,
gilli:obscured.fractions,hasinger:absorption.update}, and accounts for
most of the observed X-ray background luminosity
\citep{setti:1989.xrb.from.agn,madau:1994.agn.xrb,
comastri:1995.agn.xrb,treister:obscured.frac.z.evol,gilli:obscured.fractions}.
It may dominate the bright end of the infrared luminosity function as
well \citep{sanders96:ulirgs.mergers,komossa:ngc6240,
ptak:2003.chandra.ulirg.nuclei,
hickox:bootes.obscured.agn,daddi:2007.high.compton.thick.pops,
alexander:2008.compton.thick.z2.qsos,hopkins:sb.ir.lfs}.  The
abundance of obscured quasars remains a major uncertainty in
reconciling synthesis models of AGN populations with the BH mass
function today, and (by implication) understanding the radiative
efficiencies of quasars \citep{salucci:bhmf,yutremaine:bhmf,
hopkins:bol.qlf,shankar:bol.qlf}.  Various specific galaxy populations
(for example EXOs, XBONGs, ULIRGs, SMGs) commonly host obscured AGN
\citep{xbongs,georgantopoulos:xbong.dilution,max:2005.6240.core,mainieri:2005.exos,
alexander:xray.smgs,daddi:2007.high.compton.thick.pops,alexander:2008.compton.thick.z2.qsos,
riechers:qso.host.wet.merger.remnant.z4,
trump:lowl.agn.dilution,georgakakis:xr.agn.host.morph,
nardini:2009.agn.vs.sb.contrib.in.ulirgs}.  And
theoretical models have long predicted that in violent events such as
galaxy mergers, there should be a transition from an early, ``buried''
accretion stage corresponding to e.g.\ ``warm'' ULIRGs and similar
galaxies, to an at least partially un-obscured phase in which the AGN
removes dust and gas and is visible as a bright quasar
\citep{sanders88:quasars,sanders88:warm.ulirgs,
king:msigma.superfb.1,dimatteo:msigma,hopkins:qso.all,
hopkins:red.galaxies,hopkins:faint.slope,
hopkins:groups.qso,hopkins:msigma.limit,hopkins:ir.lfs,hopkins:sb.ir.lfs,
younger:warm.ulirg.evol}.

Yet AGN obscuration remains poorly understood.  The most popular
models invoke a torus-shaped ``donut'' of obscuring material, on
scales anywhere from $\sim0.1-100\,$pc, to explain most of the heavily
obscured AGN population \citep{antonucci:agn.unification.review,
lawrence:receding.torus}.  If one empirically assumes unification of
obscured and unobscured AGN, then a number of the properties of the
torus can be inferred: scale radii somewhere in the range above, and
scale heights $h/R$ of order $\sim1/3$
\citep{risaliti:seyfert.2.nh.distrib}.  The observed distributions of
quasar and AGN column densities, and their detailed SED properties,
place strong constraints on the densities, structure, and column
densities within the obscuring material, with typical column densities
as high as $\sim10^{26}\,{\rm cm^{-2}}$ through the edge-on plane of
the material.  And direct observations are beginning to probe these
scales, through combinations of diverse techniques such as adaptive
optics and maser observations
\citep{greenhill:circinus.acc.disk,jaffe:ngc1068.torus.properties,
mason:ngc1068.torus.obs,sanchez:circinus.torus.mass,
davies:3227.torus.mass.and.sfr,
krips:nuclear.disk.torus.obs.seyferts,davies:sfr.properties.in.torus,
hicks:obs.torus.properties,ramosalmeida:pc.scale.torus.emission}, 
giving constraints on the 
kinematics, gas and dynamical masses, and star formation rates 
at these scales. 
Indeed, this simple model of obscuration has proven successful 
at explaining a large number of AGN observables,
and the torus forms the basis of most models uniting Type 1 and 
Type 2 AGN.

These successes should not mask the fact that the torus remains a {\em
phenomenological} model. The simple ``donut'' picture is just a toy
model -- there are a large and growing number of un-ambiguous cases
where it fails, whether in predicting detailed radiative transfer
properties coming from the microphysical gas structure
\citep{mason:ngc1068.torus.obs,elitzur:torus.wind,mor:2009.torus.structure.from.fitting.obs},
or where the implied torus properties would involve bizarre radii and/or 
dust temperatures \citep{kuraszkiewicz:2000.hblr.nlagn,tran:hblr.2,
page:2004.xr.qso.strong.submm,
stevens:xray.qso.hosts,ramosalmeida:pc.scale.torus.emission}, 
or where it is simply clear that the dominant obscuration 
is isotropic, or time dependent, or comes from much larger scales \citep[e.g.\ those 
associated with circumnuclear starbursts and/or the host galaxy; see][]{soifer84b,
scoville86,sanders88:quasars,sanders88:warm.ulirgs,
zakamska:qso.hosts,liu:2009.z2.qso.hosts.mergers,
donley:2005.qso.xr.hosts,rigby:qso.hosts,schartmann:2005.torus.modeling,
hatziminaoglou:2009.torus.properties.inferred.obs,
rowanrobinson:xr.ir.comparison.of.torii,
martinez-sansigre:high.qsos.in.submm,
lagos:2011.agn.gal.orientation}.

Without a {\em physical} model for the origin and evolution of nuclear gas inflows, 
a large number of questions remain unanswered. 
Where do toroidal-like obscuring gas structures come from, in the first place? What determines 
their characteristic gas masses, radii, and structure? Why are such 
structures ubiquitous around AGN? Are they, in fact? 
It is also usually assumed that the torus is simply an 
obscuring ``bystander'' to the accretion event, or at most a passive fuel 
reservoir. But could the torus play some more 
critical role in the accretion process itself? 
A major long-standing puzzle is what drives and maintains the scale 
height of the torus -- simple thermal pressure would be lost to cooling 
in a time much shorter than the local dynamical time. 
A large number of torus properties have been attributed to feedback 
from either young stars or the BH accretion itself -- including the 
typical scale heights, clumping/phase structure, 
gaseous velocity dispersions, possible correlations between 
these quantities and star formation, and even the fueling rates onto the BH 
\citep[e.g.][and references therein]{wada:starburst.torus.model,
schartmann:2009.stellar.fb.effects.on.torus}. 
But it is important to recall that we do not yet understand 
the basic dynamics of gas and stars entirely in the absence of feedback!

There have been some attempts to address these from a 
physically motivated perspective, both in analytic 
and numerical work
\citep{kawakatu:disk.bhar.model,cattaneo:2005.mgr.agn.obsc,
elvis:outflow.model,hopkins:twostage.feedback,
elitzur:torus.wind,wada:torus.mol.gas.hydro.sims}.
However, analytic models are severely limited by the fact that 
the systems at these radii are highly non-linear, often 
chaotic, and not necessarily in steady state (with inflow, 
mass buildup, star formation, and feedback processes all 
competing). If one wishes to {\em simultaneously} follow 
the torus itself and the chaotic, non-symmetric gas inflows that 
form it in the first place, simulations are necessary. 
But simulations of galactic scales used to follow inflows 
and AGN obscuration have resolution of $\sim100\,$pc, much 
larger than the relevant scales here \citep{cattaneo:2005.mgr.agn.obsc,
hopkins:lifetimes.methods,hopkins:lifetimes.obscuration}. 

Although progress has been made with zoom-in refinement techniques
\citep[see e.g.][]{escala:nuclear.gas.transport.to.msigma,
levine2008:nuclear.zoom,mayer:bh.binary.sph.zoom.sim}, the
computational expense involved means that these simulations have, thus
far, only barely probed that scales of interest and, in doing so, have
made restrictive assumptions (typically explicitly turning off cooling
and/or star formation on small scales); moreover they provide only a
snapshot at one instant from the parent simulation -- they cannot
survey {\em statistical} properties of the nuclear region.
Alternatively some simulations have simply adopted an assumed
small-scale structure as an initial condition and studied the resulting
gas dynamics at small radii \citep[e.g.][]{
schartmann:2009.stellar.fb.effects.on.torus,
wada:starburst.torus.model,wada:torus.mol.gas.hydro.sims}.  A number
of important conclusions have been drawn from these studies; however,
they not only bypass the question of the obscuring material origin,
but also have thus far adopted idealized potentials, without live star
formation and/or self-gravity of the gas.  As such, the appearance and
evolution of gravitational modes is suppressed.
\citet{cuadra:binary.bh.mergers.w.gas.disks} and \citet{fukuda:nuclear.ring} show (albeit in similar idealized 
studies that neglected star formation and stellar feedback) that when included, 
gravitational torques from self-gravity are an order-of-magnitude stronger than hydrodynamic torques 
from pressure forces or viscosity; the same conclusions have been reached for 
intermediate ($\gtrsim100\,$pc-scale) bars in a large number of 
hydrodynamic simulations \citep{noguchi:merger.induced.bars.dissipationless,
noguchi:merger.induced.bars.gas.forcing,
hernquist.89,barnes.hernquist.91,
barneshernquist96,hopkins:disk.survival}, 
and follow from analytic arguments \citep[see references above 
and][]{rice:maximum.viscous.alpha,hopkins:inflow.analytics}. 

Recently, 
to understand the angular momentum transport required for massive BH
growth, we have carried out a series of numerical simulations
of inflow from galactic to BH scales
\citep{hopkins:zoom.sims}.\footnote{\label{foot:url}Movies of these
  simulations are available at \movieurltwo} By re-simulating the
central regions of galaxies, gas flows can be followed from galactic
scales of $\sim100\,$kpc to much smaller radii, with an ultimate
spatial resolution $<0.1\,$pc.  For sufficiently gas-rich disky
systems, gas inflow continues all the way to $\lesssim 0.1$ pc.  Near
the radius of influence of the BH, the systems become unstable to the
formation of lopsided, eccentric gas+stellar disks.  This eccentric
pattern is the dominant mechanism of angular momentum transport at
$\lesssim 10$ pc, and can lead to accretion rates as high as
$\sim10\,\msun\,{\rm yr^{-1}}$, sufficient to fuel the most luminous
quasars.  In addition, through this process, some of the gas
continuously turns into stars and builds up a nuclear stellar disk.
Relics of these stellar disks may be evident around BHs in nearby 
galaxies \citep{hopkins:m31.disk}, 
such as M31 and NGC4486b
\citep{lauer93,tremaine:m31.nuclear.disk.model,bender:m31.nuclear.disk.obs,
lauer:ngc4486b,lauer:centers,
  houghton:ngc1399.nuclear.disk,
  thatte:m83.double.nucleus,debattista:vcc128.binary.nucleus}. 
In this paper, we examine the possibility that the disk that drives accretion 
and accounts for these stellar relics, in its gas-rich phase, may in fact 
{\em be} the canonical torus-like obscuration region near AGN. 
If correct, this implies both an a priori understanding of torus formation 
and structure, and an entirely new paradigm in which to view the 
nature of AGN obscuration. 

Specifically, we here perform a first comparison of these hydrodynamic simulations 
with the observed properties of AGN obscuration. 
We focus on dynamical properties and quantities such as the column 
density distribution that can be robustly predicted without reference to 
higher-order radiative transfer effects (which will be investigated in future work). 
In \S~\ref{sec:sims}, we summarize the properties of the numerical 
simulations, and in \S~\ref{sec:formation} show how they naturally form torus-like 
obscuring structures. In \S~\ref{sec:vertical}, we consider the scale heights 
and vertical structure of these torii, and examine how this can 
arise independent of stellar feedback from various gravitational processes.
In \S~\ref{sec:dynamical}, we compare a number of observable dynamical 
properties of the predicted torii to nuclear-scale observations of AGN. 
We then in \S~\ref{sec:subres} consider the full column density distribution, and 
in particular how it depends on sub-grid assumptions about the clumpiness of the 
ISM phase structure on un-resolved scales. 
We use this in \S~\ref{sec:obscured.fraction} to consider the predicted obscured 
fractions as a function of AGN and galaxy properties. 
Finally, we summarize our conclusions and discuss observational tests and 
future work in \S~\ref{sec:discussion}.

\section{The Simulations}
\label{sec:sims}

The simulations described here are from a survey of multi-scale 
``zoom-in'' runs which model gas inflows and star formation from 
large galactic scales to sub-pc scales, and have been discussed in a series of papers 
\citep{hopkins:zoom.sims,hopkins:inflow.analytics,hopkins:m31.disk,
hopkins:slow.modes,hopkins:cusp.slopes}. A detailed description 
and list of simulations 
is presented in \citet{hopkins:zoom.sims}; we briefly summarize the salient properties here. 

The simulations were performed with the TreeSPH code {\small
GADGET-3} \citep{springel:gadget}; the detailed numerical methods are 
described there and in \citet{springel:entropy,springel:models}. 
The simulations include collisionless stellar disks and bulges, dark
matter halos, gas, and BHs.  For this study, we are interested in
isolating the physics of gas inflow. As a result, we do not include
explicit models for BH accretion 
feedback -- the BH's only dynamical role is via its gravitational influence on
scales $\lesssim 10\,$pc.

Because of the large dynamic range in both space and time needed
for the self-consistent simulation of galactic inflows and nuclear
disk formation, we use a ``zoom-in'' re-simulation approach.  This
begins with a large suite of simulations of galaxy-galaxy mergers, and
isolated bar-(un)stable disks.  These simulations have
$0.5\times10^{6}$ particles, corresponding to a spatial resolution of
$50\,$pc. These simulations have been described in a series of previous papers 
\citep{dimatteo:msigma,robertson:msigma.evolution,
robertson:fp,robertson:disk.formation,
cox:kinematics,younger:minor.mergers,hopkins:disk.survival}.
From this library of simulations, we select representative simulations
of gas-rich major mergers of Milky-Way mass galaxies (baryonic mass
$10^{11}\,\msun$), and their isolated but bar-unstable analogues, to
provide the basis for our re-simulations. 
The dynamics on smaller scales does not depend critically on the
details of the larger-scale dynamics.  Rather, the small-scale
dynamics depends primarily on global parameters of the system, such as
the total gas mass channeled to the center relative to the
pre-existing bulge mass.

Following gas down to the BH accretion disk requires much higher
spatial resolution than is present in the galaxy-scale simulations. We
begin by selecting snapshots from the galaxy-scale simulations at key
epochs. In each, we isolate the central $0.1-1$\,kpc region, which contains
most of the gas that has been driven in from large scales.  Typically
this is about $10^{10}\,\msun$ of gas, concentrated in a roughly
exponential profile with a scale length of $\sim0.3-0.5\,$kpc. 
From this mass distribution, we then re-populate the gas in the
central regions at much higher resolution, and simulate the dynamics
for several local dynamical times.  These simulations involve $10^{6}$
particles, with a resolution of a few pc and particle masses of
$\approx 10^{4}\,\msun$.  We have run $\sim50$ such re-simulations,
corresponding to variations in the global system properties, the model
of star formation and feedback, and the exact time in the larger-scale
dynamics at which the re-simulation occurs.
\citet{hopkins:zoom.sims} present a number of tests of this
re-simulation approach and show that it is reasonably robust for this
problem.  This is largely because, for gas-rich disky systems, the
central $\sim 300$ pc becomes strongly self-gravitating, generating
instabilities that dominate the subsequent dynamics.

These initial re-simulations capture the dynamics down to $\sim 10$
pc, still insufficient to quantitatively describe accretion onto a
central BH.  We thus repeat our re-simulation process once more, using
the central $\sim10-30\,$pc of the first re-simulations to initialize
a new set of even smaller-scale simulations.  These typically have
$\sim10^{6}-10^{7}$ particles,
a spatial resolution of $0.1\,$pc, and a particle mass
$\approx100\,\msun$. We carried out $\sim50$ such simulations to test
the robustness of our conclusions and survey the parameter space of
galaxy properties.  These final re-simulations are evolved for
$\sim10^{7}$ years -- many dynamical times at $0.1$\,pc, but
short relative to the dynamical times of the larger-scale parent
simulations. We also carried out a few extremely
  high-resolution intermediate-scale simulations, which include
  $\sim5\times10^{7}$ particles and resolve structure from $\sim$ kpc
  to $\sim0.3\,$pc -- these are slightly less high-resolution than the
  net effect of our two zoom-ins, but they obviate the need for a
  second zoom-in iteration 
  and ``bridge'' the scales of the above simulation suites. 
  The conclusions from these higher
  resolution simulations are identical.

Our simulations include gas cooling and star formation, with gas
forming stars at a rate motivated by the observed \citet{kennicutt98}
relation. Specifically, we use a star formation rate per unit volume
$\dot \rho_{\ast} \propto \rho^{3/2}$ with the normalization chosen so
that a Milky-way like galaxy has a total star formation rate of about
$1\,M_{\sun} \, {\rm yr^{-1}}$.
Varying the exact slope or normalization of this relation has no qualitative
effect on our conclusions.  However, we caution that since we do not 
resolve the scales of individual bound star-forming cores in these simulations, 
the star formation is probably more uniform over the small radii than it would 
be in a more realistic ISM model. This is unlikely to be important for 
global properties here, but may have important consequences for e.g.\ detailed 
radiative transfer effects.

Because we cannot resolve the detailed processes of supernovae
explosions, stellar winds, and radiative feedback, the effect of feedback from stars
is crudely modeled with an effective equation of state 
\citep{springel:multiphase}. In this approach, feedback is assumed to
generate a non-thermal (turbulent, in reality) sound speed that
depends on the local star formation rate, and thus the gas density. 
\citet{hopkins:zoom.sims} describe in detail the effects of different subgrid ISM sound speeds on angular momentum transport and inflow rates, and argue that observations favor effective 
turbulent speeds of $\sim 10-50 \ {\rm km\,s^{-1}}$ for densities $\sim 1-10^5$ cm$^{-3}$, respectively.
But because the real physics and their effects are uncertain, it is important to 
vary this prescription and determine which of our conclusions are sensitive to 
the assumed sub-grid properties.

Within the context of this model, we can interpolate between two
extremes using a parameter $\qeos$. At one end, the gas has an
effective sound speed of $10 \, {\rm km\,s^{-1}}$, motivated by, e.g.,
the observed turbulent velocity in atomic gas in nearby spirals or the
sound speed of low density photo-ionized gas; this is the
``no-feedback'' case with $\qeos=0$.\footnote{This is still a 
non-trivial dispersion at large radii in galaxy disks. At the scales 
we focus on here, however, this corresponds 
to sounds speeds far below the circular velocity, 
and Jeans masses $\sim100\,\msun$, our resolution limit. As such, 
allowing cooling to even lower temperatures $=10\,$K makes 
no difference beyond the $\qeos=0$ case.}
This is broadly similar to what is assumed in 
\citet{bournaud:disk.clumps.to.bulge,
teyssier:2010.clumpy.sb.in.mergers}.
The opposite extreme, $\qeos=1$,
represents the ``maximal feedback'' model of \citet{springel:models};
in this case, $100\%$ of the energy from supernovae is assumed to stir
up the ISM.  This equation of state is substantially stiffer, with
effective sound speeds as high as $\sim200\,{\rm km\,s^{-1}}$. 
This is qualitatively similar to the near-adiabatic equations-of-state 
in the BH accretion studies of \citet{mayer:bh.binary.sph.zoom.sim,
dotti:bh.binary.inspiral}.
The sound speed at scales we consider cannot meaningfully be much larger than 
this, since it is similar to the circular/escape velocity. 
By varying $\qeos$, we examine a spectrum of intermediate cases: for example, 
equations of state similar to the ``starburst'' model in \citet{klessen:2007.imf.from.turbulence} 
or the sub-GMC equation of state in \citet{spaans:2005.gmc.eos}. 
Most of our suite of simulations focuses on a wide range of 
sub-grid sound speeds $\sim20-100\,{\rm km\,s^{-1}}$, motivated
by a variety of observations of dense, star forming regions both
locally and at high redshift \citep{downes.solomon:ulirgs,
  bryant.scoville:ulirgs.co,forsterschreiber:z2.disk.turbulence,
  iono:ngc6240.nuclear.gas.huge.turbulence}, and recent 
  numerical simulations \citep{hopkins:rad.pressure.sf.fb}.

Within this range, we found little difference in the physics of
angular momentum transport or in the resulting accretion rates, gas
masses, etc.\ on the scales we consider \citep{hopkins:zoom.sims}. 
More detailed comparison with the explicit stellar feedback models presented in 
\citet{hopkins:rad.pressure.sf.fb,hopkins:fb.ism.prop,hopkins:stellar.fb.winds} 
will be the subject of future work.
Here, we will focus on the effects on the obscuring gas near the BH. 
Because we are not explicitly accounting for or resolving feedback 
processes, we do not expect these models to accurately reflect the detailed 
dynamics of gas in response to strong feedback. Rather, we wish to use 
our suite of simulations to identify behavior that is robust to the effective 
pressure or turbulent sound speed of the gas -- i.e.\ to identify robust 
aspects of the system that are present even {\em without} feedback such as 
stellar winds.

\section{Formation of The Torus}
\label{sec:formation}

\citet{hopkins:zoom.sims} show that when large-scale inflows 
are sufficient, the buildup of gas in the central regions of the galaxy 
triggers a cascade of secondary instabilities, that drive rapid inflows 
to still smaller radii and ultimately onto the BH. Around the BH radius of influence, 
these instabilities generically take the form of an $m=1$ mode 
-- a thick, eccentric, slowly precessing gas+stellar disk, in which the 
eccentric stellar pattern torques strongly on the gas, inducing shocks and inflows. 
The disk can then propagate gas inflows and the $m=1$ pattern down to 
small radii $\lesssim 0.1\,$pc, where it transitions to a traditional alpha-disk.
This should be generic to any quasi-Keplerian potential in a dissipative 
system with shocks \citep{hopkins:inflow.analytics,hopkins:slow.modes}. 

\begin{figure*}
    \centering
    \plotside{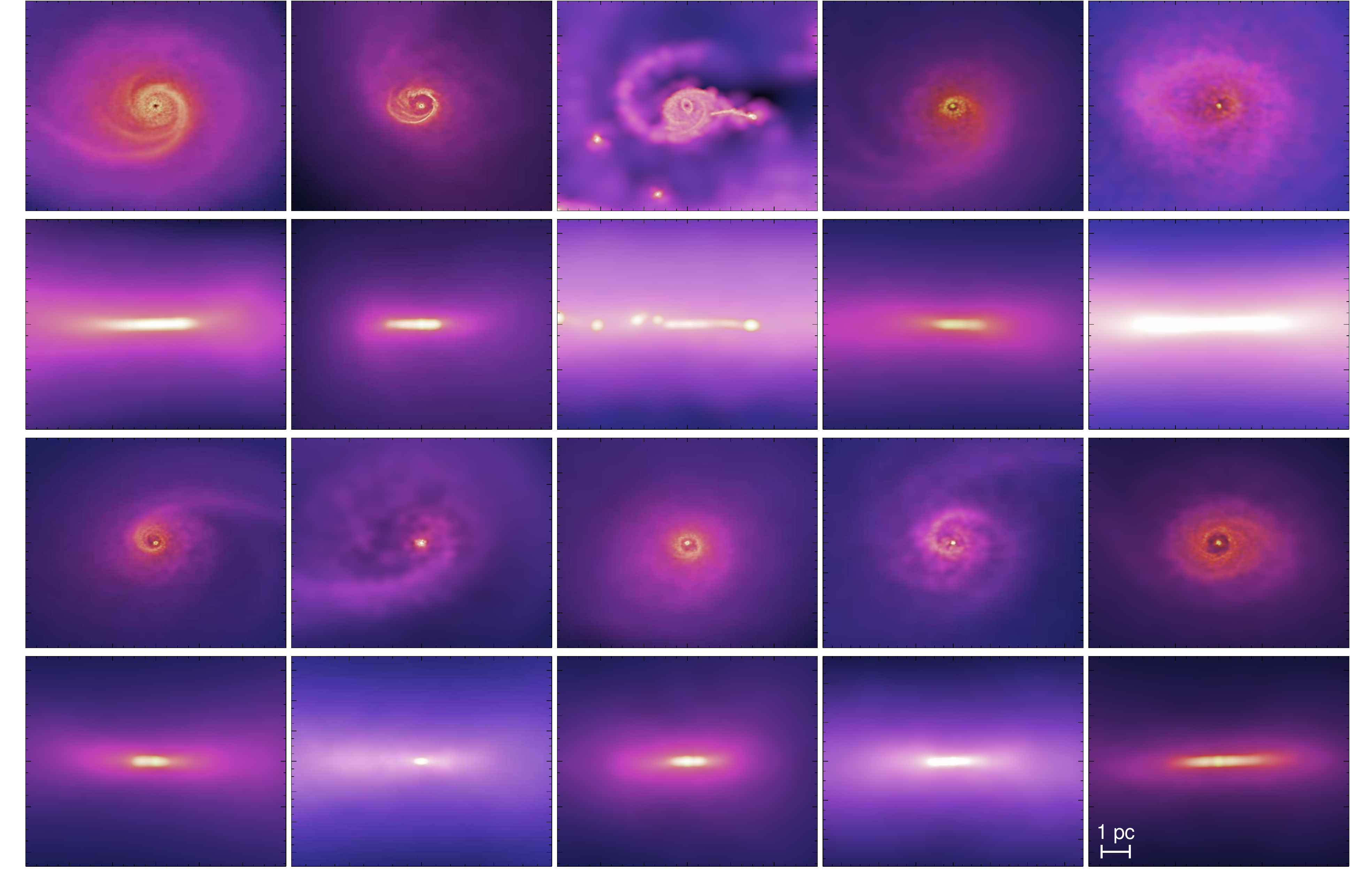}
    \caption{The face-on (x-y) and edge-on (x-z) disk structure of the nuclear disks 
    in several representative simulations. Scale is the same in all panels (lower right). 
    Each example is a simulation of the central $\sim100\,$pc of galaxy nuclei with 
    different initial large-scale galaxy properties (inflow-to-BH mass ratios, gas 
    fractions, and treatments of stellar feedback; details in text).
    Intensity encodes gas surface density (increasing from 
    $N_{H}\lesssim10^{21}\,{\rm cm^{-2}}$ to $N_{H}\gtrsim10^{25}\,{\rm cm^{-2}}$. 
    Colors encode the absolute star formation rate of the gas (increasing from blue to red/yellow).
    The formation of a lopsided, gas-rich disk is ubiquitous. 
    Regions where gas shocks (edges in this image) dissipate energy,
    leading to rapid gas inflow. Viewed edge-on, the disks are all thick, with 
    columns $\gtrsim 10^{22}\,{\rm cm^{-2}}$ to $h/R\sim$unity.
    \label{fig:torus.representative}}
\end{figure*}


\begin{figure*}
    \centering
    \plotside{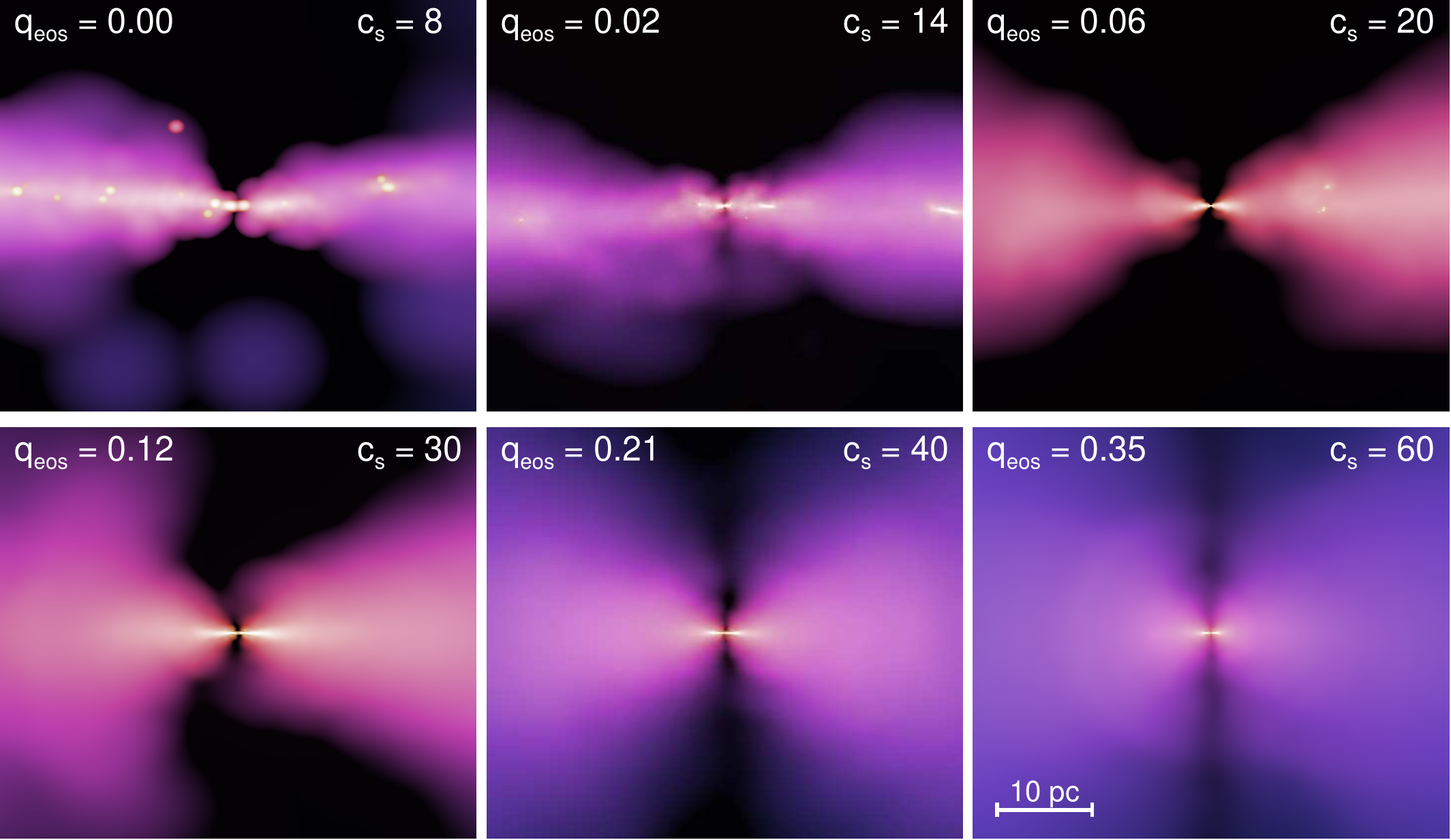}
    \caption{Edge-on gas mass distribution plotted in cylindrical coordinates $(R,\,z)$ 
    to highlight the torus-like structure of the disk, with intensity and color as 
    Figure~\ref{fig:torus.representative}. The simulations shown here 
    are a survey of $q_{\rm eos}$, which determines the effective (sub-grid) 
    pressure support of the ISM: the (mass-averaged) effective sub-grid sound 
    speed $c_{\rm s}$ is labeled in each panel. 
    Figure~\ref{fig:torus.representative}, for some of our survey of $q_{\rm eos}$. 
    As expected, the systems become more puffed-up with increasing 
    $q_{\rm eos}$ (sub-grid $c_{s}$), and for $q_{\rm eos}\gtrsim0.4$ they are 
    nearly spherical. But at small $q_{\rm eos}$, the scale heights do not 
    decrease as rapidly as $\propto c_{s}$, but approach some asymptotic 
    minimum. 
    \label{fig:edgeon.qtest}}
\end{figure*}

Figure \ref{fig:torus.representative} 
shows some illustrative examples of the 
nuclear gas disks that form around the BH radius of influence in our simulations. 
We plot gas surface density maps, with color
encoding the gas effective sound speed, from scales of $\gtrsim10\,$kpc
to $<1\,$pc.  The initial large-scale simulation in this case is a fairly gas-rich major
merger of two $\sim L_{\ast}$ galaxies (with initial bulges of mass 
$1/3$ the disk mass and BHs of mass $10^{7}\,\msun$). 
The zoom-in 
simulations were carried out just after the coalescence of the two
nuclei, which is near the peak of star formation activity, but when
the system is still quite gas rich.\footnote{The specific properties of 
each simulation are given in \citet{hopkins:zoom.sims}, 
those shown here are (top-to-bottom, left-to-right): 
Nf8h1c0thin, Nf8h1c1thin, Nf8h1c1qs, Nf8h1c1dens, Nf8h1c0 ({\em left}); 
Nf8h1c1ICs, Nf3h1c1mid, Nf2h2b2, Nf8h2b2, Nf8h2b4 ({\em right}). 
They have (respectively) 
initial gas
fractions $f_{\rm gas}\sim 0.5,\,0.6,\,0.8,\,0.8,\,0.8,\,0.75,\,0.26,\,0.20,\,0.8,\,0.8$; 
BH mass $\sim3\times10^{7}\,\msun$ and disk mass 
$\sim1.2,1.7,3.0,8.1,0.25,1.7,4.6,7.0,3.5,0.5\,\times10^{7}\,\msun$ inside $10\,$pc, 
and sub-grid sound speeds 
$c_{s}\sim35,\,20,\,40,\,50,\,10,\,40,\,30,\,25,\,25,\,20\,\,{\rm km\,s^{-1}}$.}

We both show the global structure, and 
edge-on $(R,\,z)$ disk.
The scales shown include the BH radius of influence, about $10\,$pc 
in these galaxies. 
In the face-on projection, the $m=1$ modes that form at these scales are clearly evident. 
They drive large torques on the gas, driving inflow into $\ll 0.1\,$pc 
at accretion rates as high as $10\,\msun\,{\rm yr^{-1}}$ in these simulations, sufficient to power 
the most luminous quasars (see Figures~5 \&\ 13 in \citet{hopkins:zoom.sims}).

Here, however, we note the broad resemblance of these nuclear disks 
to the canonical AGN ``torus.''
The disks are thick, with characteristic scale $\sim0.1-10\,$pc, 
gas masses $\sim M_{\rm BH}$, and scale heights of order unity.
Of course, unlike in toy models of the torus, the gas is part of a continuous 
distribution at all radii, and its structure is non-trivial.

\section{Vertical Structure: Dependence on Stellar Feedback}
\label{sec:vertical}

\subsection{Overview}
\label{sec:vertical:overview}

The major input parameter of our models is the parameterization of the effects of 
stellar feedback on the ISM. This is accomplished, here, with the parameter 
$q_{\rm eos}$ described in \S~\ref{sec:sims}, that allows us to interpolate between 
a feedback-free ISM and one with large 
non-thermal internal gas velocities and pressures driven by stellar feedback. 

The so-called torus is defined largely by its vertical structure, which 
determines the obscured fractions. To the extent that 
the amount of turbulent velocity and pressure support in the simulated gas is 
defined by a sub-resolution model, we must ask whether the 
vertical structure we see in our simulations 
is entirely a consequence of our model inputs, 
or whether there are robust statements and predictions we can make. 

We therefore consider the vertical structure in detail,
in a specific survey of $q_{\rm eos}$. 
This survey ({\bf Nf8h2b4q} in \citealt{hopkins:zoom.sims}) is a typical, canonical set of 
conditions ($3\times10^{7}\,\msun$ BH, with disk-to-BH mass ratio of 
a few inside $\sim100\,$pc initially, and initial gas fraction $\sim50\%$, 
typical of the simulations in Figure~\ref{fig:torus.representative}). 
We re-simulate the identical cases, but 
with $q_{\rm eos}=0.0$, $0.018$,  $0.06$, $0.10$, $0.12$, $0.15$, $0.21$, $0.25$, 
$0.35$, $0.60$, $1.0$. The spacing in $q_{\rm eos}$ is chosen such that 
the implied turbulent gas sound speeds are spaced over roughly equal 
logarithmic intervals from the minimum $q_{\rm eos}=0$ floor 
($10\,{\rm km\,s^{-1}}$) to the maximum $q_{\rm eos}=1$ value (which is 
density dependent, but $\sim100\,{\rm km\,s^{-1}}$ at range of interest).

Figure~\ref{fig:edgeon.qtest} shows the edge-on ($R,\,z$) gas structure, as a 
function of $q_{\rm eos}$. A few generic features stand out. 
The disks are generally thick.
At the smallest radii ($\lesssim0.1-1\,$pc), they eventually become thin, 
since the gravity from the BH becomes arbitrarily strong. 
This gives a torus-like morphology. Flares (discussed below) and lopsidedness 
(reflecting the lopsided disk mode on these scales) are not uncommon. 
As a function of $q_{\rm eos}$, we see unsurprisingly that the gas 
distribution becomes more smooth and vertically extended at higher $q_{\rm eos}$. 
For $q_{\rm eos}\gtrsim0.4$, the system is no longer really a vertically supported disk, 
but spherical -- however, as discussed in \citet{hopkins:zoom.sims}, this is likely an 
unrealistically large implicit feedback efficiency. 

The most surprising thing about Figure~\ref{fig:edgeon.qtest} is how little 
change there is as a function of $q_{\rm eos}$. For $q_{\rm eos}\sim0-0.35$, 
there is a factor of $\sim5$ change in $c_{s}$, which leads to a naive expectation of a 
factor of $\sim5-25$ change in $h/R$. We see much weaker variation.
We now consider this quantitatively.

\subsection{General Expectations}
\label{sec:vertical:general}

To inform our comparisons, consider a simple smooth, isothermal system, 
in which the self-gravity of the gas is negligible (i.e.\ the potential is dominated 
by the BH, stars, and/or dark matter). 
The equation of vertical hydrostatic equilibrium
\be
\frac{\partial P}{\partial z}=c_{s}^{2}\,\frac{1}{\rho}\,\frac{\partial \rho}{\partial z} 
= - \frac{\partial \Phi}{\partial z}
\ee
then has the trivial solution
\be
\rho(R,z) = \rho_{0}(R)\,\exp{ \left\{  c_{s}^{-2}\, [ \Phi(R,0) - \Phi(R,z) ] \right\} }\ .
\ee
For large $z/R$, this depends on the specific form of 
$\Phi$, and so on the details of the global mass distribution. 
However, if the disk is thin, i.e.\ most of the mass is at $z/R\ll1$, 
then this has a particularly simple expression. 
For any background spherical mass distribution, we 
have $\partial\Phi/\partial z = (\partial\Phi/\partial r)\,(\partial r/\partial z) = 
(V_{c}^{2}/r)\,(z/r)$, where $r^{2}=R^{2}+z^{2}$ and 
$V_{c}^{2}=G\,M_{\rm enc}(<r)/r$.
So for $z\ll R$, 
$\Phi(R,0) - \Phi(R,z) \approx G\,M_{\rm enc}(<R)\,R^{-2}\,z/2 \approx 
V_{c}^{2}\,(z/R)^{2}/2$. 

Together, this gives the especially simple solution for the density 
for a quasi-spherical potential: 
\begin{align}
\label{eqn:rhovsz}
\rho(R,z) & \approx 
\rho_{0}(R)\,\exp{ \left\{  -\frac{1}{2}\,\left( \frac{z}{h_{s}}\right)^{2} \right\} } \\ 
\label{eqn:scaleheight}
\frac{h_{s}}{R} & \equiv \frac{c_{s}}{V_{c}} = c_{s}\,\left( \frac{G\,M_{\rm enc}(<R)}{R} \right)^{-1/2} 
\end{align}

Of course, the $c_{s}$ here does not need to be thermal. Non-thermal pressure 
sources such as turbulent motions will have the same effect. So for comparison with simulations we  
should take $c_{s}\rightarrow c_{z,\,\rm eff}$, where $c_{z,\,\rm eff}^{2}=c_{s}^{2}+\sigma_{z}^{2}$ 
includes both the thermal and/or sub-resolution effective sound speed ($c_{s}$) and 
resolved turbulent vertical motions ($\sigma_{z}$). 

\begin{figure*}
    \centering
    \plotside{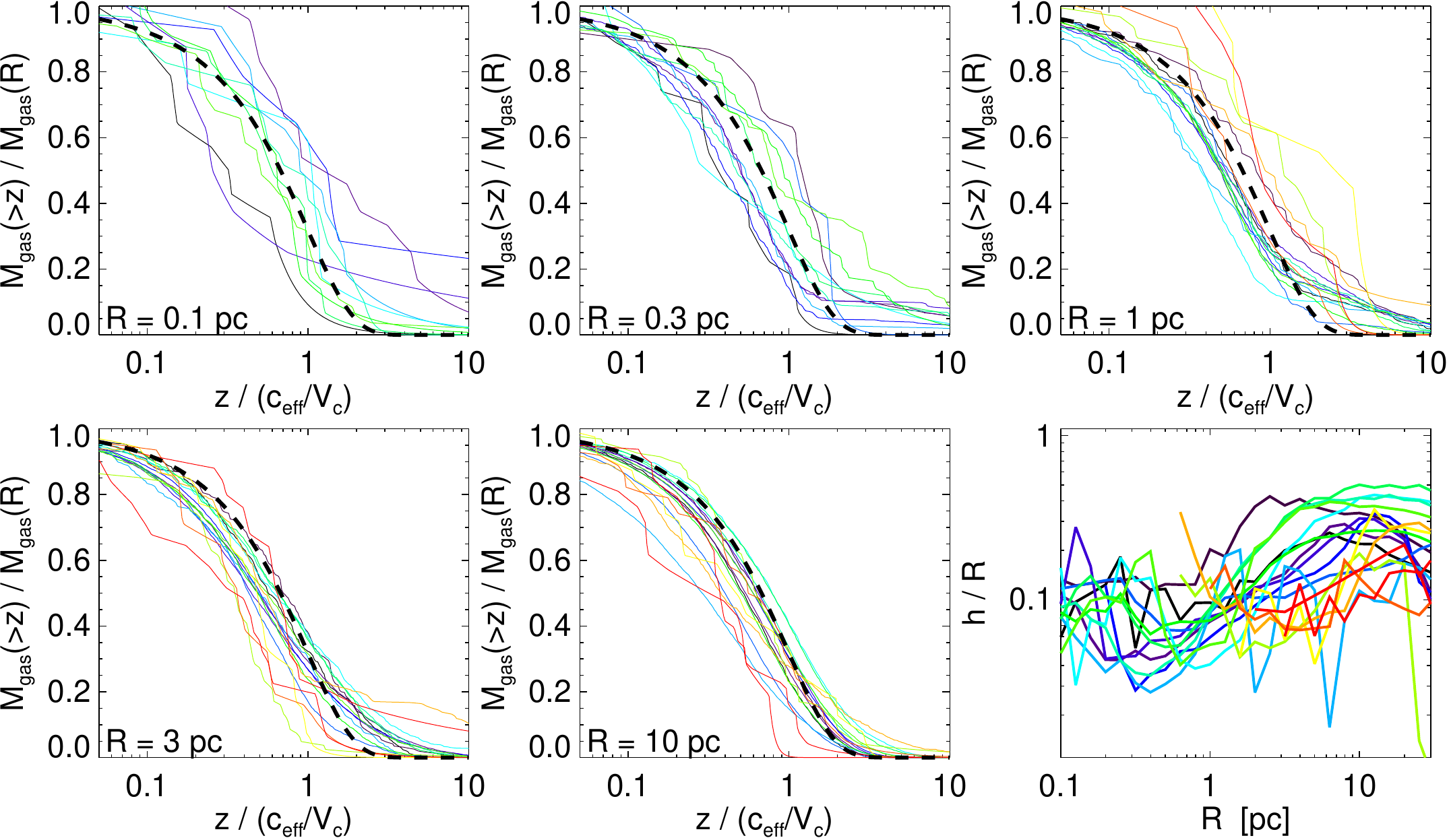}
    \caption{Vertical gas mass distribution at different radii. 
    Panels show the integrated gas mass fraction ($M_{\rm gas}(>|z|)/M_{\rm gas}$) 
    above a given height $|z|$, for gas in a narrow radial annulus about $R$ (each 
    $R$ labeled). The heights are normalized to the expected scale height 
    $z_{0} = c_{\rm eff}/V_{c}$, where $c_{\rm eff}=(c_{s}^{2}+\sigma_{z}^{2})^{1/2}$, 
    where $c_{s}$ is the sub-grid velocity dispersion (plus thermal sound speed) 
    and $\sigma_{z}$ is the resolved gas velocity dispersion. 
    Each line is a different simulation (with varied initial gas fraction, disk and BH 
    mass, sub-grid equation of state, and star formation laws); 
    shown at a randomly-chosen time near the peak of the inflow onto the BH 
    (but behavior is similar over the entire duration of the simulations). 
    Thick dashed black line is the simple Gaussian expectation for an 
    isothermal gas disk with weak self-gravity in vertical equilibrium 
    (Equation~\ref{eqn:rhovsz}). 
    {\em Bottom Right:} The scale height $h$ (best-fit dispersion $z_{0}$, fitting 
    the vertical gas distribution at each radius to a Gaussian), as a function of 
    radius, for each simulation. The scale heights are significant, and 
    the vertical behavior approximately follows the linear expectation at these 
    radii, if the full vertical dispersions are included. 
    \label{fig:torus.z.profile}}
\end{figure*}

Figure~\ref{fig:torus.z.profile} compares this expectation for 
$\rho(z)$ as a function of $c_{z,\,\rm eff}/V_{c}$ 
to the actual vertical mass distribution measured in narrow radial annuli from 
$\sim1-10\,$pc. We use the full $c_{z,\,\rm eff}$ as defined above.
The distributions are reasonably described by the above scalings, 
A gaussian core is typical, with a slightly broader (often more 
exponential) distribution at high-$z$. Remember that at sufficiently 
large $|z|$, the correct solution involves the full potential; if we account 
for this more accurately, we see similar agreement. The important point 
is that the gas does appear to be in vertical equilibrium.

\subsection{Gravitational Support}
\label{sec:vertical:gravitational}

Given the gas dispersion $c_{z,\,\rm eff}$ in the simulations, the vertical structure 
is what we would expect. 
But are these dispersions primarily sub-resolution (set by the model), 
thermal, or gravitational in origin? 

\begin{figure*}
    \centering
    \plotside{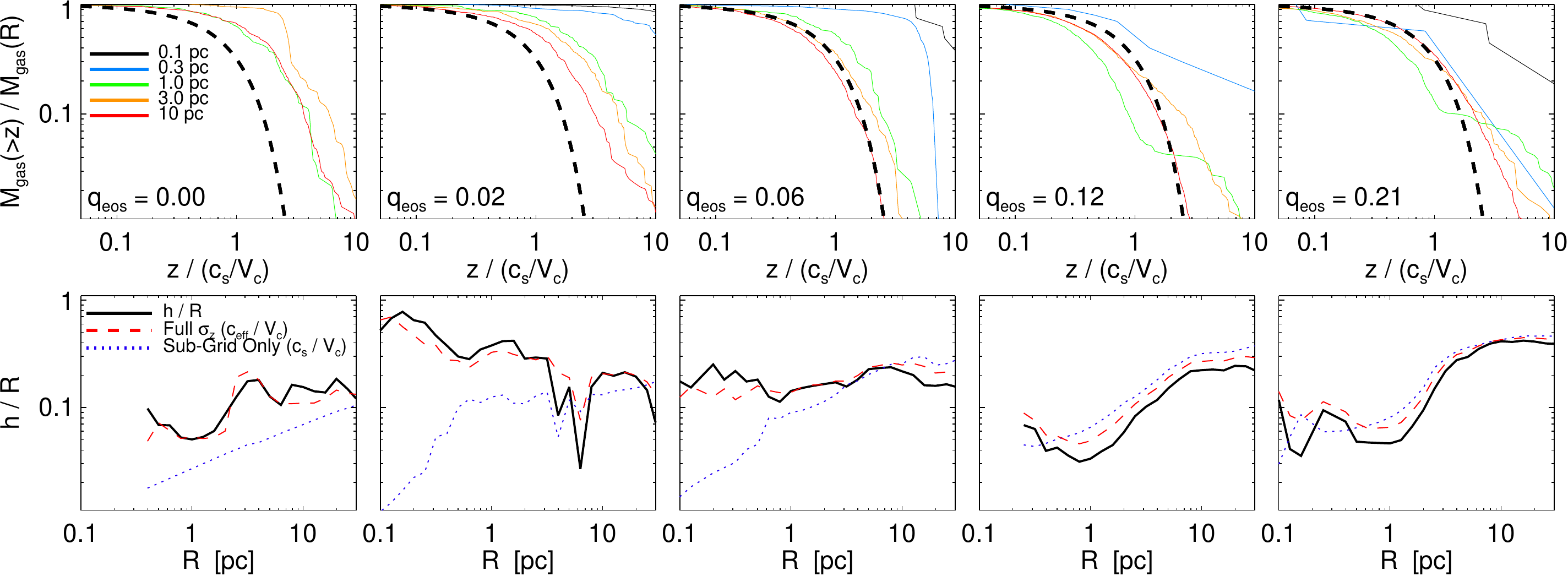}
    \caption{{\em Top:} Vertical gas distribution, in our 
    survey of $q_{\rm eos}$. Each panel shows a different simulation in the 
    $q_{\rm eos}$ survey. Each line shows the vertical gas distribution at a different 
    radius (as labeled). Here, the x-axis is scaled {\em only} by $c_{s}/V_{c}$, 
    the scale height expected if the sub-resolution (feedback-driven) 
    velocity dispersions were dominant (as compared to $c_{\rm eff}$). 
    {\em Bottom:} Gas scale height $h$ versus radius (solid black lines). 
    Dotted blue lines show the expected height if just the subgrid velocities 
    were present. 
    Dashed red lines compare the expected height including the full 
    resolved dispersion $c_{\rm eff}$. In large-$q_{\rm eos}$ systems, 
    the implicit feedback dominates the vertical support. 
    But the scale heights in low-$q_{\rm eos}$ systems are supported by large 
    {\em resolved} turbulent vertical velocities, despite the lack of feedback. 
    There is some non-feedback dispersion source in these systems. 
    \label{fig:torus.z.profile.q}}
\end{figure*}

Figure~\ref{fig:torus.z.profile.q} again considers the vertical gas profile, 
but in our survey of different $q_{\rm eos}$. 
We know from Figure~\ref{fig:torus.z.profile} that accounting for the 
{\em full} gas motion explains the observed scale heights. Therefore here 
we compare the expectation if the gas motions were purely thermal and/or 
sub-resolution -- i.e.\ $c_{z\,\rm eff}=c_{s}$, where $c_{s}$ is the sound speed 
and is dominated by the sub-resolution turbulent effective $c_{s}$ (since the 
explicit cooling time of the gas is $\sim10^{4}$ times shorter than its dynamical time). 

In the higher-$q_{\rm eos}$ (higher effective $c_{s}$) simulations, this explains most 
of the pressure support, i.e.\ the resolved turbulent dispersion $\sigma_{z}\ll c_{s}(\rm sub-grid)$. 
But at low $q_{\rm eos}$, the scale heights and $c_{z,\,\rm eff}$ do not drop 
nearly as quickly as the sub-grid $c_{s}$ alone. There is some non-thermal, resolved 
gravitational process giving rise to minimum scale heights and vertical dispersions.

What, then, dominates the effective vertical ``heating'' in the torus region?

\subsubsection{Clump-Clump Encounters}
\label{sec:vertical:gravitational:clumps}

It has been proposed that two-body scattering between dense molecular clumps 
in the gas could maintain the observed scale heights \citep{krolik:clumpy.torii,
nayakshin:forced.stochastic.accretion.model,hobbs:turbulence.agn.feeding}.
However, we find these effects are negligible in our simulations. 

Consider clumps within the plane of a disk. Scattering a 
clump to large $V_{z}\sim V_{c}$ requires both (a) an encounter between two 
clumps with {\em relative} velocity $\gtrsim V_{c}$, and (b) an encounter 
within an impact parameter $b$ such that $G\,M_{\rm cl}/b\,V_{c}\gtrsim V_{c}$. 
The mean time per clump between such encounters is just  
$\tau^{-1}\sim f(V_{c}/\sigma)\,n_{\rm cl}\,b^{2}\,V_{c}$, 
where $n_{\rm cl}$ is the volume density of clumps and 
$f(V_{c}/\sigma)$ is the fraction of the clumps moving on orbits 
with large non-circular motions ($|V-V_{c}|\gtrsim V_{c}$). 
If the system is sufficiently thin such that $b>h$, the disk thickness, then 
this becomes $\tau^{-1}\sim f(V_{c}/\sigma)\,dN_{\rm cl}/dA\,b\,h\,V_{c}$. 
Using $n_{\rm cl}=\bar{\rho}_{\rm gas}/M_{\rm cl}=\Sigma_{\rm gas}/h\,M_{\rm cl}$, 
the maximum $b$ above, and $V_{c}^{2} \sim G\,M_{\rm enc}/r$, this can be written
\be
\Omega\,\tau \sim \frac{1}{f(V_{c}/\sigma)}\,\left( \frac{M_{\rm enc}}{M_{\rm gas}} \right)\,
\left( 1 + Q\,N_{\rm cl} \right)
\ee 
where $Q$ is the Toomre $Q \sim (h/R)\,(M_{\rm gas}/M_{\rm enc})^{-1}$ and 
$N_{\rm cl}$ the total number of clumps, and the expression shown interpolates 
between the extremely-thin and thick-disk cases. 
For a Maxwellian velocity distribution, $f(V_{c}/\sigma) \sim \exp{\{-(V_{c}/\sigma)^{2}/2\}}$. 
Since both $f(V_{c}/\sigma)$ and $M_{\rm gas}/M_{\rm enc} \sim M_{\rm gas}/M_{\rm BH}$ 
are small at this radius, collisions require many dynamical times. 
But any induced vertical heating will relax away in just a single or couple dynamical 
times, since the cooling time is much shorter than the dynamical time. 
So without continuous energy input to 
drive large dispersions -- which is essentially the problem we wished to solve in 
the first place -- this mechanism fails. 

Moreover, if star formation occurs with some efficiency relative to the dynamical 
time ($\dot{\rho}_{\ast} \sim \epsilon \rho\,\sqrt{G\,\rho}$, with $\epsilon\sim1-10\%$), 
then using the fact that any clump must have $\rho \gtrsim M_{\rm enc}/R^{3}$ to 
avoid tidal destruction, clump-clump {\em gas} heating must occur faster than the 
gas exhaustion timescale in a clump, requiring
\be 
\left( \frac{M_{\rm gas}}{M_{\rm enc}} \right) \gg \epsilon\,(1+Q\,N_{\rm cl})\,
\exp{\left\{\frac{1}{2}\left( \frac{V_{c}}{\sigma} \right)^{2} \right\}}
\ee
Even for $\sigma\sim V_{c}$ (which begs the question) and an extremely thin disk 
$Q\,N_{\rm cl}\lesssim1$, this requires $M_{\rm gas}/M_{\rm enc}\gg0.1$, 
which is not satisfied at the inner radii $\lesssim10\,$pc.

\subsubsection{Twists and Misalignment}
\label{sec:vertical:gravitational:twists}

\begin{figure}
    \centering
    \plotone{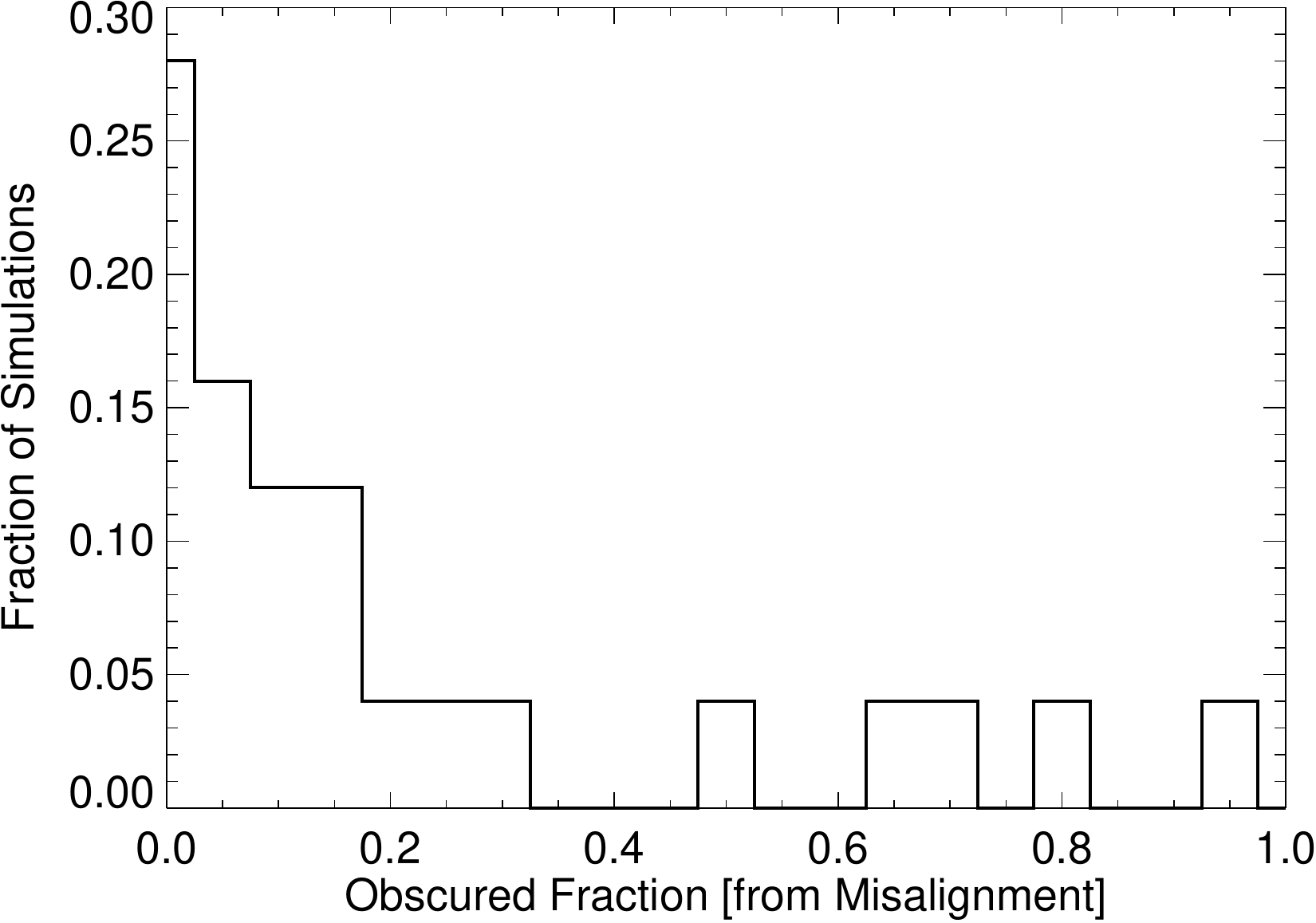}
    \caption{Contribution of disk mis-alignments and twists (as opposed to 
    disk {\em thickness}) to obscuration. For each simulation in the sample 
    shown in Figure~\ref{fig:torus.z.profile}, we calculate the (time-averaged) fraction of sightlines 
    towards the BH which would be obscured ($N_{H}>10^{22}\,{\rm cm^{-2}}$) assuming 
    the disk is razor-thin at every radial annulus $\Delta r$ (using the angular momentum vector of 
    gas in $\Delta r$ to define the disk plane). If the disk plane (angular momentum direction) 
    were constant with radius, this would be zero, but if the plane tilts as a function of radius, 
    it can be non-zero (a $180^{\circ}$ flip at some radius, 
    maintained for the duration of the simulation, would give an obscured fraction of unity). 
    Although there are significant mis-alignments \citep[see][]{hopkins:agn.alignment}, 
    and in a few systems they can account for obscured fractions $\gtrsim50\%$, 
    they would only give an integrated obscured fraction of $\sim25\%$ if the 
    disk were thin at all radii. And even simulated cases with no ``twists'' still have 
    large obscured fractions and $h/R$ in Figure~\ref{fig:torus.z.profile}.
    The torus must actually be thick to match observations -- so 
    some process must explain large scale-heights in the gas.
    \label{fig:alignment.obscuration}}
\end{figure}

Another possibility is that large covering factors are maintained by virtue of the fact 
that the nuclear disk is mis-aligned with the larger-scale inflow/bar/disk. 
This is particularly interesting because observations 
find relatively little correlation between the axes of AGN (traced by 
jets or the torus) and the inclination of the host galaxy 
(e.g.\ \citealt{keel:1980.seyfert.vs.galaxy.inclination,lawrence:1982.torus.alignment,
schmitt:1997.radio.alignment.w.host,simcoe:1997.agn.host.alignment,kinney:2000.bh.jet.directions,
gallimore:2006.agn.outflow.gal.alignment,zhang:2009.agn.vs.hubble.type}; but 
see also \citealt{maiolino:1995.seyfert.torus.alignment,shen:2010.torus.alignment} and 
references therein). 
In a companion paper, \citet{hopkins:agn.alignment}, we show that this lack of alignment 
is reproduced in our simulations owing to two processes. First, 
occasionally the central gas supply is strongly influenced by a single or couple large clumps 
that form at large radii, fragment and sink, 
realigning the central angular momentum vector \citep[see e.g.][]{nayakshin:forced.stochastic.accretion.model}. 
Examples of this have also been seen in cosmological zoom-in simulations \citep{levine:sim.mdot.pwrspectrum}. 
Second, even in smooth flows, the secondary and tertiary gravitational instabilities 
will tend to de-couple their angular momentum from the primary (external) bar/spiral 
structure, and semi-chaotically precess or tumble in three dimensions
\citep[see][]{heller:secondary.bar.instability,shlosman:nested.bar.evol,elzant:nested.bar.scales,
maciejewski:nested.bar.models,englmaier:nested.bar.decoupling}. 

In Figure~\ref{fig:alignment.obscuration} we show how this can contribute to 
obscuration. It is straightforward to measure the axis 
of angular momentum of the disk in a radial annulus, and define the corresponding inclination 
$\Theta(r)$ (relative to the initial, uniform angular momentum axis of the entire initial disk). 
We also know the mass $\Delta M_{\rm gas}(r)$ enclosed in each annulus; we 
can simply integrate along all sightlines towards the BH, assuming 
the mass is in an axisymmetric razor-thin disk with inclination $\Theta(r)$, to obtain the column 
density $\Sigma_{\rm gas}$ along each sightline. 
If a sightline is covered by the disk at some radius, it is ``obscured.''\footnote{Technically, 
we require a column that translates to $N_{H}>10^{22}\,{\rm cm^{-2}}$, but 
because of our razor-thin assumption, this is almost identical to being covered by the disk.}

We consider these assumptions because they effectively 
define a minimum obscured fraction stemming purely from twists and misalignments. 
This fraction can be considerable, but there is a broad range in different simulations -- 
many systems have only $<20\%$ covering fractions, but 
there is a long tail towards near full covering (anti-alignment of the central 
and outer disks). Integrated over all simulations and snapshots, the average covering fraction is $\approx25\%$. 
A warped or twisted disk can therefore yield large covering angles towards 
the BH even when the disk itself is thin.

However, this is not the full story. First of all, the covering fractions of $\approx25\%$ are 
still significantly lower than the total covering fraction of obscuration in the simulations, by a 
factor of at least $\sim2$. Moreover the cases with weak twists ($\ll20\%$ covering in 
Figure~\ref{fig:alignment.obscuration}) still exhibit large obscured fractions and thick disks. 
The key point is that the vertical density 
distribution in Figures~\ref{fig:torus.z.profile}-\ref{fig:torus.z.profile.q} shows that 
we must explain the actual {\em thickness}, not just the orientation of the disks. 
This is true for observations as well -- empirical modeling of the hot dust continua indicates 
that the obscuring region must be geometrically thick, not just a 
misaligned larger-scale thin disk \citep[e.g.][and references therein]{deo:2009.ir.seyfert.continuum}.
A time-dependent twist can, in principle, ``pump up'' vertical motions, but fast cooling times make it 
difficult to sustain a large scale height anywhere except close to the location of 
the twist (where the pumping occurs). Some mechanism that pumps vertical motion throughout
the disk, on a timescale comparable to the local dynamical time, is required.

\subsubsection{Bending Modes}
\label{sec:vertical:gravitational:bending}

Bending modes can provide an efficient channel 
for ``heating'' the torus. 
Their behavior is particularly interesting in response to 
``slow modes'' in a quasi-Keplerian potential. 
Consider a general bending mode 
\begin{align}
h(R,\phi,t) & =H(R)\,\exp{\left\{i\,\left(\omega_{b}\,t-m_{b}\,\phi\right) \right\}} \\  
H(r) & = h_{0}(R)\,\exp{\left\{i\,\int^{R}k_{b}(R^{\prime}){\rm d}R^{\prime} \right\}}
\end{align}
in a system that includes some 
quasi-spherical component (BH+bulge+halo) and a thin disk 
with surface density $\Sigma_{d}$, angular (vertical) frequency $\Omega$ 
($\nu$), 
and velocity dispersions in the radial, azimuthal, and vertical 
directions $\sigma_{r}$, $\sigma_{\phi}$, $\sigma_{z}$. 
The value $k_{b}$ is the radial wavenumber of the bending mode, and 
$m_{b}$ is its azimuthal wavenumber. 
In the WKB regime, if $\sigma_{r}^{2}\gtrsim\sigma_{\phi}^{2}$, 
the dispersion relation can be written 
\be
(\omega_{b}-m_{b}\,\Omega)^{2} = \nu^{2} + 2\pi\,G\,\Sigma_{d}\,|k_{b}| + 
(\sigma_{z}^{2}-\sigma_{r}^{2})\,k_{b}^{2}
\label{eqn:bending.dispersion}
\ee 
\citep{kulsrud:bending.modes.70,kulsrud:firehose.instability,
mark:slab.bending.instabilities,poliachenko:firehose.instability}.
\footnote{Note that the $\sigma_{r}^{2}$ that appears in 
Equation~\ref{eqn:bending.dispersion} is not technically a dispersion 
(that being defined $\langle v^{2} \rangle - \langle v \rangle^{2}$), but 
the mean $\langle v_{r}^{2} \rangle$. Thus streaming/bulk motion in the 
radial direction is affected just as much as random motions about some 
mean $v_{r}$ (important for our purposes, since gas parcels being collisional 
tend to move in coherent streaming motion).}

If $\sigma_{r}$ is sufficiently large, the system is vulnerable to the so-called 
``firehose'' instability and bending modes will be self-excited. 
However, it is unusual to see such large $\sigma_{r}$ (and 
$\sigma_{r}^{2}>\sigma_{\phi}^{2}$) in disks. Even with large $\sigma_{r}$, the fact 
that $\sigma_{r}<V_{c}(r)$ for any meaningful ``disk'' 
means that usually, when the self-gravity of the disk 
is small compared to the background potential, the system is stable. 
And in even in self-gravitating disks with large $\sigma_{r}$, it typically 
takes only a small $v_{z}$ to stabilize them, so the induced $h/R$ is 
not large. 

However, consider the special case of interest here, where the disk is quasi-Keplerian 
and has a large lopsided mode driving accretion. 
The system is (initially) a thin 
disk in the quasi-Keplerian potential of a BH -- i.e.\ to lowest order, the parameters 
are those of a pure Keplerian potential, with some correction terms 
that scale with $M_{d}/\mbh$ of $\mathcal{O}(\epsilon)\ll1$. 
As discussed above, and in previous works, the disk develops 
a gravitational instability {\em in the disk plane} (the standard density 
waves of spiral/bar/etc. modes), which we can describe by e.g.\ the perturbed density field 
$\Sigma_{1}(R,t)=|a|\,\Sigma_{0}(R)\,\exp{ \{i\,(\int^{R}k_{p}(R^{\prime}){\rm d}R^{\prime} + 
\omega_{p}\,t - m_{p}\,\phi  ) \}}$. Here $|a|$ is the effective mode amplitude in the 
density field at $R$, and the properties $\omega_{p}$, $m_{p}$, and 
$k_{p}$ refer to the frequencies and wavenumbers of this, in-plane mode 
(independent from the $\omega_{b}$, $m_{b}$, and $k_{b}$ of the bending mode). 
The fact that the potential is quasi-Keplerian, i.e.\ has 
$\Omega\approx\kappa$, favors (and supports for 
long periods of time) global, ``slow'' $m=1$ modes -- 
modes with $m_{p}=1$, $\omega_{p} \sim \epsilon\,\Omega \ll \Omega$, 
and $|k_{p}\,R|\sim1$. These are the lopsided/eccentric modes that 
we see above. 
The potential of the BH+disk system is 
\begin{equation}
\Phi(r) = -\frac{G\,\mbh}{r} + \Phi_{d}(r)
\end{equation}
and it is useful to define the parameter 
\begin{align}
\varomega &\equiv \frac{\Omega^{2}-\kappa^{2}}{2\,\Omega} 
= - \frac{1}{2\,\Omega}\,\left( \frac{2}{r}\,\frac{d}{dr} + \frac{d^{2}}{dr^{2}} \right)\Phi_{d}\ .
\end{align}

To first order in $\epsilon$, then, 
the WKB dispersion relation of such modes in a cold ($c_{s}\ll V_{c}$) disk is 
\be 
\omega_{p} = \varomega + \pi\,G\,\Sigma_{d}\,|k_{p}|\,\Omega^{-1}
\ee
\citep{tremaine:slow.keplerian.modes}.
The equations of motion for the perturbed 
velocity 
${\bf v} = R\,\Omega\,\hat{\phi} + v_{r}\,\hat{R} + v_{\phi}\,\hat{\phi}$ 
become, at this order, 
\begin{align}
v_{r} &= -\frac{i}{2\,(\omega_{p}-\varomega)}\,\left(  
\frac{{d}\Phi_{1}}{{d}r} + \frac{2\,\Phi_{1}}{r}\right) = 
-\frac{\Sigma_{1}}{\Sigma_{d}}\,\Omega\,|k_{p}|^{-1} \\ 
v_{\phi} &=\frac{i}{2}\,v_{r}
\end{align}
where we have used the WKB relation $\Phi_{1}\approx-2\pi\,G\,|k_{p}|^{-1}\,\Sigma_{1}$. 
Since $\sigma_{r}^{2}=\langle |v_{r}|^{2} \rangle$ and $V_{c}=\Omega\,R$, this becomes 
just 
\be
\sigma_{r} = |a|\,|k_{p}\,R|^{-1}\,V_{c}
\ee
with $\sigma_{r}^{2} = 4\,\sigma_{\phi}^{2}$. 
And recall for our simulations, the magnitude $|a|$ in the disk is order unity during the 
active phases of BH growth \citep{hopkins:zoom.sims}. 

This is the important point -- because of cancellations that occur (essentially the 
entire system is near-resonance), the radial induced velocities from the $m=1$ 
mode are quite large, $\sim V_{c}$, {\em independent} of how small the ratio 
$M_{d}/\mbh$ may be. 

Now return to the dispersion relation for bending modes. For the non-disk (Keplerian) 
part of the potential, 
$\nu^{2}=\Omega^{2}$. 
To leading (zeroth) order in $\epsilon\sim M_{d}/M_{\rm BH}$, then, the dispersion relation becomes 
\be 
\label{eqn:dispersion.rel.bending.2}
\left(\frac{\omega_{b}}{\Omega} - m_{b}  \right)^{2} = 
1 +  \left[  \left(\frac{h}{R}\right)^{2} - |a|^{2}\,|k_{p}\,R|^{-2} \right]\,|k_{b}\,R|^{2}
\ee
where we have defined $h/R=\sigma_{z}/V_{c}$. 

Recall, $|a|\,|k_{p}\,R|^{-1}\sim1$ for the lopsided disk mode, and 
$|k_{b}\,R|\gg1$ in the WKB limit -- thus, whenever the disk is thin, 
{\em the radial motions induced by the $m_{p}=1$ eccentric disk 
excite bending modes}. 
These bending modes will grow on the local dynamical timescale, 
since $\omega_{b}\sim \Omega$. Compare the slow modes in the plane, 
which have $\omega_{p}\sim \epsilon\,\Omega$ and can be 
long-lived. 

The bending mode, of course, creates some non-trivial vertical motion. 
The growth of the mode will saturate when it drives a $\sigma_{z}$ 
sufficiently large so as to reach the marginal stability condition of 
Equation~\ref{eqn:dispersion.rel.bending.2} (Im$(\omega_{b})=0$), 
which we can write as   
\begin{align}
\left(\frac{h}{R}\right) &\rightarrow |k_{b}\,R|^{-1}\,\left( |a|^{2}\,|k_{p}\,R|^{-2}\,|k_{b}\,R|^{2} -1 \right)^{1/2} \\ 
&\approx \frac{|a|}{|k_{p}\,R|} \sim |a| 
\end{align}
where the second equality uses $|k_{b}\,R|\gtrsim1$. 
In short, tightly-wound bending modes arise, and saturate 
$v_{z}$ at a large fraction of $V_{c}$, such that $h/R$ is driven to 
order unity wherever the eccentric mode persists, 
{\em independent} of the degree of self-gravity of the disk!

\begin{figure}
    \centering
    \plotone{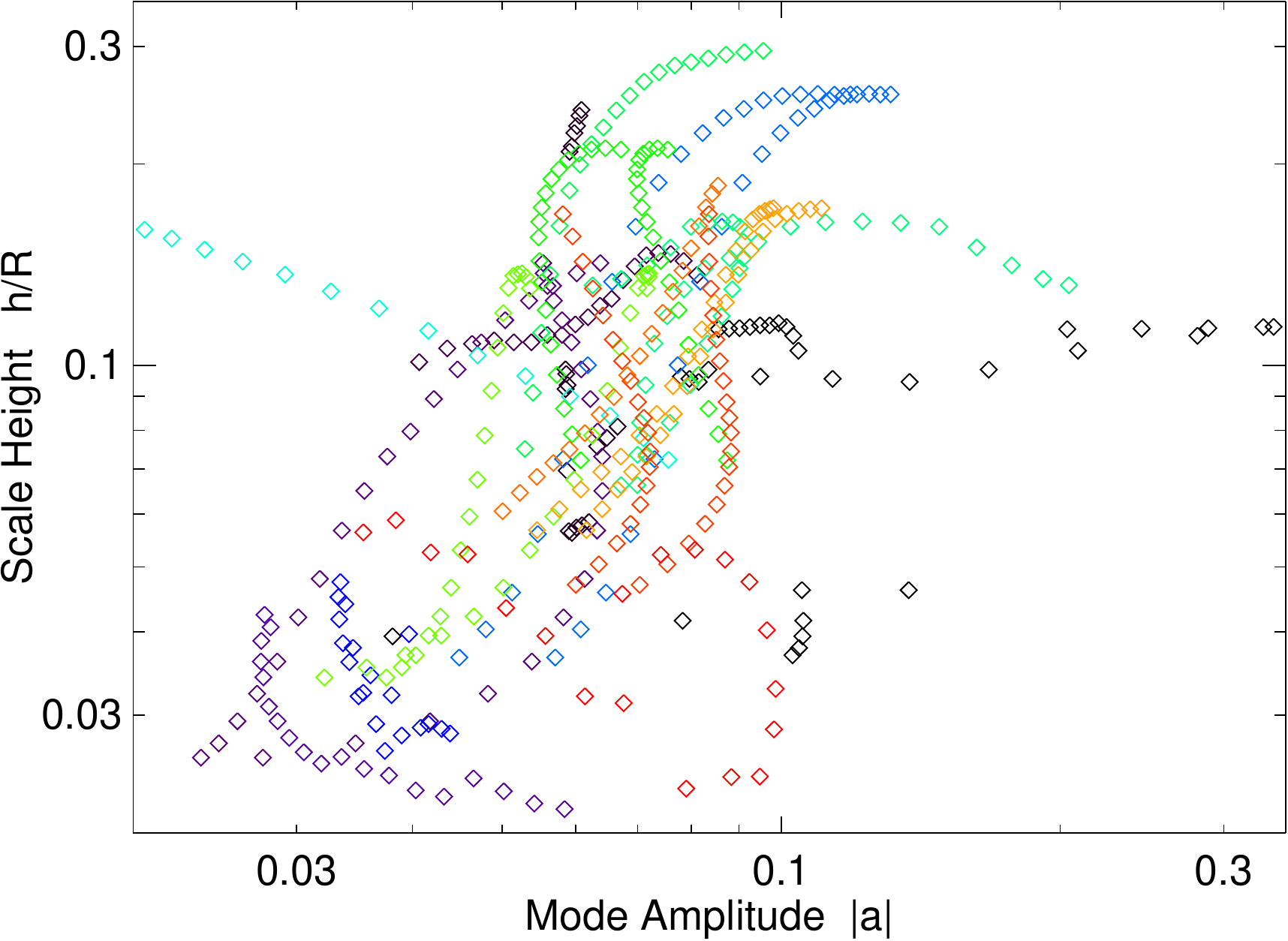}
    \caption{Comparison of the scale-height $h$ (as in Figure~\ref{fig:torus.z.profile}) 
    to the $m=1$ eccentric disk mode amplitude $|a|\equiv |\Sigma_{1}/\Sigma$, 
    at different radii (each point samples radii evenly in $\log{R}$ from 
    $R=1-10\,$pc), for different simulations (each color denotes a different simulation, 
    near its peak activity). All simulations shown are ``cold'' (i.e.\ have weak 
    stellar feedback assumed in their sub-resolution prescription, specifically 
    $q_{\rm eos}<0.1$), so that the scale height is dominated by {\em resolved} 
    turbulent vertical motions, not the feedback model input. 
    In these cases, there is a correlation of the form $h/R\sim|a|$, 
    corresponding to the prediction from an $h/R$ pumped-up by 
    bending modes, themselves excited by radial motions from the 
    in-plane eccentric mode. The torus height can be continuously 
    sustained by exchange of energy from the eccentric/lopsided 
    disk mode that powers the accretion onto the BH. 
    \label{fig:h.vs.a}}
\end{figure}

In Figure~\ref{fig:h.vs.a}, we check whether this prediction at all describes 
our simulations. We compare the scale height $h/R$ to the measured mode 
amplitude $|a|$ of the in-plane $m=1$ mode, at a random time during the active 
phase, for each of a subset of our simulations. We chose only the simulations 
for which $q_{\rm eos}\le0.1$, where we can confirm that the sub-grid assumed 
$c_{s}$ does not dominate $c_{\rm eff}$ or the vertical scale height on the 
scales we measure (see Figure~\ref{fig:torus.z.profile.q}). 
We sample both quantities at even intervals in $\log{R}$ from 
$R=0.3-10\,$pc. There is, unsurprisingly, large scatter, but a correlation is significant 
at $>3\,\sigma$ and  consistent with 
$h/R\sim|a|$ over most of the simulated range. That the relation is not exactly linear 
at the high-$h/R$ end and shows considerable scatter is expected, both 
because of contributions from $k_{b}$ and $k_{\rho}$ in the derivations above, 
non-linear effects (especially at $h/R$ and/or $|a|$ $\gtrsim0.1$), and 
some non-zero support from $c_{s}$. But it is quite unlikely that 
this relation would arise accidentally -- after all, for otherwise equal properties, 
a lower-$h/R$ disk is actually more gravitationally unstable, so if anything 
we would naively expect the inverse of the observed correlation.

\section{Basic Dynamical Properties of the ``Torus''}
\label{sec:dynamical}

\begin{figure*}
    \centering
    \plotside{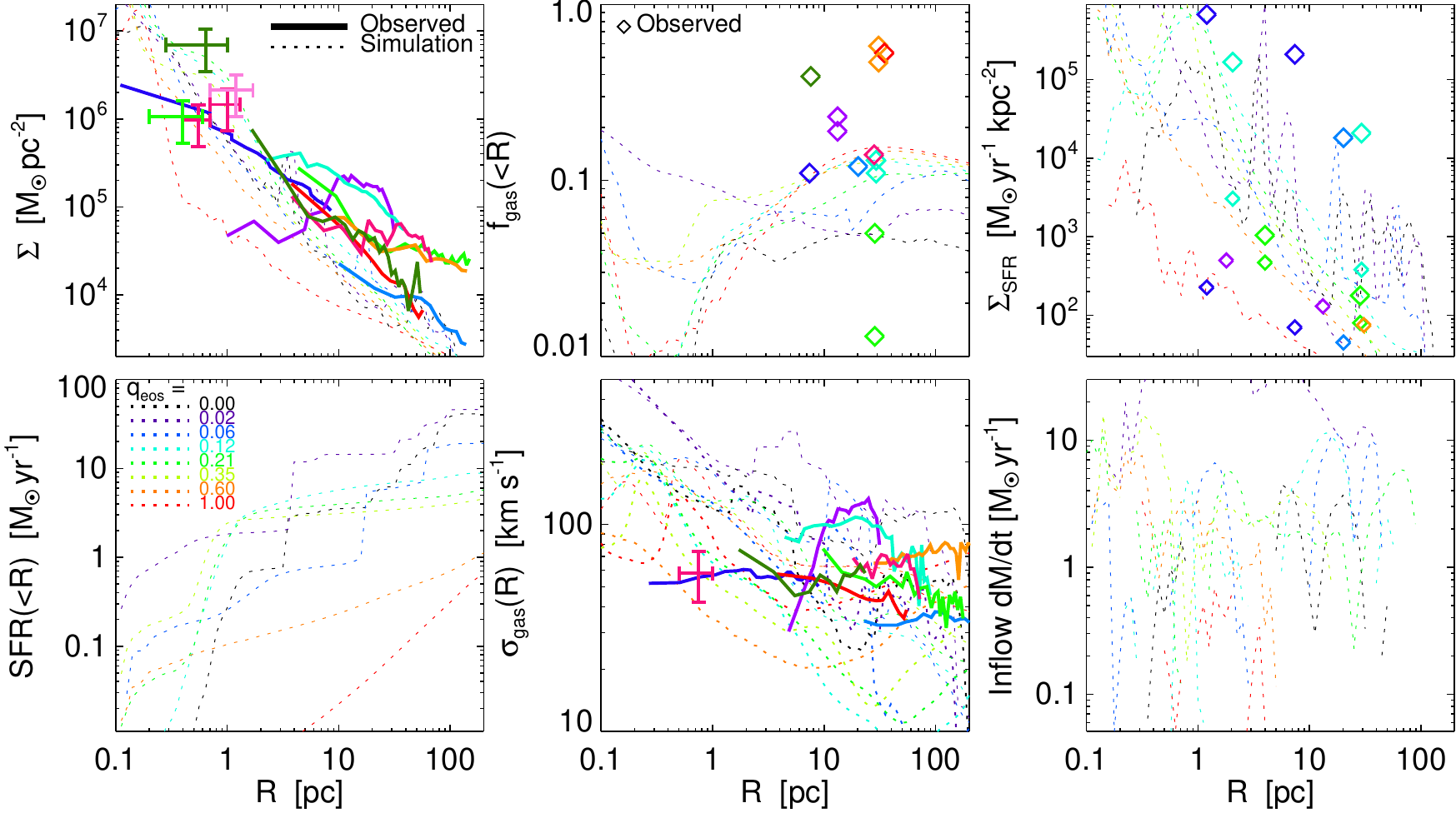}
    \caption{Azimuthally-averaged nuclear disk properties versus radius. 
    Each dotted line is a simulation with different galaxy and inflow properties, 
    but one where {\em some} nuclear inflow is excited. 
    We chose a random time near the peak of activity for each simulation to show 
    here, but the results are similar over the entire active phase of each. 
    Here for a set of simulations with identical initial conditions but 
    varied $q_{\rm eos}$, the parameter describing our effective 
    equation of state and stellar feedback model. 
    Lines range from black ($q_{\rm eos}=0$; effective $c_{s}=10\,{\rm km\,s^{-1}}$) 
    to red ($q_{\rm eos}=1$; effective $c_{s}=100\,{\rm km\,s^{-1}}$), 
    evenly spaced in $\log{c_{s}}$.    
    {\em Top Left:} Surface mass density profiles. 
    Points with error bars show constraints from AGN maser disks 
    \citep[NGC 3079;][magenta]{kondratko:3079.acc.disk.maser}, 
    \citep[NGC 3393;][dark green]{kondratko:ngc3393.acc.disk.maser}, 
    \citep[NGC 1068;][pink]{lodato:1068.maser}, 
    \citep[Circinus;][light green]{greenhill:circinus.acc.disk}. 
    Solid lines show constraints from adaptive-optics (AO) measurements of 
    AGN 
    (NGC 1068, 1097, 3227, 3783, 4051, 4151, 6814, 7469, Circinus; in  
    violet, green, cyan, blue, dark green, magenta, red, orange, dark blue, respectively) 
    \citep{davies:sfr.properties.in.torus} and \citet{hicks:obs.torus.properties}. 
    {\em Top Center:} Gas fraction ($M_{\rm gas}(<R)/(M_{\ast}(<R)+M_{\rm gas}(<R))$). 
    Diamonds are the AO systems (same color styles), at the resolution limits 
    for complementary constraints used to derive $f_{\rm gas}$. 
    {\em Top Right:} SFR surface density. 
    For each AO system where a measurement is available, two points are shown. 
    The first is the current SFR (small points), second is the peak SFR (larger points), 
    both estimated from the fits to the SFR history inside the minimum and 
    maximum observed radii in \citet{davies:sfr.properties.in.torus}. 
    {\em Bottom Left:} Integrated SFR$(<R)$. 
    {\em Bottom Center:} Vertical velocity dispersion of the gas. Dotted lines 
    show both the resolved and sub-grid assumed dispersions (added in quadrature). 
    {\em Bottom Right:} Instantaneous inflow rate through $R$, generated by 
    gravitational torques from the eccentric disk structure. 
    \label{fig:torus.structure.profiles}}
\end{figure*}

Thus far, we've focused on the origin of torus structural properties 
in simulations. We now examine these properties in more detail and 
compare to observations. Figure~\ref{fig:torus.structure.profiles} 
shows a number of (azimuthally averaged) properties of the nuclear 
gas, as a function of radius. 
We plot the gas surface mass density, 
gas fraction, SFR, vertical gas velocity dispersion, and gas inflow rate $\dot{M}$ 
(here defined so positive is inflow). 
The velocity dispersion includes both resolved and sub-resolution components, 
i.e.\ $c_{\rm eff}^{2} \equiv c_{s}^{2}+\sigma_{z}^{2}$, where 
$c_{s}$ is the sub-grid implied sound speed (plus any thermal components) and 
$\sigma_{z}$ is resolved vertical dispersion. 

We show this for our suite of simulations from Figure~\ref{fig:torus.z.profile.q} 
in which we systematically vary the 
sub-grid equation of state (via the parameter $\qeos$). 
For each, we select a random snapshot near the peak of inflow activity. 
Because the global properties -- gas density profiles, inflow rates, 
circular velocities, etc -- 
are primarily set by global gravitational torques \citep[see][]{hopkins:inflow.analytics}, 
the parameter $\qeos$ does not appear have a dramatic qualitative effect on these 
properties. The primary effect is to determine the efficiency of fragmentation, 
which in turn changes the variability and global efficiency of star formation and 
gas exhaustion. 
If we consider the wider range of simulations shown in Figure~\ref{fig:torus.z.profile}, 
which vary the initial gas fractions, bulge-to-disk, and BH-to-disk mass ratios, 
we find a similar range in the predicted properties. 

In more detail, \citet{hopkins:cusp.slopes} show that the 
surface density profiles that arise are a natural consequence of the 
dynamics of tidal torques from the $m=1$ lopsided disk instabilities. Specifically, 
the perturbation dynamics set a robust range of ``quasi-equilibrium'' 
profiles in which the gas mass density remains quasi-steady state 
over the active phase so long as there is sufficient initial inflow to trigger the 
process. If the profile is a power law $\Sigma\propto R^{-\eta}$, then 
this range is $1/2\lesssim\eta\lesssim1$, 
similar to that seen in ``cuspy'' ellipticals.

The SFR surface density follows simply from the assumed local relation between star 
formation efficiency and dynamical time: in the simulations, $\dot{\rho}_{\ast}\propto \rho^{3/2}$. 
Competition between gas inflows and SF sets the gas fractions, 
although these evolve significantly via depletion.

There are some observations to which we can compare. 
Water masers have been observed and used to map the inner disk structure around AGN in 
a few nearby galaxies \citep{greenhill:4945.maser,greenhill:circinus.acc.disk,
braatz:agn.maser.search,henkel:agn.masers,
kondratko:agn.masers,kondratko:agn.masers.2,
kondratko:ngc3393.acc.disk.maser}. 
These are sensitive to densities 
$\sim10^{8}-10^{9.5}\,{\rm cm^{-3}}$ (typically $\sim0.1-1\,$pc). 
At larger radii, interferometry has also been used to image the molecular and HI gas in the 
nuclei of some nearby systems \citep{lonsdale:vlbi.agn.cores,
schinnerer:interfer.obs.1068,combes:liner.nuclear.ring,
garcia.burillo:torques.in.agn.nuclei.obs.maps.no.inflow,
schinnerer:submm.merger.w.compact.mol.gas}. Complemented with adaptive-optics 
imaging of nearby nuclei, this gives constraints on the gas+stellar dynamics, 
and information on the star formation history 
\citep{kuntschner:ell.ages,davies:3227.torus.mass.and.sfr,
sanchez:circinus.torus.mass,davies:sfr.properties.in.torus,hicks:obs.torus.properties}. 

We compile these observations and compare to our 
simulations in Figure~\ref{fig:torus.structure.profiles}. Most of the observed systems 
have BHs with broadly similar masses to our $\sim3\times10^{7}\,\msun$. 
We plot the observations at all radii 
available. The maser observations are shown as points with error bars 
for resolved properties of disks outside the minimum radius enclosing the BH. 
The larger-scale surface densities mapped from the gas velocity fields with 
VLBI are shown as solid lines. 
For constraints involving 
stars (gas fractions, SFR), the 
VLBI+AO constraints are shown as diamonds, at the minimum resolved radii 
of the AO observations. The nuclear SF history is modeled for several cases 
in \citet{davies:sfr.properties.in.torus}; we show their estimated {\em current} 
SFR both at the innermost radii where stellar light is measured and 
at the outer radii where the integrated light is used to determine the 
SFH. We also show their estimated maximum SFR of each observed burst 
from the fitted SFH within the observed radius. 
In all cases the observations broadly bracket the simulations, albeit 
with larger uncertainties in $f_{\rm gas}$ and the SFH. 

Of course, since these properties all scale with the dynamical properties of the 
system, they are all mutually correlated. A Kennicutt-Schmidt type law similar 
to that observed \citep[$\Sigma_{\rm SF}\propto\Sigma_{\rm gas}^{\eta}$; 
for the nuclear-scale observations see][]{hicks:obs.torus.properties} is effectively built into our 
simulations by sub-grid assumption.\footnote{Both the observed and simulated 
Kennicutt-type laws appear to have an index closer to $\eta\sim1.7$ rather than the canonical 
$\eta\sim1.4$. In the simulations, this is because we assume a local $\dot{\rho}_{\ast}\propto \rho^{3/2}$, 
and for the simple case of a gas disk contracting at constant $h/R$ this 
predicts $\eta=1.75$.}
We have discussed extensively the gravitational origin of the dispersion ($\sigma$). But both 
$\sigma$ and $\Sigma$ are related to $V_{c}$, for obvious dynamical reasons, 
and increase at smaller radii and/or in more massive/dense systems. And $\Sigma$ is 
tied to $\Sigma_{\rm SF}$ via the Kennicutt relation. We therefore predict a 
relation between $\Sigma_{\rm SF}$ and $\sigma$ for purely gravitational dynamic reasons. 
In the past, such a correlation has been interpreted as evidence of stellar feedback driving 
the observed dispersions -- we find this may not be necessary.

\section{The Column Density Distribution: To Clump or Not to Clump?}
\label{sec:subres}

Thus far, all of our analysis has concerned global properties of 
the simulated torii, which we have reason to believe should be robust to the exact micro-structure 
of gas on unresolved scales. However, sub-resolution structure can be important in 
calculating the column densities observed towards the BH. 
We therefore consider this now with two simple sub-resolution models.

\subsection{The No-Substructure Case: Smooth Torii}
\label{sec:subres:smooth}

One extreme is trivial: we simply take the gas distribution exactly 
as-is from the simulations, without any assumed sub-grid substructure. 
The column density along a given line-of-sight at each time can then be simply 
determined \citep[following][]{hopkins:lifetimes.letter}. We generate 
$\sim1000$ radial lines-of-sight (rays) uniformly spaced in solid angle and 
with its origin at the BH, and integrate the line-of-sight density until 
outside the galaxy. 

This assumption maximizes obscuration, since locking mass up in 
sub-resolution clumps would confine mass to smaller covering fractions 
\citep[see the discussion from simulations in][]{hopkins:lifetimes.methods}.

\subsection{The Clumpy Torus}
\label{sec:subres:clumpy}

In fact, we know that there must be sub-structure in the gas, because cooling and 
star formation occur. Most of the mass in the ISM is probably locked into dense cold clumps. 
Unfortunately our simulation, limited by the physics included, does not predict the 
clump properties but only indirectly assumes an effective ISM state. 
However, with some simple assumptions, we can construct a sub-resolution estimate 
of all the relevant clump properties, without the introduction of any tunable parameters. 

Assume temporarily that most of the mass in the ISM is locked 
into $N_{\rm cl}$ dense clumps, with median mass $M_{\rm cl}$, 
size $R_{\rm cl}$, and mean density 
$\rho_{\rm cl}= M_{\rm cl}/(4\pi/3)\,R_{\rm cl}^{3}$. 
Define the density contrast $\rho_{\rm cl}=x\,\bar{\rho}$, 
with respect to the volume-average background density 
$\bar{\rho}$. 
We make two assumptions, both 
just at the order-of-magnitude level: that the clumps are 
quasi-virial, and that they are in pressure equilibrium with the 
external medium. The first implies 
that whatever supports the clump generates an 
effective pressure $P_{\rm cl}\sim \rho_{\rm cl}V_{\rm cl}^{2}$ 
where $V_{\rm cl}^{2}\sim G\,M_{\rm cl}/R_{\rm cl}$. 
But this is just $P_{\rm cl} = G\,\Sigma_{\rm cl}^{2}$, 
where $\Sigma_{\rm cl}$ is the column density through the clump $\sim \rho_{\rm cl}\,R_{\rm cl}$. 
To within a factor of two or so, this is even true for clumps in free-fall collapse, so is 
likely to be robust. 
We know the external effective (volume-average) pressure of the medium, 
$P_{\rm eff}$ -- this is just the volume-average pressure used for all SPH calculations. 
It is straightforward to then set $P_{\rm cl}\sim P_{\rm eff}$, and obtain 
\begin{equation}
\Sigma_{\rm cl} = \sqrt{P_{\rm eff}/G}\ .
\end{equation}
Pressure equilibrium is a less certain assumption, but if we were to 
force a mass-radius or linewidth-radius 
relation similar to the observed Larson's laws 
in GMCs ($\sigma \propto R^{1/2}$), we would obtain the same 
dimensional scalings.\footnote{These clouds cannot, however, 
simply follow an extrapolation of the local GMC scalings. 
The local GMC size-mass relation implies an 
approximately constant clump surface density $\Sigma\sim 10^{22}\,{\rm cm^{-2}}$. 
But this is much less than the mean surface density of gas already at 
these radii, so any substructure must obey a relation at least different in normalization. 
} Assuming that clumps follow the Jeans mass and radius in a self-regulating 
$Q=1$ disk actually also results in the same dimensional scalings, so it may be 
robust in a variety of regimes. 

The probability of a path length $\Delta r$ intersecting a 
cloud is given by $p=(N_{\rm cl}/V_{\rm tot})\,\sigma_{\rm cl}\,\Delta r$, 
where $V_{\rm tot}$ is the total volume, and $\sigma_{\rm cl} \sim \pi\,R_{\rm cl}^{2}$ 
the clump cross section. 
But since $N_{\rm cl}\sim M_{\rm tot}/M_{\rm cl}$, this 
simply reduces to 
\begin{equation}
\label{eqn:clumpprob}
p_{\rm cl}\sim \bar{\rho}\,\Sigma_{\rm cl}^{-1}\,\Delta r = 
\bar{\rho}\,(P_{\rm eff}/G)^{-1/2}\,\Delta r\ .
\end{equation}

The only two quantities we ultimately care about, the probability of 
intersecting a clump, and the clump column, have the useful feature that the 
clump density contrast and number of clumps completely cancel out. 
Thus, for {\em any} system where the mass is concentrated in 
quasi-virial, pressure-equilibrium clumps, we can determine the column density 
distribution and probability of sightlines seeing clumps based only on reference 
to well-determined volume-average gas properties in the simulations 
($\bar{\rho}$ and $P_{\rm eff}$). Of course, the external pressure is set in part 
by our adjustable $\qeos$, so it is 
important to examine the consequences of that choice.
Higher-order detailed radiative transfer effects 
will depend on the specific clump sizes and other internal properties, 
but these are not our focus here. 
Because of the cancellation of the exact size and density contrast 
(and correspondingly clump mass), the above relations hold 
for an arbitrary spectrum of clump masses, sizes, and/or densities.

The column density along a given line-of-sight can 
then be integrated outward from the BH. 
For each integration step $\Delta r$ along the ray (taken to be increments of 
$\epsilon\,h_{\rm sml}$, where $\epsilon\sim0.01\ll1$ and 
$h_{\rm sml}$ is the local smoothing length at each point), we determine 
the probability $p_{\rm cl}$ of intersecting a clump, and probabilistically 
assign the ray a collision or not. If there is a collision, the integrated column 
is increased by $\Sigma_{\rm cl}$. If not, the column is integrated through the 
``diffuse'' (non-clump) phase of the ISM. The mass fraction in this phase 
(i.e.\ mass fraction {\em not} in star-forming clumps) is determined 
implicitly in the GADGET code \citep[see][]{springel:multiphase}, 
but is always small and should have near-unity volume filling factor. 

Whether or not these assumptions are justified in detail, this provides a 
useful toy model, and we show that it can account for a number of observations. 
Moreover, on galactic scales, the assumptions above have been 
borne out by a large number of independent observations 
\citep{larson:gmc.scalings,wardthompson94:protostellar.core.sizes,
scoville:gmc.properties,solomon:gmc.scalings,
rosolowsky:m31.gmcs,fuller:cloud.size.linewidth,andre:cloud.size,blitz:h2.pressure.corr}. 
Of course, such clumps as observed locally could not survive the tidal forces 
near a supermassive BH. 
But even on nuclear starburst scales, 
it appears that the star formation efficiency per clump dynamical time is low, implying they 
must be quasi-virial and not wildly out of pressure equilibrium
\citep{tan:mol.cloud.formation.times,krumholz:sf.eff.in.clouds}. Similar constraints 
come from clump structure in the narrow-line region \citep[e.g.][]{crenshaw:nlr,rice:nlr.kinematics}. 
And the fact that similar dimensional scalings arise from Jeans considerations 
implies they are likely to be generic to within factors of a few. 
Finally, we note that the dynamic range in column density is so large 
that violations of the above assumptions would have to be more than order-of-magnitude 
in order to qualitatively affect our conclusions.

\subsection{Column Densities: Model and Observations}
\label{sec:subres:columns}

\begin{figure}
    \centering
    \plotone{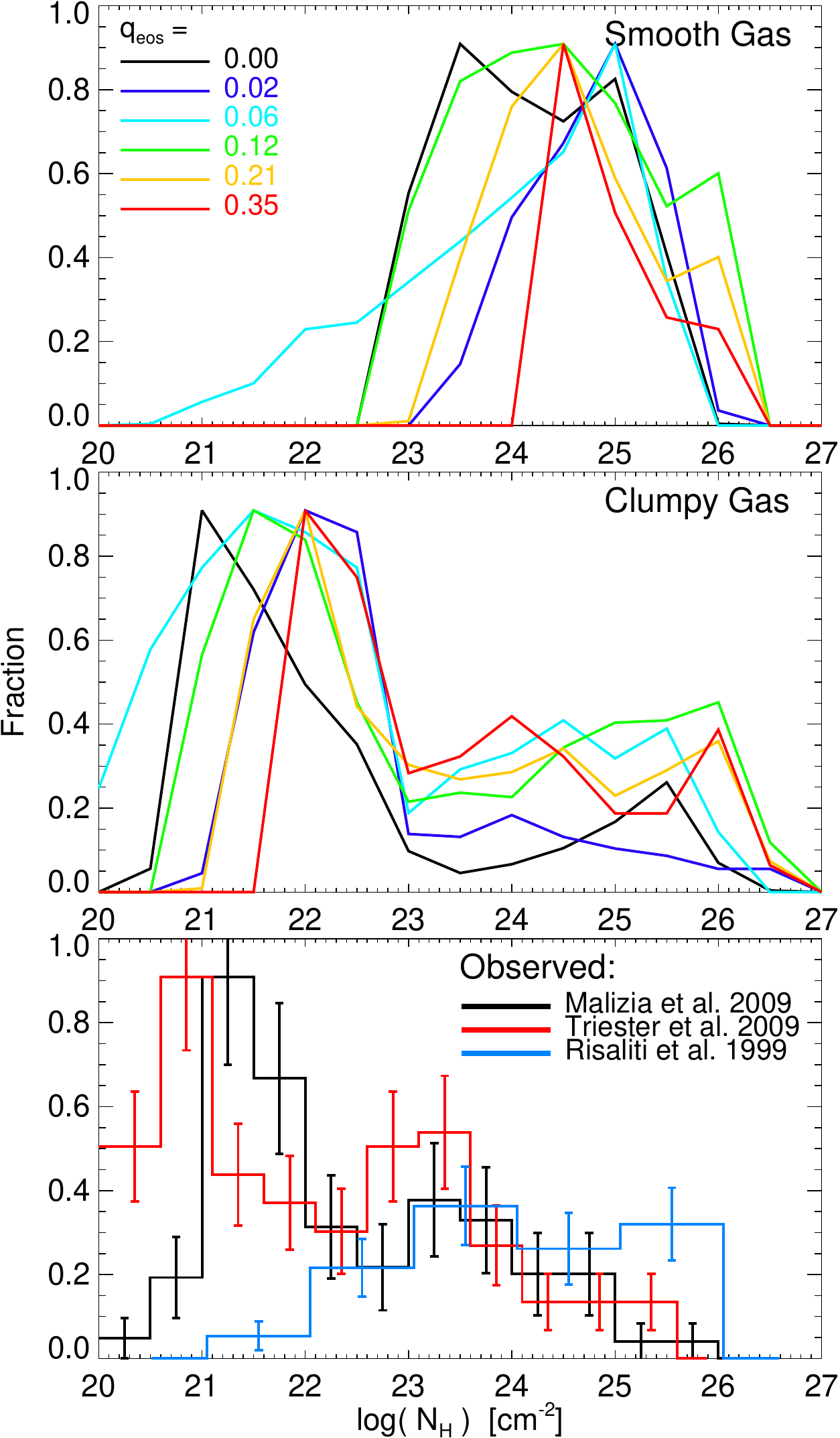}
    \caption{{\em Top:} Column density distribution predicted 
    by our simulations. Each line is a 
    simulation with different sub-grid equation of state (i.e.\ 
    feedback/pressure support in the gas), as in Figure~\ref{fig:torus.structure.profiles}. 
    Here, we assume there is no sub-structure in the gas (i.e.\ gas is 
    perfectly smooth below our resolution limits). 
    The obscuration is clearly over-predicted.
    {\em Middle:} The same, but assuming the gas is clumpy. 
    The clumps are assumed to be quasi-virial and in pressure 
    equilibrium with the outside medium -- this completely determines 
    the predicted $N_{\rm H}$ distribution with {\em no} free parameters. 
    {\em Bottom:} Observational estimates of the column density distribution, 
    from \citet[][black]{malizia:integral.obscured.agn.column.dist}, 
    \citet[][red]{treister:compton.thick.fractions}, and \citet[][blue]{risaliti:seyfert.2.nh.distrib}. 
    The cutoff in these samples at $N_{\rm H}\gtrsim10^{26}\,{\rm cm^{-2}}$ is a 
    selection effect. 
    Allowing for a simple clumpy gas model, without any tunable 
    parameters, provides a good match to observed $N_{\rm H}$ distributions.
    Because of the effects of gravitational maintenance of 
    $h/R$, and the global similarity of mass distributions 
    shown in Figure~\ref{fig:torus.structure.profiles}, 
    there is relatively little dependence of the column 
    density distribution of $q_{\rm eos}$. 
    \label{fig:nh.dist}}
\end{figure}

Figure~\ref{fig:nh.dist} compares the resulting column density 
distribution, for simulations with varied $q_{\rm eos}$ 
(each sampled at a random time near their peak of 
accretion). We compare the distribution from the ``smooth''  
and ``clumpy'' torus models above, and that observed. 
Because of the dynamic range in $N_{\rm H}$ predicted, we are specifically 
interested in comparison with samples sensitive to Compton-thick populations. 
We compile the (estimated intrinsic) distribution of column densities determined from 
the INTEGRAL/IBIS AGN sample 
of \citet[][$20-40$\,keV]{malizia:integral.obscured.agn.column.dist}, 
the predominantly SWIFT/BAT sample 
of \citet[][$\sim100\,$keV]{treister:compton.thick.fractions}, and 
the nearby OIII sample in \citet{risaliti:seyfert.2.nh.distrib} 
(this is a Type 2-only sample, so we normalize to their estimated total fraction of 
Type 2 AGN). The latter sample is most complete at the highest columns 
$>10^{25}\,{\rm cm^{-2}}$; none are sensitive to AGN with 
$N_{\rm H}>10^{26}\,{\rm cm^{-2}}$. At lower column densities, 
these are consistent with a wide variety of hard X-ray observations 
from e.g.\ Chandra and XMM
\citep{ueda03:qlf,lafranca:hx.qlf,
silverman:hx.spacedensity.ldde,hasinger:absorption.update}. 
And more recent, independent analysis of larger SWIFT/BAT samples 
also agrees well \citep{burlon:2011.xr.column.dist.veryhardxr}.

Unsurprisingly, the predicted columns in the smooth torus model 
are uniformly large, in conflict with the observations. 
This is not a problem of there being ``too much'' gas -- 
recall that the actual total gas masses and gas densities predicted 
at these scales agreed well with those in observed AGN 
(Figure~\ref{fig:torus.structure.profiles}). What this shows is that it is not possible to 
reconcile the observed central masses, gas densities, and/or SFRs of 
AGN with their obscured fractions, without invoking some small-scale 
gas clumping. The problem cannot simply be that systems are observed 
at different states either -- as pointed out in 
\citet{hicks:obs.torus.properties}, 
several observed optically un-obscured AGN have 
instantaneous near line-of-sight volume-averaged 
gas densities in $<1-10\,$pc that should naively imply columns of 
$N_{\rm H}\sim10^{25-26}\,{\rm cm^{-2}}$, similar to our 
predictions here without sub-resolution clumping. 
And indeed direct observations on this scale have argued for 
such clumping \citep{risaliti:nh.column.variability,mason:ngc1068.torus.obs,
sanchez:circinus.torus.mass,
nenkova:clumpy.torus.model.2,ramosalmeida:pc.scale.torus.emission,
hoenig:clumpy.torus.modeling,deo:2011.z2.clumpy.torii}.

The column density distribution predicted by the clumpy torus model, 
on the other hand, agrees well with that observed. 
The basic features are easily understood: the small mass fraction 
in the diffuse ISM phase shifts the main peak in the $N_{H}$ 
distribution to lower values. The tail towards larger $N_{H}$ is 
caused by obscuration by clumps. The relative ``flatness'' of the 
tail is broadly expected for vertical profiles similar to those in 
Figure~\ref{fig:torus.z.profile}.

Although the systems plotted 
differ in some subtle details,  
there is little dependence on the parameterization of 
stellar feedback (our $q_{\rm eos}$ parameter). 
Why should the column density distribution be so insensitive to stellar feedback? 
Most important are the factors discussed in \S~\ref{sec:vertical:gravitational}, i.e.\ 
the contribution of gravitational heating which keeps the disks somewhat 
puffed up, and means that the gaseous scale height does not scale as 
strongly with $q_{\rm eos}$ as might otherwise be expected. 

There are also two handy `conspiracies,' in the clumpy torus scenario, which 
make the predicted column density distribution primarily a function of global, 
rather than local parameters. 
In the (near-polar) regime where $p_{\rm cl}\ll 1$, 
it is quite difficult in any model to obtain a column density radically different 
from those shown. 
This is because, even if all the mass is locked in cold clumps, 
a column of at least $\sim10^{21-22}\,{\rm cm^{-2}}$ will arise just from 
diffuse, non star-forming galactic gas on much larger scales 
\citep[see][]{hopkins:lifetimes.obscuration,
hopkins:qso.all}. 
We do see some systematic difference in the 
lowest columns seen, because the exact mass in the ``diffuse phase'' 
depends on the sub-grid model -- but for almost any reasonable model this mass is small, 
so these differences are all in the un-obscured 
range (and therefore dominated by  or comparable to galaxy-scale effects). 
In the opposite (near-disk plane) regime, where $p_{\rm cl}>1$, 
the total column encountered is $\sim \Sigma_{\rm cl}\,p_{\rm cl}\sim \bar{\rho}\,\Delta{r}$ -- 
i.e.\ in the optically thick regime the column density is simply the same 
as that of the ``average medium,'' independent of the gas properties or 
phase structure so long as the global dynamical properties predicted are similar
(physically, this simply represents where clouds will begin to 
overlap, thus making a more uniform molecular medium). 
This is true even if we discard our assumptions of virial and/or pressure equilibrium. 
It is only in the intermediate column regime (which interpolates broadly between the two, 
so we do not expect any features or particular sensitivity to appear) where 
the detailed assumed model of clump properties makes some difference.

\begin{figure}
    \centering
    \plotone{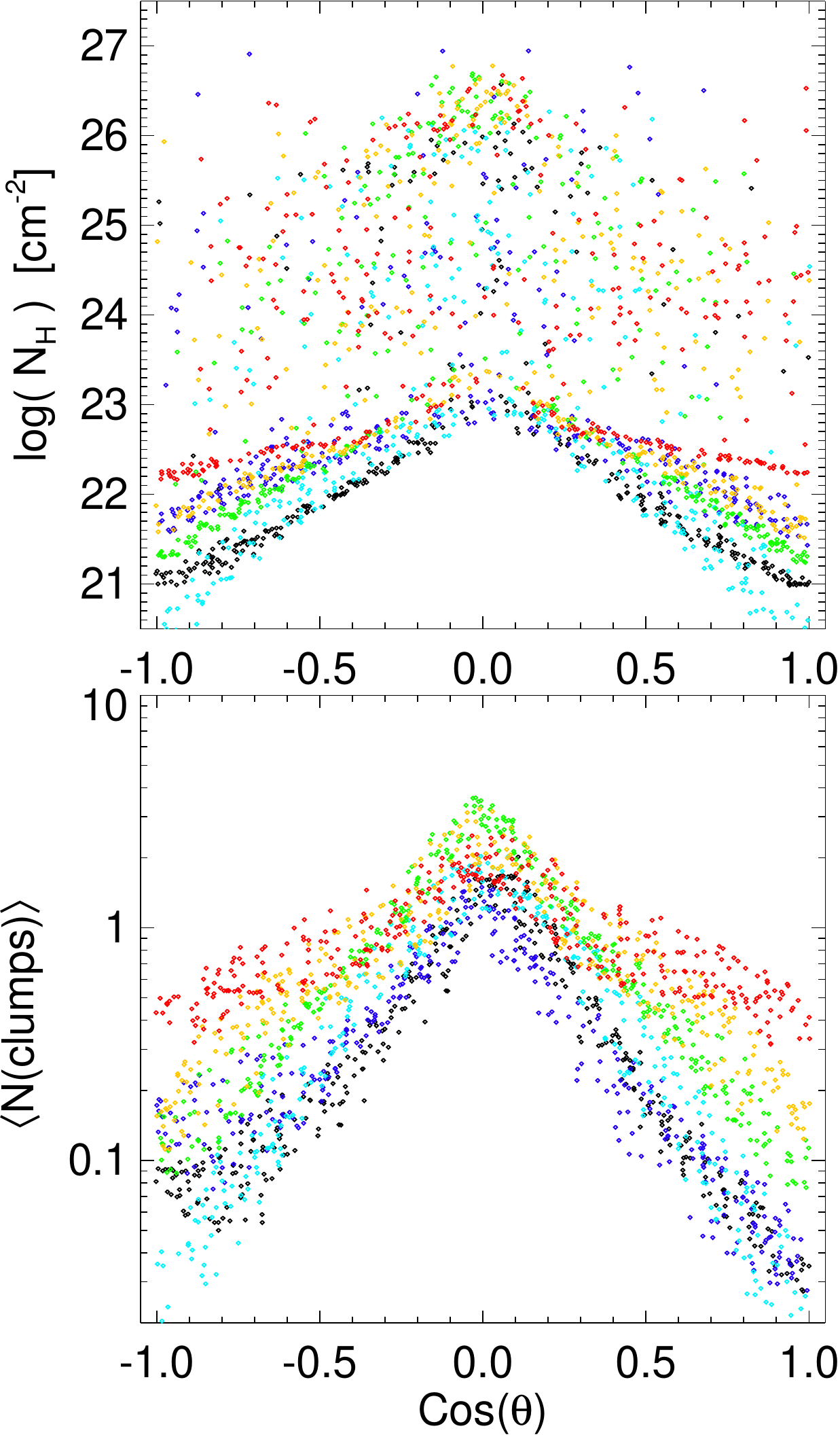}
    \caption{Obscuration properties from the clumpy-torus model 
    in Figure~\ref{fig:nh.dist}, versus viewing angle 
    ($\cos{\theta}=0$ is edge-on; $\cos{\theta}=\pm1$ is polar). 
    For $q_{\rm eos}$ survey.
    {\em Top:} Integrated column. Dense clusters of points at 
    lower $N_{\rm H}$ pass through the diffuse ISM only, 
    points scattered to higher $N_{\rm H}$ encounter clumps 
    along the line-of-sight. The appearance of bimodality is somewhat 
    artificial (see Figure~\ref{fig:nh.dist}; for most simulations, there 
    are not actually two peaks). Each point is a sightline to the BH 
    in a given simulation. 
    {\em Bottom:} Expectation value for the number of clumps 
    encountered along a given sightline. The clump number density 
    is always Poisson, with $\sim$a few clumps along the typical 
    edge-on line of sight, and rapidly declines above a scale-height of 
    $h/R\sim1/2$. 
    \label{fig:column.vs.theta}}
\end{figure}

Figure~\ref{fig:column.vs.theta} illustrates how the column density 
varies with inclination angle, $\theta$ (for the ``clumpy'' scenario). 
Qualitatively, the behavior is expected: columns increase 
towards the disk plane. There is, however, significant scatter 
in the column density at a given $\theta$, even within a 
given simulation at a given time. 
Strikingly similar results are seen in simulations by \citet{wada:torus.mol.gas.hydro.sims}, 
despite including a very different model for stellar feedback, 
and ignoring the role of self-gravity. 
We also show the expectation value of the number of clumps encountered 
along each sightline. 
As expected, this increases along the disk plane. 
Pole-on, it is $\sim0.1\ll1$ in almost all cases. 
Edge-on, it typically reaches $\sim$ a few. 

These values are consistent with various indirect constraints 
from attempts to model AGN SEDs 
\citep{mason:ngc1068.torus.obs,shi:silicate.contraints.on.torii,
thompson:dust.em.from.unobscured.agn,ramosalmeida:pc.scale.torus.emission,
mor:2009.torus.structure.from.fitting.obs,hoenig:clumpy.torus.modeling,
nenkova:clumpy.torus.model.1,nenkova:clumpy.torus.model.2}. 
Almost universally, these studies have found that a similar clumpy torus 
is required, with a number of clumps of order several along the edge-on lines 
of sight, characteristic locations/outer radii of most of the clumps from 
$\sim1-100$\,pc from the BH, and (where constrained) radial 
clump distributions with roughly power-law scaling 
$dp/dr \sim r^{-1}$. We find, for our typical gas surface density 
profiles $\Sigma_{\rm gas}\propto r^{-(0.5-1)}$, 
a $dp/dr \sim r^{-(0.7-1.2)}$ over the dynamic range of interest here. 

The number of clumps can be crudely estimated 
from Eqn.~\ref{eqn:clumpprob}. It is straightforward to show that 
this equation reduces to $\langle N_{\rm clumps} \rangle = \int p_{\rm cl}\,dr 
\sim (R/h)\,Q^{-1/2}$ where $h\approx c_{s}/\Omega$ 
is the scale height of the torus and $Q$ is the 
usual Toomre $Q$. 
For a self-regulating disk, therefore, with $Q\sim1$, we naturally expect 
$\langle N_{\rm clumps}\rangle \sim (h/R)^{-1} \sim$ a few. 
The same scaling pertains if we discard pressure 
equilibrium and instead assume clumps are characteristically Jeans-scale in a 
$Q\sim1$ disk (since then the scale of clumps within $R$ is $\sim h$). 

The characteristic value of a few clumps is also interesting because 
it implies that one is almost always in the Poisson regime. 
This has several implications. 
First, there should be a large scatter between the column 
observed and actual viewing angle, consistent with a wide variety of 
observations (see references above). 
Second, clumping has a number of important radiative transfer effects, 
which will be discussed in subsequent work. 
Third, this allows for highly variable obscuration. A clump moving through the 
line of sight can lead to variation in the column density by several orders 
of magnitude. 
The detailed variability will depend on the clump size spectrum 
and other properties, but the maximal variability timescale should scale as 
$\sim R_{\rm cl}/R\,\Omega(R)$; since most of these clumps are 
at $\sim0.1-1\,$pc, the constraint that clumps not be tidally shredded 
($\rho_{\rm cl}\gtrsim M_{\rm BH}/R^{3}$, and $N_{\rm cl}>1$) sets an 
upper limit to the variability timescale of $\sim5-100\,$yr (for $0.1-1\,$pc), 
for a $10^{8}\,\msun$ BH. For partial obscuration, a more realistic 
clump density contrast and/or larger clump number, the obscuration could vary 
on a timescale $0.01-0.1$ times this (i.e.\ months-year). Such rapid, extreme variability 
in X-ray obscuration has been seen in several AGN 
\citep{risaliti:nh.column.variability,risaliti:2005.nh.variability.1365,
matt:2003.yearly.compton.thick.variability,lamer:2003.agn.transmission.event.mol.cloud,
guainazzi:2005.agn.rate.of.obscuration.variability,
fruscione:abs.by.warped.disk,immler:2003.mrk.6.yr.variability}.

\section{The Obscured Fraction and Torus Properties}
\label{sec:obscured.fraction}

We now examine how the column density distribution depends on 
global properties. 
For the sake of comparison with observations, 
we parameterize the distribution by means of 
the ``obscured fraction'': specifically, the fraction above a given 
column density $N_{\rm H}>10^{22}\,{\rm cm^{-2}}$ 
(a value typically adopted in observational studies). 
Henceforth, we ignore the ``smooth torus'' model -- it does not agree with 
observations and gives uninteresting (always near-unity) obscured fractions.

\begin{figure*}
    \centering
    \plotside{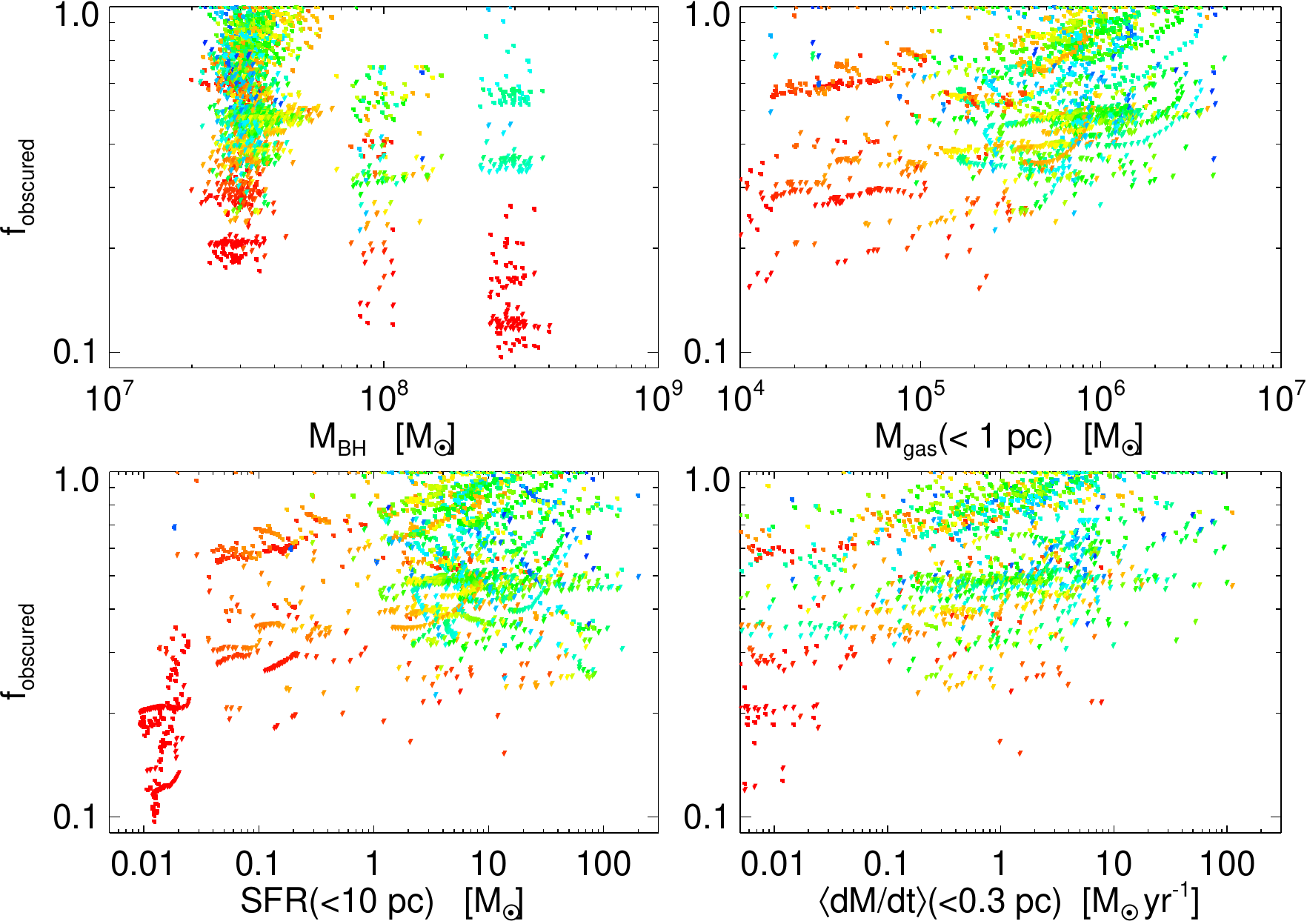}
    \caption{Obscured fraction (fraction of sightlines 
    that encounter a column density $N_{\rm H}>10^{22}\,{\rm cm^{-2}}$) 
    for our simulations as a function of various properties. 
    Each simulation is sampled at several random times both before, 
    during, and after its period of peak activity, and we show results 
    from all simulations with different initial conditions and 
    $q_{\rm eos}$ values. 
    Points are colored by the instantaneous gas fraction 
    inside of $10\,$pc (from red at $f_{\rm gas}<0.05$ to 
    dark blue at $f_{\rm gas}>0.8$). 
    {\em Top Left:} Versus $M_{\rm BH}$ (we add $\sim0.1\,$dex 
    scatter in $M_{\rm BH}$ so the points can be distinguished). 
    For otherwise the same conditions 
    (e.g.\ $M_{\rm gas}$), $f_{\rm obscured}$ declines with $M_{\rm BH}$ 
    as approximately $\propto M_{\rm BH}^{-1/2}$.
    {\em Top Right:} Versus $M_{\rm gas}$ inside $1\,$pc, at fixed $M_{\rm BH}$. 
    Obscured fractions increase weakly ($\propto M_{\rm gas}^{1/4}$) with gas mass.
    {\em Bottom Left:} Versus the star formation rate inside $10\,$pc. 
    There is a weak increase in $f_{\rm obscured}$ with the circum-nuclear star formation 
    rate, driven by the dependence on $M_{\rm gas}$ 
    (slightly weaker $\propto \dot{M}_{\ast}^{0.15-0.2}$, because $\dot{M}_{\ast}^{0.15-0.2}$ 
    is super-linear in $M_{\rm gas}$).
    {\em Bottom Right:} Versus 
    inflow rate into small radii, at fixed $M_{\rm BH}=3\times10^{7}\,\msun$ 
    (if this continued to arbitrarily small radii, this would be proportional to 
    the Eddington ratio $\dot{M}_{\rm BH}/\dot{M}_{\rm Edd}=L/L_{\rm Edd}$).
    The dependence here is significant, but very weak ($\propto \dot{M}^{0.1}$).
    \label{fig:fobsc.vs.prop}}
\end{figure*}

Figure~\ref{fig:fobsc.vs.prop} compares the obscured fraction in the 
``clumpy'' model with a number of nuclear properties. For each simulation, 
we measure the relevant properties at randomly sampled times and viewing 
angles. We show the obscuration versus total mass inside some small 
radius (essentially, the BH mass), versus gas mass, versus the nuclear SFR, 
and versus the BH Eddington ratio. 

Unsurprisingly, $f_{\rm obsc}$ increases with the gas mass inside a small 
radius $<1\,$pc. Note, however, that the correlation is weak: $f_{\rm obsc}\propto M_{\rm gas}^{1/4}$. 
The midplane columns should increase more rapidly with $M_{\rm gas}$, but these are 
already optically thick -- the obscured fraction grows slowly with the fraction of sightlines above 
the disk that (at higher column) become optically thick.
More interesting is the correlation this implies -- $f_{\rm obsc}$ also 
increases with the nuclear SFR. Behavior along these lines has been observed at a wide variety of 
scales -- Type 2 AGN are more likely to be found in more rapidly star-forming 
hosts, and/or hosts with younger stellar populations \citep{brotherton99:postsb.qso,
canalizostockton01:postsb.qso.mergers,yip:qso.eigenspectra,
jahnke:qso.host.sf,zakamska:qso.hosts,
nandra:qso.host.colors,silverman:qso.hosts}. 
The observational correlation appears to be particularly strong when the 
{\em nuclear} stellar populations are isolated \citep{shi:qso.host.sf.lf,
wang:seyfert.nuclear.sf, imanishi:2002.nuclear.sfr.vs.agn,imanishi:2004.nuclear.sfr.vs.agn,
davies:sfr.properties.in.torus}. Note that the SFRs inside of 
$\sim10\,$pc can reach large $\sim10\,\msun\,{\rm yr}$ values; however, 
as shown in \citet{hopkins:zoom.sims} (Fig.~14), this is correlated with the BH inflow rates 
on these scales as $\dot{M}_{\ast}\sim\dot{M}_{\rm BH}$ 
(again, both tracing the gas mass supply) -- for a less 
extreme quasar the ``zero point'' expected inside $\sim10\,$pc would be more 
like $\sim0.1-1\,\msun\,{\rm yr^{-1}}$. 
Conversely, ULIRGs and mergers with more pronounced star formation 
in their nuclei are more likely to host obscured Seyferts or quasars, 
whereas those with slightly older populations are more likely to exhibit 
Type 1 signatures \citep{farrah:qso.vs.sf.sed.fitting,farrah:qso.seds.in.warm.ulirgs,
sanders:agn.vs.sf.in.ulirgs,guyon:qso.hosts.ir,dasyra:ulirg.qso.spectroscopy,
yuan:ulirgs.are.agn.sb.composites.sequence}.

Most likely, at least some of this trend owes to the role of AGN feedback 
in clearing away some of the gas and dust \citep[see e.g.][]{sanders88:quasars,
hopkins:lifetimes.letter,hopkins:lifetimes.obscuration,hopkins:sb.agn.delay,
granato:sam,narayanan:co.outflows}, 
but it can simply arise as we see here from the larger gas and dust supply 
``burying'' the AGN until star formation exhausts much of that material. 
We stress that this is a true {\em nuclear}-scale ($<10\,$pc) correlation here, and the 
nuclear SF contributes negligibly ($\lesssim 0.3\,\%$) to the total SFR; 
there is not necessarily any predicted correlation between the AGN obscuration 
and the total/large-scale galaxy SFR. 

Given similar gas properties, the obscured fraction 
decreases with BH mass. 
This is expected because the BH gravity provides a stabilizing force 
that tries to ``flatten'' the torus (for fixed 
gas properties, the disk $h/R \propto M_{\rm BH}^{-1/2}$). 
If more luminous AGN are, on average, 
more massive BHs, then this suggests an inverse correlation 
between QSO luminosity and obscured fractions. 
Indeed, the existence of such an apparent correlation is 
well-established \citep{hill96:sf.abs.in.radio.gal,
  simpson99:thermal.imaging.of.radio.gal,
  willott00:optical.qso.frac.vs.l,simpson00:ir.photometry.radio.qsos,
  steffen03:agn.obsc.vs.l.z,ueda03:qlf,grimes04:obsc.frac.vs.l,hasinger04:obsc.frac.and.xrb,
  sazonov.revnivtsev04:local.hx.qlf,barger:qlf,
  simpson:type1.frac,hao:local.lf,gilli:obscured.fractions,
  hickox:bootes.obscured.agn,hasinger:absorption.update}. 
However, it is still unclear precisely how much of this 
correlation owes to alternative possibilities such as simple dilution 
by the host galaxy and/or differences in the 
Eddington ratio distribution and accretion state \citep[for a detailed 
discussion, see][and references therein]{hopkins:seyfert.bimodality}. 
Moreover, without a full cosmological model to predict e.g.\ the 
distribution of active BH masses and Eddington ratios, we 
cannot forward model the BH {\em luminosity} distribution to 
construct a direct comparison with observations. But the predicted 
scaling here is not especially strong; 
it may well be that additional physics is needed to recover the full 
observed correlation -- most commonly, AGN feedback is invoked 
to ``blow away'' some of the torus in the most luminous systems 
(see references above).

\section{Discussion}
\label{sec:discussion}

We have studied AGN obscuration in a series of multi-scale hydrodynamic 
simulations that can self-consistently follow gas from 
$>10\,$kpc galactic scales to $<0.1\,$pc. 
These simulations include the full self-gravity of stars and 
gas (along with BHs and dark matter), 
gas cooling, and star formation, along with varied prescriptions 
for feedback from young stars; these are all critical 
to the behavior we see, and have not before been 
simultaneously modeled on nuclear scales. 
In these simulations, inflows from large scales, when sufficiently large, 
lead to a cascade of instabilities on small scales, 
ultimately yielding large nuclear gas masses and 
accretion rates onto the AGN. The scenario is qualitatively 
similar to the ``bars within bars'' model, but there is a high 
degree of variability and morphological diversity at each stage 
(with spirals, bars, clumps, flocculent structures, all present 
and alternatively powering inflows and outflows) -- 
a more apt description would be ``stuff within stuff'' 
\citep{hopkins:zoom.sims}. Once gas nears the radius of influence of the 
BH, it generically forms an unstable $m=1$ mode (a lopsided or
eccentric disk) that slowly precesses about the BH
\citep{hopkins:m31.disk}.  The stellar and gas disk precess
differently, leading to strong gravitational torques that can drive
accretion rates of up to $\sim10\,\msun\,{\rm yr^{-1}}$ onto the BH.

In this paper, we show that these nuclear, lopsided disks in fact
naturally account for the long-invoked ``toroidal obscuring region''
used to explain the obscuration of Type 2 AGN. Up to now, these models
have been essentially phenomenological -- we show for the first time
the formation of sub-pc scale obscuring structures from galaxy-scale
inflows, and in a suite of $\sim100$ simulations show that they are
quite {generic} and arise ubiquitously with this inflow scenario.  We
show that the global dynamical properties -- gas and stellar
densities, density profiles, kinematics, gas fractions, and star
formation rates, agree well with observations of AGN obscuring regions
from scales as small as $0.1\,$pc to $\gtrsim100\,$pc.

This implies a fundamentally new paradigm in which to view the
obscuring region or ``torus.''  Far from being a passive bystander or
simple fuel reservoir for the accretion process, it is itself the {\em
driver} of that accretion. The torus is the gravitational structure on
scales within the BH radius of influence that torques the gas and
forces continuous gas inflow onto the BH.  The same lopsided modes
that drive accretion can also provide the scale height, column density
distribution, and characteristic gas properties of the structure.

As such, the predicted torii have non-trivial substructure: both
small-scale clumping in the gas (discussed below), global $m=1$
patterns, and warps/twists arising from bending modes at a range of
radii.  On large scales, the $m=1$ modes tend to manifest as
lopsided/eccentric disks, or one-armed spirals; on small scales, they
become more tightly wound spirals.  Their typical amplitude in surface
density is expected to be at the $\sim10$'s of percent level
\citep[see][]{hopkins:zoom.sims}; the amplitude of induced
non-circular velocities and corresponding magnitude of ``offsets'' of
the BH from the spatial center of the galaxy nuclei (in units of the
BH radius of influence) are about the same.  Maser observations may
show indications of asymmetry in the structure around nuclei
\citep{schinnerer:interfer.obs.1068,greenhill:circinus.acc.disk,kondratko:3079.acc.disk.maser,fruscione:abs.by.warped.disk,kondratko:agn.masers.2};
it is also possible that measurements of the velocity structure of
e.g.\ molecular emission lines from the torus region may be able to
measure such asymmetries in the near future.

We have argued that a large number of obscuration properties
traditionally associated with ``feedback'' processes from AGN and star
formation may, in fact, be explained by purely gravitational physics.
Even in the absence of feedback, properly including the full
self-gravity of gas and stars leads to disks with large $h/R$,
sufficient to account for the observed column density distribution.
This arises because of a combination of large-scale warps and twists
(for example, where the lopsided disk mode meets an outer bar) and
bending modes within the disk itself. The latter will, even in the
absence of any large-scale twists or warps, tend to pump up $h/R$
wherever the eccentric disk mode is excited until an order unity
$h/R\sim|a|$ is reached. Since bending modes are fast modes (pattern
speed $\sim \Omega$), this can continuously transfer energy from the
orbital motion to vertical motions on a single dynamical time,
maintaining vertical scale heights even when the cooling time is
arbitrarily short.

These warps and twists also naturally lead to the observed lack of
correlation between nuclear-scale disk inclination angles and those of
their parent/host galaxies. This will be even more prominent in
systems which are driven on large scales by mergers, but can occur
even in entirely secularly fueled AGN. They also account for observed
gas velocity dispersions in AGN nuclei, and the correlations between
those dispersions and quantities such as the local gas mass, star
formation rate, and mass in young stars (all via their inherent
dynamical correlations, not via any feedback channel). The efficiency
of gravitational torques and induced inflow also naturally leads to
convergence in nuclear gas masses and density profiles, leaving relic
``cusps'' similar to those observed
\citep{hopkins:cusp.slopes}.

These mechanisms can naturally explain observed global quantities such
as the gas scale heights, masses, and density profiles. However,
modeling the actual obscured fraction of AGN requires a more explicit
model for the actual sub-structure on Jeans mass scales and well
below, in the ISM surrounding black holes.  We show in fact that any
model which matches the observed dynamical properties (particularly
global gas masses), but assumes ``smooth'' gas (uniformly distributed,
say, out to some scale height corresponding to the average obscured
fraction), will simultaneously fail to explain the observed column
density distributions.  A natural explanation for this discrepancy is
that the gas is clumpy on multiple scales, broadening the column
density distribution along all sightlines.  There must, in fact, be
structure on the relevant scales, since we know there is star
formation at these radii (so some gas must reside in dense, tidally
bound star-forming clumps).  Unfortunately, our present models do not
explicitly resolve the necessary physics of star formation and GMC
formation/destruction via stellar feedback needed in order to
explicitly simulate the sub-structure of the gas down to these scales.

However, we find that we can 
obtain predicted column density distributions in good agreement with those observed if 
we assume that whatever sub-grid clumps exist obey a couple of basic assumptions: 
namely that they are (at least at the order-of-magnitude level) near both virial and pressure 
equilibrium (or, instead of pressure equilibrium, that they are Jeans-scale in a 
$Q\sim1$ self-regulating disk). 
These assumptions are sufficient to (statistically) predict the column density distribution 
that would be observed, regardless of the actual clump mass spectrum and physical origin 
(and without any adjustable parameter introduced). 
The predictions agree quite 
well with the column density distribution of both un-obscured, obscured, and Compton-thick AGN. 
This suggests that these basic properties should still hold for substructure in these regions, 
and that -- if so -- the uncertain quantities in our simulations (such as the feedback 
prescription and star formation recipe), make no dramatic difference in the column 
density distribution, since its key properties are set by the basic dynamics above. 
Essentially, if these assumptions hold, the distribution of observed column densities towards AGN is 
itself a natural consequence of gravitational clumping at the Jeans length/mass in a self-gravitating, 
globally $Q\sim1$ disk -- no exotic wind physics (driven by either stars or AGN) need to necessarily 
be invoked.

Higher-order probes of the structure in this region, for example 
studying the clump properties (their sizes and masses), 
constraining the ratio of stellar feedback to dynamical support 
in driving scale heights, and making predictions for line 
structure and other effects that might be used to probe the sub-structure 
and lopsided precession that power accretion, will require 
detailed treatment of the radiative transfer from the accretion disk 
through the circumnuclear region. This will be the subject of a 
future paper, and should enable a host of new predictions for 
comparison with future observations. 

Another important next step will be the inclusion of realistic,
physically motivated feedback models. Coupling our simulations with
radiative transfer will be a major advance.  Although this approach
will not be strictly self-consistent, we will, for the first time, be
able to examine how radiation pressure impacts inflowing and
star-forming gas using a realistic description of multi-scale AGN gas
distributions from $\sim0.1-1000\,$pc scales.  In particular, to study
where the photon momentum is absorbed
\citep[compare][]{murray:momentum.winds,
ciotti:2010.mom.agn.fb,hopkins:rad.pressure.sf.fb}, how radiation
pressure profiles vary throughout the gas, how photon diffusion may
affect the role of feedback \citep{thompson:rad.pressure}, and whether
a realistic clumpy gas medium suppresses or enhances the efficiency of
feedback-induced ``shutdown'' in star formation
\citep{hall:2007.agn.outflow.substructure,
hopkins:twostage.feedback,tortora:2009.agn.jet.fb.and.ell.colors}. 
In bright quasars and/or nuclear starbursts, the gas structure may well be modified 
not just locally (in terms of its clumpiness or sub-grid pressure support), but 
globally by strong outflows driven, for example, by radiation pressure 
\citep[see][and references therein]{debuhr:2010.mom.fb.bhgrowth}. 
Even in the regime where some material is being expelled at the 
escape velocity, it is difficult to alter many of the basic dynamical properties 
of the gas (total mass enclosed and its relation to inflow rates, obscured fractions, etc) 
at the order-of-magnitude level \citep[see e.g.][]{marconi:2008.rad.pressure.virial.masses}, 
but may well make a large contribution to 
the observed scale height of obscuring material and can be critical to understanding 
how AGN self-regulate, why torii exhibit complex sub-structure, and 
perhaps scalings of obscured fraction with luminosity and/or redshift.

We have focused here on small-scale obscuration, at radii 
traditionally associated with the AGN ``torus.'' We stress, however, that this 
does {\em not} mean that there is a single object that accounts for the 
obscuration of all systems. As is evident in all of our comparisons, the gas 
distribution is truly continuous. 
Of course, there will be gas on small scales near the BH whenever 
it is active, which can occult and obscure different emission regions. 
This may take the form of an AGN wind, especially in high Eddington ratio 
systems \citep[see e.g.][]{elvis:outflow.model,elitzur:torus.wind}.

There is also well-resolved gas from the host galaxies in these
systems.  The latter, on say $>100\,$pc scales, is not likely to be
Compton-thick, simply because the characteristic Jeans scales,
etc. are too large. However, this can easily dominate the production
of more moderate column densities $N_{\rm H}\sim10^{22}\,{\rm
cm^{-2}}$.  This ``host galaxy obscuration'' is especially important
in the early phases of inflow forming a kpc-scale starburst, as the
central kpc may be isotropically enshrouded in dust for a time
$\sim10^{8}\,$yr.  Such obscuration as it arises in simulations of
galaxy mergers has been discussed at length in e.g.\
\citet{hopkins:lifetimes.obscuration,hopkins:qso.all,hopkins:seyferts,
hayward:2011.smg.merger.rt},
and we refer to those papers for more details. 
Observations have also made it clear that a significant fraction of 
obscuration must come from host galaxies \citep[especially 
in starbursts and edge-on disks; see e.g.][]{zakamska:qso.hosts,rigby:qso.hosts,
martinez-sansigre:high.qsos.in.submm,
lagos:2011.agn.gal.orientation}
Since our current simulations are the first to simultaneously resolve both the
nuclear scales where very large columns arise, and the galaxy scales
where more moderate but potentially more isotropic (or at least
differently-oriented) columns can arise, it will be interesting to
investigate the relative contributions to obscuration from different
scales, as a function of evolutionary stage and galaxy/BH properties.
Because the AGN spectrum changes as it moves through the inner
obscuring regions, this will require a full treatment of radiative
transfer, as described above.

\acknowledgments 
We thank Eliot Quataert, Nadia Zakamska, Jenny Greene, 
Patrik Jonsson, and Josh Younger for helpful discussions 
in the development of this work. 
Support for PFH was provided by the Miller Institute for Basic Research 
in Science, University of California Berkeley. 
DN and LH acknowledge support from the NSF via grant AST-1009452.
\\

\bibliography{/Users/phopkins/Documents/lars_galaxies/papers/ms}

\begin{thebibliography}{252}
\expandafter\ifx\csname natexlab\endcsname\relax\def\natexlab#1{#1}\fi

\bibitem[{{Alexander} {et~al.}(2005){Alexander}, {Bauer}, {Chapman}, {Smail},
  {Blain}, {Brandt}, \& {Ivison}}]{alexander:xray.smgs}
{Alexander}, D.~M., {Bauer}, F.~E., {Chapman}, S.~C., {Smail}, I., {Blain},
  A.~W., {Brandt}, W.~N., \& {Ivison}, R.~J. 2005, \apj, 632, 736

\bibitem[{{Alexander} {et~al.}(2008){Alexander}, {Chary}, {Pope}, {Bauer},
  {Brandt}, {Daddi}, {Dickinson}, {Elbaz}, \&
  {Reddy}}]{alexander:2008.compton.thick.z2.qsos}
{Alexander}, D.~M., {et~al.} 2008, \apj, 687, 835

\bibitem[{{Aller} \& {Richstone}(2007)}]{aller:mbh.esph}
{Aller}, M.~C., \& {Richstone}, D.~O. 2007, \apj, 665, 120

\bibitem[{{Andre} {et~al.}(1996){Andre}, {Ward-Thompson}, \&
  {Motte}}]{andre:cloud.size}
{Andre}, P., {Ward-Thompson}, D., \& {Motte}, F. 1996, \aap, 314, 625

\bibitem[{{Antonucci}(1993)}]{antonucci:agn.unification.review}
{Antonucci}, R. 1993, \araa, 31, 473

\bibitem[{{Barger} \& {Cowie}(2005)}]{barger:qlf}
{Barger}, A.~J., \& {Cowie}, L.~L. 2005, \apj, 635, 115

\bibitem[{{Barnes} \& {Hernquist}(1996)}]{barneshernquist96}
{Barnes}, J.~E., \& {Hernquist}, L. 1996, \apj, 471, 115

\bibitem[{{Barnes} \& {Hernquist}(1991)}]{barnes.hernquist.91}
{Barnes}, J.~E., \& {Hernquist}, L.~E. 1991, \apjl, 370, L65

\bibitem[{{Bender} {et~al.}(2005)}]{bender:m31.nuclear.disk.obs}
{Bender}, R., {et~al.} 2005, \apj, 631, 280

\bibitem[{{Blitz} \& {Rosolowsky}(2006)}]{blitz:h2.pressure.corr}
{Blitz}, L., \& {Rosolowsky}, E. 2006, \apj, 650, 933

\bibitem[{Bournaud {et~al.}(2007)Bournaud, Elmegreen, \&
  Elmegreen}]{bournaud:disk.clumps.to.bulge}
Bournaud, F., Elmegreen, B.~G., \& Elmegreen, D.~M. 2007, The Astrophysical
  Journal, 670, 237

\bibitem[{{Braatz} {et~al.}(2004){Braatz}, {Henkel}, {Greenhill}, {Moran}, \&
  {Wilson}}]{braatz:agn.maser.search}
{Braatz}, J.~A., {Henkel}, C., {Greenhill}, L.~J., {Moran}, J.~M., \& {Wilson},
  A.~S. 2004, \apjl, 617, L29

\bibitem[{{Brotherton} {et~al.}(1999)}]{brotherton99:postsb.qso}
{Brotherton}, M.~S., {et~al.} 1999, \apjl, 520, L87

\bibitem[{{Bryant} \& {Scoville}(1999)}]{bryant.scoville:ulirgs.co}
{Bryant}, P.~M., \& {Scoville}, N.~Z. 1999, \aj, 117, 2632

\bibitem[{{Burlon} {et~al.}(2011){Burlon}, {Ajello}, {Greiner}, {Comastri},
  {Merloni}, \& {Gehrels}}]{burlon:2011.xr.column.dist.veryhardxr}
{Burlon}, D., {Ajello}, M., {Greiner}, J., {Comastri}, A., {Merloni}, A., \&
  {Gehrels}, N. 2011, \apj, 728, 58

\bibitem[{{Canalizo} \&
  {Stockton}(2001)}]{canalizostockton01:postsb.qso.mergers}
{Canalizo}, G., \& {Stockton}, A. 2001, \apj, 555, 719

\bibitem[{{Cattaneo} {et~al.}(2005){Cattaneo}, {Combes}, {Colombi}, {Bertin},
  \& {Melchior}}]{cattaneo:2005.mgr.agn.obsc}
{Cattaneo}, A., {Combes}, F., {Colombi}, S., {Bertin}, E., \& {Melchior}, A.
  2005, \mnras, 359, 1237

\bibitem[{{Cattaneo} {et~al.}(2009){Cattaneo}, {Faber}, {Binney}, {Dekel},
  {Kormendy}, {Mushotzky}, {Babul}, {Best}, {Br{\"u}ggen}, {Fabian}, {Frenk},
  {Khalatyan}, {Netzer}, {Mahdavi}, {Silk}, {Steinmetz}, \&
  {Wisotzki}}]{cattaneo:2009.bhfb.ell.quenching}
{Cattaneo}, A., {et~al.} 2009, \nat, 460, 213

\bibitem[{{Ciotti} {et~al.}(2010){Ciotti}, {Ostriker}, \&
  {Proga}}]{ciotti:2010.mom.agn.fb}
{Ciotti}, L., {Ostriker}, J.~P., \& {Proga}, D. 2010, \apj, 717, 708

\bibitem[{{Comastri} {et~al.}(1995){Comastri}, {Setti}, {Zamorani}, \&
  {Hasinger}}]{comastri:1995.agn.xrb}
{Comastri}, A., {Setti}, G., {Zamorani}, G., \& {Hasinger}, G. 1995, \aap, 296,
  1

\bibitem[{{Combes} {et~al.}(2004){Combes}, {Garc{\'{\i}}a-Burillo}, {Boone},
  {Hunt}, {Baker}, {Eckart}, {Englmaier}, {Leon}, {Neri}, {Schinnerer}, \&
  {Tacconi}}]{combes:liner.nuclear.ring}
{Combes}, F., {et~al.} 2004, \aap, 414, 857

\bibitem[{{Cox} {et~al.}(2006){Cox}, {Dutta}, {Di Matteo}, {Hernquist},
  {Hopkins}, {Robertson}, \& {Springel}}]{cox:kinematics}
{Cox}, T.~J., {Dutta}, S.~N., {Di Matteo}, T., {Hernquist}, L., {Hopkins},
  P.~F., {Robertson}, B., \& {Springel}, V. 2006, \apj, 650, 791

\bibitem[{{Crenshaw} {et~al.}(2000)}]{crenshaw:nlr}
{Crenshaw}, D.~M., {et~al.} 2000, \aj, 120, 1731

\bibitem[{{Croton} {et~al.}(2006)}]{croton:sam}
{Croton}, D.~J., {et~al.} 2006, \mnras, 365, 11

\bibitem[{Cuadra {et~al.}(2009)Cuadra, Armitage, Alexander, \&
  Begelman}]{cuadra:binary.bh.mergers.w.gas.disks}
Cuadra, J., Armitage, P.~J., Alexander, R.~D., \& Begelman, M.~C. 2009, \mnras,
  393, 1423

\bibitem[{{Daddi} {et~al.}(2007){Daddi}, {Alexander}, {Dickinson}, {Gilli},
  {Renzini}, {Elbaz}, {Cimatti}, {Chary}, {Frayer}, {Bauer}, {Brandt},
  {Giavalisco}, {Grogin}, {Huynh}, {Kurk}, {Mignoli}, {Morrison}, {Pope}, \&
  {Ravindranath}}]{daddi:2007.high.compton.thick.pops}
{Daddi}, E., {et~al.} 2007, \apj, 670, 173

\bibitem[{{Dasyra} {et~al.}(2006){Dasyra}, {Tacconi}, {Davies}, {Genzel},
  {Lutz}, {Naab}, {Sanders}, {Veilleux}, \&
  {Baker}}]{dasyra:ulirg.qso.spectroscopy}
{Dasyra}, K.~M., {et~al.} 2006, New Astronomy Review, 50, 720

\bibitem[{{Davies} {et~al.}(2007){Davies}, {S{\'a}nchez}, {Genzel}, {Tacconi},
  {Hicks}, {Friedrich}, \& {Sternberg}}]{davies:sfr.properties.in.torus}
{Davies}, R.~I., {S{\'a}nchez}, F.~M., {Genzel}, R., {Tacconi}, L.~J., {Hicks},
  E.~K.~S., {Friedrich}, S., \& {Sternberg}, A. 2007, \apj, 671, 1388

\bibitem[{{Davies} {et~al.}(2006){Davies}, {Thomas}, {Genzel}, {S{\'a}nchez},
  {Tacconi}, {Sternberg}, {Eisenhauer}, {Abuter}, {Saglia}, \&
  {Bender}}]{davies:3227.torus.mass.and.sfr}
{Davies}, R.~I., {et~al.} 2006, \apj, 646, 754

\bibitem[{{Debattista} {et~al.}(2006){Debattista}, {Ferreras}, {Pasquali},
  {Seth}, {De Rijcke}, \& {Morelli}}]{debattista:vcc128.binary.nucleus}
{Debattista}, V.~P., {Ferreras}, I., {Pasquali}, A., {Seth}, A., {De Rijcke},
  S., \& {Morelli}, L. 2006, \apjl, 651, L97

\bibitem[{{Debuhr} {et~al.}(2011){Debuhr}, {Quataert}, \&
  {Ma}}]{debuhr:2010.mom.fb.bhgrowth}
{Debuhr}, J., {Quataert}, E., \& {Ma}, C. 2011, \mnras, 412, 1341

\bibitem[{{Deo} {et~al.}(2009){Deo}, {Richards}, {Crenshaw}, \&
  {Kraemer}}]{deo:2009.ir.seyfert.continuum}
{Deo}, R.~P., {Richards}, G.~T., {Crenshaw}, D.~M., \& {Kraemer}, S.~B. 2009,
  \apj, 705, 14

\bibitem[{{Deo} {et~al.}(2011){Deo}, {Richards}, {Nikutta}, {Elitzur},
  {Gallagher}, {Ivezic}, \& {Hines}}]{deo:2011.z2.clumpy.torii}
{Deo}, R.~P., {Richards}, G.~T., {Nikutta}, R., {Elitzur}, M., {Gallagher},
  S.~C., {Ivezic}, Z., \& {Hines}, D. 2011, \apj, in press, arXiv:1101.2855

\bibitem[{{Di Matteo} {et~al.}(2005){Di Matteo}, {Springel}, \&
  {Hernquist}}]{dimatteo:msigma}
{Di Matteo}, T., {Springel}, V., \& {Hernquist}, L. 2005, \nat, 433, 604

\bibitem[{{Donley} {et~al.}(2005){Donley}, {Rieke}, {Rigby}, \&
  {P{\'e}rez-Gonz{\'a}lez}}]{donley:2005.qso.xr.hosts}
{Donley}, J.~L., {Rieke}, G.~H., {Rigby}, J.~R., \& {P{\'e}rez-Gonz{\'a}lez},
  P.~G. 2005, \apj, 634, 169

\bibitem[{{Dotti} {et~al.}(2009){Dotti}, {Ruszkowski}, {Paredi}, {Colpi},
  {Volonteri}, \& {Haardt}}]{dotti:bh.binary.inspiral}
{Dotti}, M., {Ruszkowski}, M., {Paredi}, L., {Colpi}, M., {Volonteri}, M., \&
  {Haardt}, F. 2009, \mnras, 396, 1640

\bibitem[{{Downes} \& {Solomon}(1998)}]{downes.solomon:ulirgs}
{Downes}, D., \& {Solomon}, P.~M. 1998, \apj, 507, 615

\bibitem[{{El-Zant} \& {Shlosman}(2003)}]{elzant:nested.bar.scales}
{El-Zant}, A.~A., \& {Shlosman}, I. 2003, \apjl, 595, L41

\bibitem[{{Elitzur} \& {Shlosman}(2006)}]{elitzur:torus.wind}
{Elitzur}, M., \& {Shlosman}, I. 2006, \apjl, 648, L101

\bibitem[{{Elvis}(2000)}]{elvis:outflow.model}
{Elvis}, M. 2000, \apj, 545, 63

\bibitem[{{Englmaier} \& {Shlosman}(2004)}]{englmaier:nested.bar.decoupling}
{Englmaier}, P., \& {Shlosman}, I. 2004, \apjl, 617, L115

\bibitem[{{Escala}(2007)}]{escala:nuclear.gas.transport.to.msigma}
{Escala}, A. 2007, \apj, 671, 1264

\bibitem[{{Farrah} {et~al.}(2003){Farrah}, {Afonso}, {Efstathiou},
  {Rowan-Robinson}, {Fox}, \& {Clements}}]{farrah:qso.vs.sf.sed.fitting}
{Farrah}, D., {Afonso}, J., {Efstathiou}, A., {Rowan-Robinson}, M., {Fox}, M.,
  \& {Clements}, D. 2003, \mnras, 343, 585

\bibitem[{{Farrah} {et~al.}(2005){Farrah}, {Surace}, {Veilleux}, {Sanders}, \&
  {Vacca}}]{farrah:qso.seds.in.warm.ulirgs}
{Farrah}, D., {Surace}, J.~A., {Veilleux}, S., {Sanders}, D.~B., \& {Vacca},
  W.~D. 2005, \apj, 626, 70

\bibitem[{{Feoli} \& {Mancini}(2009)}]{feoli:bhfp.1}
{Feoli}, A., \& {Mancini}, L. 2009, \apj, 703, 1502

\bibitem[{{Ferrarese} \& {Merritt}(2000)}]{FM00}
{Ferrarese}, L., \& {Merritt}, D. 2000, \apjl, 539, L9

\bibitem[{{F{\"o}rster Schreiber}
  {et~al.}(2006)}]{forsterschreiber:z2.disk.turbulence}
{F{\"o}rster Schreiber}, N.~M., {et~al.} 2006, \apj, 645, 1062

\bibitem[{{Fruscione} {et~al.}(2005){Fruscione}, {Greenhill}, {Filippenko},
  {Moran}, {Herrnstein}, \& {Galle}}]{fruscione:abs.by.warped.disk}
{Fruscione}, A., {Greenhill}, L.~J., {Filippenko}, A.~V., {Moran}, J.~M.,
  {Herrnstein}, J.~R., \& {Galle}, E. 2005, \apj, 624, 103

\bibitem[{{Fukuda} {et~al.}(2000){Fukuda}, {Habe}, \&
  {Wada}}]{fukuda:nuclear.ring}
{Fukuda}, H., {Habe}, A., \& {Wada}, K. 2000, \apj, 529, 109

\bibitem[{{Fuller} \& {Myers}(1992)}]{fuller:cloud.size.linewidth}
{Fuller}, G.~A., \& {Myers}, P.~C. 1992, \apj, 384, 523

\bibitem[{{Gallimore} {et~al.}(2006){Gallimore}, {Axon}, {O'Dea}, {Baum}, \&
  {Pedlar}}]{gallimore:2006.agn.outflow.gal.alignment}
{Gallimore}, J.~F., {Axon}, D.~J., {O'Dea}, C.~P., {Baum}, S.~A., \& {Pedlar},
  A. 2006, \aj, 132, 546

\bibitem[{Garc{\'\i}a-Burillo {et~al.}(2005)Garc{\'\i}a-Burillo, Combes,
  Schinnerer, Boone, \&
  Hunt}]{garcia.burillo:torques.in.agn.nuclei.obs.maps.no.inflow}
Garc{\'\i}a-Burillo, S., Combes, F., Schinnerer, E., Boone, F., \& Hunt, L.~K.
  2005, Astronomy and Astrophysics, 441, 1011

\bibitem[{{Gebhardt} {et~al.}(2000)}]{Gebhardt00}
{Gebhardt}, K., {et~al.} 2000, \apjl, 539, L13

\bibitem[{Georgakakis {et~al.}(2009)Georgakakis, Coil, Laird, Griffith, Nandra,
  Lotz, Pierce, Cooper, Newman, \& Koekemoer}]{georgakakis:xr.agn.host.morph}
Georgakakis, A., {et~al.} 2009, eprint arXiv, 0904, 3747, accepted for
  publication in MNRAS

\bibitem[{{Georgantopoulos} \&
  {Georgakakis}(2005)}]{georgantopoulos:xbong.dilution}
{Georgantopoulos}, I., \& {Georgakakis}, A. 2005, \mnras, 358, 131

\bibitem[{{Gilli} {et~al.}(2007){Gilli}, {Comastri}, \&
  {Hasinger}}]{gilli:obscured.fractions}
{Gilli}, R., {Comastri}, A., \& {Hasinger}, G. 2007, \aap, 463, 79

\bibitem[{{Granato} {et~al.}(2004){Granato}, {De Zotti}, {Silva}, {Bressan}, \&
  {Danese}}]{granato:sam}
{Granato}, G.~L., {De Zotti}, G., {Silva}, L., {Bressan}, A., \& {Danese}, L.
  2004, \apj, 600, 580

\bibitem[{{Greenhill} {et~al.}(1997){Greenhill}, {Moran}, \&
  {Herrnstein}}]{greenhill:4945.maser}
{Greenhill}, L.~J., {Moran}, J.~M., \& {Herrnstein}, J.~R. 1997, \apjl, 481,
  L23+

\bibitem[{{Greenhill} {et~al.}(2003)}]{greenhill:circinus.acc.disk}
{Greenhill}, L.~J., {et~al.} 2003, \apj, 590, 162

\bibitem[{{Grimes} {et~al.}(2004){Grimes}, {Rawlings}, \&
  {Willott}}]{grimes04:obsc.frac.vs.l}
{Grimes}, J.~A., {Rawlings}, S., \& {Willott}, C.~J. 2004, \mnras, 349, 503

\bibitem[{{Guainazzi} {et~al.}(2005){Guainazzi}, {Fabian}, {Iwasawa}, {Matt},
  \& {Fiore}}]{guainazzi:2005.agn.rate.of.obscuration.variability}
{Guainazzi}, M., {Fabian}, A.~C., {Iwasawa}, K., {Matt}, G., \& {Fiore}, F.
  2005, \mnras, 356, 295

\bibitem[{{Guyon} {et~al.}(2006){Guyon}, {Sanders}, \&
  {Stockton}}]{guyon:qso.hosts.ir}
{Guyon}, O., {Sanders}, D.~B., \& {Stockton}, A. 2006, \apjs, 166, 89

\bibitem[{{Hall} {et~al.}(2007){Hall}, {Sadavoy}, {Hutsemekers}, {Everett}, \&
  {Rafiee}}]{hall:2007.agn.outflow.substructure}
{Hall}, P.~B., {Sadavoy}, S.~I., {Hutsemekers}, D., {Everett}, J.~E., \&
  {Rafiee}, A. 2007, \apj, 665, 174

\bibitem[{{Hao} {et~al.}(2005)}]{hao:local.lf}
{Hao}, L., {et~al.} 2005, \aj, 129, 1795

\bibitem[{{Hasinger}(2004)}]{hasinger04:obsc.frac.and.xrb}
{Hasinger}, G. 2004, Nuclear Physics B Proceedings Supplements, 132, 86

\bibitem[{{Hasinger}(2008)}]{hasinger:absorption.update}
---. 2008, \aap, 490, 905

\bibitem[{{Hasinger} {et~al.}(2005){Hasinger}, {Miyaji}, \&
  {Schmidt}}]{hasinger05:qlf}
{Hasinger}, G., {Miyaji}, T., \& {Schmidt}, M. 2005, \aap, 441, 417

\bibitem[{{Hatziminaoglou} {et~al.}(2009){Hatziminaoglou}, {Fritz}, \&
  {Jarrett}}]{hatziminaoglou:2009.torus.properties.inferred.obs}
{Hatziminaoglou}, E., {Fritz}, J., \& {Jarrett}, T. 2009, \mnras, accepted,
  arXiv:0907.2389 [astro-ph]

\bibitem[{{Hayward} {et~al.}(2011){Hayward}, {Kere{\v s}}, {Jonsson},
  {Narayanan}, {Cox}, \& {Hernquist}}]{hayward:2011.smg.merger.rt}
{Hayward}, C.~C., {Kere{\v s}}, D., {Jonsson}, P., {Narayanan}, D., {Cox},
  T.~J., \& {Hernquist}, L. 2011, \apj, in press [arXiv:1101.0002]

\bibitem[{Heller {et~al.}(2001)Heller, Shlosman, \&
  Englmaier}]{heller:secondary.bar.instability}
Heller, C., Shlosman, I., \& Englmaier, P. 2001, The Astrophysical Journal,
  553, 661

\bibitem[{{Henkel} {et~al.}(2005){Henkel}, {Peck}, {Tarchi}, {Nagar}, {Braatz},
  {Castangia}, \& {Moscadelli}}]{henkel:agn.masers}
{Henkel}, C., {Peck}, A.~B., {Tarchi}, A., {Nagar}, N.~M., {Braatz}, J.~A.,
  {Castangia}, P., \& {Moscadelli}, L. 2005, \aap, 436, 75

\bibitem[{{Hernquist}(1989)}]{hernquist.89}
{Hernquist}, L. 1989, \nat, 340, 687

\bibitem[{{Hickox} {et~al.}(2007)}]{hickox:bootes.obscured.agn}
{Hickox}, R.~C., {et~al.} 2007, \apj, 671, 1365

\bibitem[{{Hicks} {et~al.}(2009){Hicks}, {Davies}, {Malkan}, {Genzel},
  {Tacconi}, {S{\'a}nchez}, \& {Sternberg}}]{hicks:obs.torus.properties}
{Hicks}, E.~K.~S., {Davies}, R.~I., {Malkan}, M.~A., {Genzel}, R., {Tacconi},
  L.~J., {S{\'a}nchez}, F.~M., \& {Sternberg}, A. 2009, \apj, 696, 448

\bibitem[{{Hill} {et~al.}(1996){Hill}, {Goodrich}, \&
  {Depoy}}]{hill96:sf.abs.in.radio.gal}
{Hill}, G.~J., {Goodrich}, R.~W., \& {Depoy}, D.~L. 1996, \apj, 462, 163

\bibitem[{{Hobbs} {et~al.}(2010){Hobbs}, {Nayakshin}, {Power}, \&
  {King}}]{hobbs:turbulence.agn.feeding}
{Hobbs}, A., {Nayakshin}, S., {Power}, C., \& {King}, A. 2010, \mnras, in
  press, arXiv:1001.3883

\bibitem[{{Hoenig} \& {Kishimoto}(2009)}]{hoenig:clumpy.torus.modeling}
{Hoenig}, S.~F., \& {Kishimoto}, M. 2009, \aap, in press, arXiv:0909.4539

\bibitem[{{Hopkins}(2010)}]{hopkins:slow.modes}
{Hopkins}, P.~F. 2010, \mnras, in press, arXiv:1009.4702 [astro-ph]

\bibitem[{{Hopkins}(2011)}]{hopkins:sb.agn.delay}
---. 2011, \mnras, in press, arXiv:1101.4230

\bibitem[{{Hopkins} {et~al.}(2008{\natexlab{a}}){Hopkins}, {Cox}, {Kere{\v s}},
  \& {Hernquist}}]{hopkins:groups.ell}
{Hopkins}, P.~F., {Cox}, T.~J., {Kere{\v s}}, D., \& {Hernquist}, L.
  2008{\natexlab{a}}, \apjs, 175, 390

\bibitem[{{Hopkins} {et~al.}(2009{\natexlab{a}}){Hopkins}, {Cox}, {Younger}, \&
  {Hernquist}}]{hopkins:disk.survival}
{Hopkins}, P.~F., {Cox}, T.~J., {Younger}, J.~D., \& {Hernquist}, L.
  2009{\natexlab{a}}, \apj, 691, 1168

\bibitem[{{Hopkins} \& {Elvis}(2010)}]{hopkins:twostage.feedback}
{Hopkins}, P.~F., \& {Elvis}, M. 2010, \mnras, 401, 7

\bibitem[{{Hopkins} \& {Hernquist}(2006)}]{hopkins:seyferts}
{Hopkins}, P.~F., \& {Hernquist}, L. 2006, \apjs, 166, 1

\bibitem[{{Hopkins} \& {Hernquist}(2010)}]{hopkins:sb.ir.lfs}
---. 2010, \mnras, 402, 985

\bibitem[{{Hopkins} {et~al.}(2005{\natexlab{a}}){Hopkins}, {Hernquist}, {Cox},
  {Di Matteo}, {Martini}, {Robertson}, \&
  {Springel}}]{hopkins:lifetimes.methods}
{Hopkins}, P.~F., {Hernquist}, L., {Cox}, T.~J., {Di Matteo}, T., {Martini},
  P., {Robertson}, B., \& {Springel}, V. 2005{\natexlab{a}}, \apj, 630, 705

\bibitem[{{Hopkins} {et~al.}(2005{\natexlab{b}}){Hopkins}, {Hernquist}, {Cox},
  {Di Matteo}, {Robertson}, \& {Springel}}]{hopkins:lifetimes.obscuration}
{Hopkins}, P.~F., {Hernquist}, L., {Cox}, T.~J., {Di Matteo}, T., {Robertson},
  B., \& {Springel}, V. 2005{\natexlab{b}}, \apj, 632, 81

\bibitem[{{Hopkins} {et~al.}(2006{\natexlab{a}}){Hopkins}, {Hernquist}, {Cox},
  {Di Matteo}, {Robertson}, \& {Springel}}]{hopkins:qso.all}
---. 2006{\natexlab{a}}, \apjs, 163, 1

\bibitem[{{Hopkins} {et~al.}(2008{\natexlab{b}}){Hopkins}, {Hernquist}, {Cox},
  \& {Kere{\v s}}}]{hopkins:groups.qso}
{Hopkins}, P.~F., {Hernquist}, L., {Cox}, T.~J., \& {Kere{\v s}}, D.
  2008{\natexlab{b}}, \apjs, 175, 356

\bibitem[{{Hopkins} {et~al.}(2006{\natexlab{b}}){Hopkins}, {Hernquist}, {Cox},
  {Robertson}, {Di Matteo}, \& {Springel}}]{hopkins:faint.slope}
{Hopkins}, P.~F., {Hernquist}, L., {Cox}, T.~J., {Robertson}, B., {Di Matteo},
  T., \& {Springel}, V. 2006{\natexlab{b}}, \apj, 639, 700

\bibitem[{{Hopkins} {et~al.}(2007{\natexlab{a}}){Hopkins}, {Hernquist}, {Cox},
  {Robertson}, \& {Krause}}]{hopkins:bhfp.theory}
{Hopkins}, P.~F., {Hernquist}, L., {Cox}, T.~J., {Robertson}, B., \& {Krause},
  E. 2007{\natexlab{a}}, \apj, 669, 45

\bibitem[{{Hopkins} {et~al.}(2007{\natexlab{b}}){Hopkins}, {Hernquist}, {Cox},
  {Robertson}, \& {Krause}}]{hopkins:bhfp.obs}
---. 2007{\natexlab{b}}, \apj, 669, 67

\bibitem[{{Hopkins} {et~al.}(2006{\natexlab{c}}){Hopkins}, {Hernquist}, {Cox},
  {Robertson}, \& {Springel}}]{hopkins:red.galaxies}
{Hopkins}, P.~F., {Hernquist}, L., {Cox}, T.~J., {Robertson}, B., \&
  {Springel}, V. 2006{\natexlab{c}}, \apjs, 163, 50

\bibitem[{{Hopkins} {et~al.}(2005{\natexlab{c}}){Hopkins}, {Hernquist},
  {Martini}, {Cox}, {Robertson}, {Di Matteo}, \&
  {Springel}}]{hopkins:lifetimes.letter}
{Hopkins}, P.~F., {Hernquist}, L., {Martini}, P., {Cox}, T.~J., {Robertson},
  B., {Di Matteo}, T., \& {Springel}, V. 2005{\natexlab{c}}, \apjl, 625, L71

\bibitem[{{Hopkins} {et~al.}(2009{\natexlab{b}}){Hopkins}, {Hickox},
  {Quataert}, \& {Hernquist}}]{hopkins:seyfert.bimodality}
{Hopkins}, P.~F., {Hickox}, R., {Quataert}, E., \& {Hernquist}, L.
  2009{\natexlab{b}}, \mnras, 398, 333

\bibitem[{{Hopkins} {et~al.}(2006{\natexlab{d}}){Hopkins}, {Narayan}, \&
  {Hernquist}}]{hopkins:old.age}
{Hopkins}, P.~F., {Narayan}, R., \& {Hernquist}, L. 2006{\natexlab{d}}, \apj,
  643, 641

\bibitem[{{Hopkins} \& {Quataert}(2010{\natexlab{a}})}]{hopkins:zoom.sims}
{Hopkins}, P.~F., \& {Quataert}, E. 2010{\natexlab{a}}, \mnras, 407, 1529

\bibitem[{{Hopkins} \& {Quataert}(2010{\natexlab{b}})}]{hopkins:m31.disk}
---. 2010{\natexlab{b}}, \mnras, 405, L41

\bibitem[{{Hopkins} \&
  {Quataert}(2011{\natexlab{a}})}]{hopkins:inflow.analytics}
---. 2011{\natexlab{a}}, \mnras, 415, 1027

\bibitem[{{Hopkins} \& {Quataert}(2011{\natexlab{b}})}]{hopkins:cusp.slopes}
---. 2011{\natexlab{b}}, \mnras, 411, L61

\bibitem[{{Hopkins} {et~al.}(2011{\natexlab{a}}){Hopkins}, {Quataert}, \&
  {Murray}}]{hopkins:rad.pressure.sf.fb}
{Hopkins}, P.~F., {Quataert}, E., \& {Murray}, N. 2011{\natexlab{a}}, \mnras,
  in press [arXiv:1101.4940]

\bibitem[{{Hopkins} {et~al.}(2011{\natexlab{b}}){Hopkins}, {Quataert}, \&
  {Murray}}]{hopkins:stellar.fb.winds}
---. 2011{\natexlab{b}}, \mnras, in press, arXiv:1110.4638 [astro-ph]

\bibitem[{{Hopkins} {et~al.}(2011{\natexlab{c}}){Hopkins}, {Quataert}, \&
  {Murray}}]{hopkins:fb.ism.prop}
---. 2011{\natexlab{c}}, \mnras, in press, arXiv:1110.4636 [astro-ph]

\bibitem[{{Hopkins} {et~al.}(2007{\natexlab{c}}){Hopkins}, {Richards}, \&
  {Hernquist}}]{hopkins:bol.qlf}
{Hopkins}, P.~F., {Richards}, G.~T., \& {Hernquist}, L. 2007{\natexlab{c}},
  \apj, 654, 731

\bibitem[{{Hopkins} {et~al.}(2006{\natexlab{e}}){Hopkins}, {Robertson},
  {Krause}, {Hernquist}, \& {Cox}}]{hopkins:msigma.limit}
{Hopkins}, P.~F., {Robertson}, B., {Krause}, E., {Hernquist}, L., \& {Cox},
  T.~J. 2006{\natexlab{e}}, \apj, 652, 107

\bibitem[{{Hopkins} {et~al.}(2010){Hopkins}, {Younger}, {Hayward}, {Narayanan},
  \& {Hernquist}}]{hopkins:ir.lfs}
{Hopkins}, P.~F., {Younger}, J.~D., {Hayward}, C.~C., {Narayanan}, D., \&
  {Hernquist}, L. 2010, \mnras, 402, 1693

\bibitem[{{Hopkins} {et~al.}(2011{\natexlab{d}})}]{hopkins:agn.alignment}
{Hopkins}, P.~F., {et~al.} 2011{\natexlab{d}}, \mnras, in prep

\bibitem[{{Houghton} {et~al.}(2006){Houghton}, {Magorrian}, {Sarzi}, {Thatte},
  {Davies}, \& {Krajnovi{\'c}}}]{houghton:ngc1399.nuclear.disk}
{Houghton}, R.~C.~W., {Magorrian}, J., {Sarzi}, M., {Thatte}, N., {Davies},
  R.~L., \& {Krajnovi{\'c}}, D. 2006, \mnras, 367, 2

\bibitem[{{Imanishi}(2002)}]{imanishi:2002.nuclear.sfr.vs.agn}
{Imanishi}, M. 2002, \apj, 569, 44

\bibitem[{{Imanishi} \& {Wada}(2004)}]{imanishi:2004.nuclear.sfr.vs.agn}
{Imanishi}, M., \& {Wada}, K. 2004, \apj, 617, 214

\bibitem[{{Immler} {et~al.}(2003){Immler}, {Brandt}, {Vignali}, {Bauer},
  {Crenshaw}, {Feldmeier}, \& {Kraemer}}]{immler:2003.mrk.6.yr.variability}
{Immler}, S., {Brandt}, W.~N., {Vignali}, C., {Bauer}, F.~E., {Crenshaw},
  D.~M., {Feldmeier}, J.~J., \& {Kraemer}, S.~B. 2003, \aj, 126, 153

\bibitem[{{Iono} {et~al.}(2007)}]{iono:ngc6240.nuclear.gas.huge.turbulence}
{Iono}, D., {et~al.} 2007, \apj, 659, 283

\bibitem[{Jaffe {et~al.}(2004)}]{jaffe:ngc1068.torus.properties}
Jaffe, W., {et~al.} 2004, Nature, 429, 47

\bibitem[{{Jahnke} {et~al.}(2004){Jahnke}, {Kuhlbrodt}, \&
  {Wisotzki}}]{jahnke:qso.host.sf}
{Jahnke}, K., {Kuhlbrodt}, B., \& {Wisotzki}, L. 2004, \mnras, 352, 399

\bibitem[{{Kawakatu} \& {Wada}(2008)}]{kawakatu:disk.bhar.model}
{Kawakatu}, N., \& {Wada}, K. 2008, \apj, 681, 73

\bibitem[{{Keel}(1980)}]{keel:1980.seyfert.vs.galaxy.inclination}
{Keel}, W.~C. 1980, \aj, 85, 198

\bibitem[{{Kennicutt}(1998)}]{kennicutt98}
{Kennicutt}, Jr., R.~C. 1998, \apj, 498, 541

\bibitem[{{King}(2003)}]{king:msigma.superfb.1}
{King}, A. 2003, \apjl, 596, L27

\bibitem[{{Kinney} {et~al.}(2000){Kinney}, {Schmitt}, {Clarke}, {Pringle},
  {Ulvestad}, \& {Antonucci}}]{kinney:2000.bh.jet.directions}
{Kinney}, A.~L., {Schmitt}, H.~R., {Clarke}, C.~J., {Pringle}, J.~E.,
  {Ulvestad}, J.~S., \& {Antonucci}, R.~R.~J. 2000, \apj, 537, 152

\bibitem[{{Klessen} {et~al.}(2007){Klessen}, {Spaans}, \&
  {Jappsen}}]{klessen:2007.imf.from.turbulence}
{Klessen}, R.~S., {Spaans}, M., \& {Jappsen}, A. 2007, \mnras, 374, L29

\bibitem[{{Komossa} {et~al.}(2003){Komossa}, {Burwitz}, {Hasinger}, {Predehl},
  {Kaastra}, \& {Ikebe}}]{komossa:ngc6240}
{Komossa}, S., {Burwitz}, V., {Hasinger}, G., {Predehl}, P., {Kaastra}, J.~S.,
  \& {Ikebe}, Y. 2003, \apjl, 582, L15

\bibitem[{{Kondratko} {et~al.}(2005){Kondratko}, {Greenhill}, \&
  {Moran}}]{kondratko:3079.acc.disk.maser}
{Kondratko}, P.~T., {Greenhill}, L.~J., \& {Moran}, J.~M. 2005, \apj, 618, 618

\bibitem[{{Kondratko} {et~al.}(2006{\natexlab{a}}){Kondratko}, {Greenhill}, \&
  {Moran}}]{kondratko:agn.masers.2}
---. 2006{\natexlab{a}}, \apj, 652, 136

\bibitem[{{Kondratko} {et~al.}(2008){Kondratko}, {Greenhill}, \&
  {Moran}}]{kondratko:ngc3393.acc.disk.maser}
---. 2008, \apj, 678, 87

\bibitem[{{Kondratko} {et~al.}(2006{\natexlab{b}})}]{kondratko:agn.masers}
{Kondratko}, P.~T., {et~al.} 2006{\natexlab{b}}, \apj, 638, 100

\bibitem[{{Kormendy} \& {Richstone}(1995)}]{KormendyRichstone95}
{Kormendy}, J., \& {Richstone}, D. 1995, \araa, 33, 581

\bibitem[{{Krips} {et~al.}(2007)}]{krips:nuclear.disk.torus.obs.seyferts}
{Krips}, M., {et~al.} 2007, \aap, 468, L63

\bibitem[{{Krolik} \& {Begelman}(1988)}]{krolik:clumpy.torii}
{Krolik}, J.~H., \& {Begelman}, M.~C. 1988, \apj, 329, 702

\bibitem[{{Krumholz} \& {Tan}(2007)}]{krumholz:sf.eff.in.clouds}
{Krumholz}, M.~R., \& {Tan}, J.~C. 2007, \apj, 654, 304

\bibitem[{{Kulsrud} \& {Mark}(1970)}]{kulsrud:bending.modes.70}
{Kulsrud}, R.~M., \& {Mark}, J. 1970, \apj, 160, 471

\bibitem[{{Kulsrud} {et~al.}(1971){Kulsrud}, {Mark}, \&
  {Caruso}}]{kulsrud:firehose.instability}
{Kulsrud}, R.~M., {Mark}, J.~W.~K., \& {Caruso}, A. 1971, \apss, 14, 52

\bibitem[{{Kuntschner} {et~al.}(2001){Kuntschner}, {Lucey}, {Smith}, {Hudson},
  \& {Davies}}]{kuntschner:ell.ages}
{Kuntschner}, H., {Lucey}, J.~R., {Smith}, R.~J., {Hudson}, M.~J., \& {Davies},
  R.~L. 2001, \mnras, 323, 615

\bibitem[{{Kuraszkiewicz} {et~al.}(2000){Kuraszkiewicz}, {Wilkes}, {Czerny}, \&
  {Mathur}}]{kuraszkiewicz:2000.hblr.nlagn}
{Kuraszkiewicz}, J., {Wilkes}, B.~J., {Czerny}, B., \& {Mathur}, S. 2000, \apj,
  542, 692

\bibitem[{{La Franca} {et~al.}(2005)}]{lafranca:hx.qlf}
{La Franca}, F., {et~al.} 2005, \apj, 635, 864

\bibitem[{{Lagos} {et~al.}(2011){Lagos}, {Padilla}, {Strauss}, {Cora}, \&
  {Hao}}]{lagos:2011.agn.gal.orientation}
{Lagos}, C.~D.~P., {Padilla}, N.~D., {Strauss}, M.~A., {Cora}, S.~A., \& {Hao},
  L. 2011, \mnras, 414, 2148

\bibitem[{{Lamer} {et~al.}(2003){Lamer}, {Uttley}, \&
  {McHardy}}]{lamer:2003.agn.transmission.event.mol.cloud}
{Lamer}, G., {Uttley}, P., \& {McHardy}, I.~M. 2003, \mnras, 342, L41

\bibitem[{{Larson}(1981)}]{larson:gmc.scalings}
{Larson}, R.~B. 1981, \mnras, 194, 809

\bibitem[{{Lauer} {et~al.}(1993)}]{lauer93}
{Lauer}, T.~R., {et~al.} 1993, \aj, 106, 1436

\bibitem[{{Lauer} {et~al.}(1996)}]{lauer:ngc4486b}
---. 1996, \apjl, 471, L79+

\bibitem[{{Lauer} {et~al.}(2005)}]{lauer:centers}
---. 2005, \aj, 129, 2138

\bibitem[{{Lawrence}(1991)}]{lawrence:receding.torus}
{Lawrence}, A. 1991, \mnras, 252, 586

\bibitem[{{Lawrence} \& {Elvis}(1982)}]{lawrence:1982.torus.alignment}
{Lawrence}, A., \& {Elvis}, M. 1982, \apj, 256, 410

\bibitem[{{Levine} {et~al.}(2010){Levine}, {Gnedin}, \&
  {Hamilton}}]{levine:sim.mdot.pwrspectrum}
{Levine}, R., {Gnedin}, N.~Y., \& {Hamilton}, A.~J.~S. 2010, \apj, 716, 1386

\bibitem[{{Levine} {et~al.}(2008){Levine}, {Gnedin}, {Hamilton}, \&
  {Kravtsov}}]{levine2008:nuclear.zoom}
{Levine}, R., {Gnedin}, N.~Y., {Hamilton}, A.~J.~S., \& {Kravtsov}, A.~V. 2008,
  \apj, 678, 154

\bibitem[{{Liu} {et~al.}(2009){Liu}, {Zakamska}, {Greene}, {Strauss}, {Krolik},
  \& {Heckman}}]{liu:2009.z2.qso.hosts.mergers}
{Liu}, X., {Zakamska}, N.~L., {Greene}, J.~E., {Strauss}, M.~A., {Krolik},
  J.~H., \& {Heckman}, T.~M. 2009, \apj, in press, arXiv:0907.3491

\bibitem[{{Lodato} \& {Bertin}(2003)}]{lodato:1068.maser}
{Lodato}, G., \& {Bertin}, G. 2003, \aap, 398, 517

\bibitem[{{Lonsdale} {et~al.}(2003){Lonsdale}, {Lonsdale}, {Smith}, \&
  {Diamond}}]{lonsdale:vlbi.agn.cores}
{Lonsdale}, C.~J., {Lonsdale}, C.~J., {Smith}, H.~E., \& {Diamond}, P.~J. 2003,
  \apj, 592, 804

\bibitem[{{Maciejewski} \&
  {Athanassoula}(2008)}]{maciejewski:nested.bar.models}
{Maciejewski}, W., \& {Athanassoula}, E. 2008, \mnras, 389, 545

\bibitem[{{Madau} {et~al.}(1994){Madau}, {Ghisellini}, \&
  {Fabian}}]{madau:1994.agn.xrb}
{Madau}, P., {Ghisellini}, G., \& {Fabian}, A.~C. 1994, \mnras, 270, L17+

\bibitem[{{Magorrian} {et~al.}(1998)}]{magorrian}
{Magorrian}, J., {et~al.} 1998, \aj, 115, 2285

\bibitem[{{Mainieri} {et~al.}(2005){Mainieri}, {Rosati}, {Tozzi}, {Bergeron},
  {Gilli}, {Hasinger}, {Nonino}, {Lehmann}, {Alexander}, {Idzi}, {Koekemoer},
  {Norman}, {Szokoly}, \& {Zheng}}]{mainieri:2005.exos}
{Mainieri}, V., {et~al.} 2005, \aap, 437, 805

\bibitem[{{Maiolino} \& {Rieke}(1995)}]{maiolino:1995.seyfert.torus.alignment}
{Maiolino}, R., \& {Rieke}, G.~H. 1995, \apj, 454, 95

\bibitem[{{Malizia} {et~al.}(2009){Malizia}, {Stephen}, {Bassani}, {Bird},
  {Panessa}, \& {Ubertini}}]{malizia:integral.obscured.agn.column.dist}
{Malizia}, A., {Stephen}, J.~B., {Bassani}, L., {Bird}, A.~J., {Panessa}, F.,
  \& {Ubertini}, P. 2009, \mnras, 399, 944

\bibitem[{{Marconi} {et~al.}(2008)}]{marconi:2008.rad.pressure.virial.masses}
{Marconi}, A., {et~al.} 2008, \apj, 678, 693

\bibitem[{{Mark}(1971)}]{mark:slab.bending.instabilities}
{Mark}, J.~W.~K. 1971, \apj, 169, 455

\bibitem[{{Martinez-Sansigre} {et~al.}(2009){Martinez-Sansigre}, {Karim},
  {Schinnerer}, {Omont}, {Smith}, {Wu}, {Hill}, {Kloeckner}, {Lacy},
  {Rawlings}, \& {Willott}}]{martinez-sansigre:high.qsos.in.submm}
{Martinez-Sansigre}, A., {et~al.} 2009, \apj, in press, arXiv:0910.1099

\bibitem[{{Mason} {et~al.}(2006){Mason}, {Geballe}, {Packham}, {Levenson},
  {Elitzur}, {Fisher}, \& {Perlman}}]{mason:ngc1068.torus.obs}
{Mason}, R.~E., {Geballe}, T.~R., {Packham}, C., {Levenson}, N.~A., {Elitzur},
  M., {Fisher}, R.~S., \& {Perlman}, E. 2006, \apj, 640, 612

\bibitem[{{Matt} {et~al.}(2003){Matt}, {Guainazzi}, \&
  {Maiolino}}]{matt:2003.yearly.compton.thick.variability}
{Matt}, G., {Guainazzi}, M., \& {Maiolino}, R. 2003, \mnras, 342, 422

\bibitem[{{Max} {et~al.}(2005){Max}, {Canalizo}, {Macintosh}, {Raschke},
  {Whysong}, {Antonucci}, \& {Schneider}}]{max:2005.6240.core}
{Max}, C.~E., {Canalizo}, G., {Macintosh}, B.~A., {Raschke}, L., {Whysong}, D.,
  {Antonucci}, R., \& {Schneider}, G. 2005, \apj, 621, 738

\bibitem[{{Mayer} {et~al.}(2007){Mayer}, {Kazantzidis}, {Madau}, {Colpi},
  {Quinn}, \& {Wadsley}}]{mayer:bh.binary.sph.zoom.sim}
{Mayer}, L., {Kazantzidis}, S., {Madau}, P., {Colpi}, M., {Quinn}, T., \&
  {Wadsley}, J. 2007, Science, 316, 1874

\bibitem[{{Miyaji} {et~al.}(2001){Miyaji}, {Hasinger}, \&
  {Schmidt}}]{miyaji01:sx.qlf}
{Miyaji}, T., {Hasinger}, G., \& {Schmidt}, M. 2001, \aap, 369, 49

\bibitem[{{Mor} {et~al.}(2009){Mor}, {Netzer}, \&
  {Elitzur}}]{mor:2009.torus.structure.from.fitting.obs}
{Mor}, R., {Netzer}, H., \& {Elitzur}, M. 2009, \apj, in press, arXiv:0907.1654

\bibitem[{{Murray} {et~al.}(2005){Murray}, {Quataert}, \&
  {Thompson}}]{murray:momentum.winds}
{Murray}, N., {Quataert}, E., \& {Thompson}, T.~A. 2005, \apj, 618, 569

\bibitem[{{Nandra} {et~al.}(2005){Nandra}, {Laird}, \&
  {Steidel}}]{nandra05:z3.faint.qlf}
{Nandra}, K., {Laird}, E.~S., \& {Steidel}, C.~C. 2005, \mnras, 360, L39

\bibitem[{{Nandra} {et~al.}(2007)}]{nandra:qso.host.colors}
{Nandra}, K., {et~al.} 2007, \apjl, 660, L11

\bibitem[{{Narayanan} {et~al.}(2006){Narayanan}, {Cox}, {Robertson},
  {Dav{\'e}}, {Di Matteo}, {Hernquist}, {Hopkins}, {Kulesa}, \&
  {Walker}}]{narayanan:co.outflows}
{Narayanan}, D., {et~al.} 2006, \apjl, 642, L107

\bibitem[{{Nardini} {et~al.}(2009){Nardini}, {Risaliti}, {Salvati}, {Sani},
  {Watabe}, {Marconi}, \&
  {Maiolino}}]{nardini:2009.agn.vs.sb.contrib.in.ulirgs}
{Nardini}, E., {Risaliti}, G., {Salvati}, M., {Sani}, E., {Watabe}, Y.,
  {Marconi}, A., \& {Maiolino}, R. 2009, \mnras, 399, 1373

\bibitem[{{Nayakshin} \&
  {King}(2007)}]{nayakshin:forced.stochastic.accretion.model}
{Nayakshin}, S., \& {King}, A. 2007, \mnras, in press, arXiv:0705.1686

\bibitem[{{Nenkova} {et~al.}(2008{\natexlab{a}}){Nenkova}, {Sirocky},
  {Ivezi{\'c}}, \& {Elitzur}}]{nenkova:clumpy.torus.model.1}
{Nenkova}, M., {Sirocky}, M.~M., {Ivezi{\'c}}, {\v Z}., \& {Elitzur}, M.
  2008{\natexlab{a}}, \apj, 685, 147

\bibitem[{{Nenkova} {et~al.}(2008{\natexlab{b}}){Nenkova}, {Sirocky},
  {Nikutta}, {Ivezi{\'c}}, \& {Elitzur}}]{nenkova:clumpy.torus.model.2}
{Nenkova}, M., {Sirocky}, M.~M., {Nikutta}, R., {Ivezi{\'c}}, {\v Z}., \&
  {Elitzur}, M. 2008{\natexlab{b}}, \apj, 685, 160

\bibitem[{{Noguchi}(1987)}]{noguchi:merger.induced.bars.dissipationless}
{Noguchi}, M. 1987, \mnras, 228, 635

\bibitem[{{Noguchi}(1988)}]{noguchi:merger.induced.bars.gas.forcing}
---. 1988, \aap, 203, 259

\bibitem[{{Page} {et~al.}(2004){Page}, {Stevens}, {Ivison}, \&
  {Carrera}}]{page:2004.xr.qso.strong.submm}
{Page}, M.~J., {Stevens}, J.~A., {Ivison}, R.~J., \& {Carrera}, F.~J. 2004,
  \apjl, 611, L85

\bibitem[{{Poliachenko}(1977)}]{poliachenko:firehose.instability}
{Poliachenko}, V.~L. 1977, Soviet Astronomy Letters, 3, 51

\bibitem[{{Ptak} {et~al.}(2003){Ptak}, {Heckman}, {Levenson}, {Weaver}, \&
  {Strickland}}]{ptak:2003.chandra.ulirg.nuclei}
{Ptak}, A., {Heckman}, T., {Levenson}, N.~A., {Weaver}, K., \& {Strickland}, D.
  2003, \apj, 592, 782

\bibitem[{{Ramos Almeida} {et~al.}(2009){Ramos Almeida}, {Levenson}, {Rodriguez
  Espinosa}, {Alonso Herrero}, {Asensio Ramos}, {Radomski}, {Packham},
  {Fisher}, \& {Telesco}}]{ramosalmeida:pc.scale.torus.emission}
{Ramos Almeida}, C., {et~al.} 2009, \apj, in press [arXiv:0906.5368]

\bibitem[{{Rice} {et~al.}(2006){Rice}, {Martini}, {Greene}, {Pogge}, {Shields},
  {Mulchaey}, \& {Regan}}]{rice:nlr.kinematics}
{Rice}, M.~S., {Martini}, P., {Greene}, J.~E., {Pogge}, R.~W., {Shields},
  J.~C., {Mulchaey}, J.~S., \& {Regan}, M.~W. 2006, \apj, 636, 654

\bibitem[{{Rice} {et~al.}(2005){Rice}, {Lodato}, \&
  {Armitage}}]{rice:maximum.viscous.alpha}
{Rice}, W.~K.~M., {Lodato}, G., \& {Armitage}, P.~J. 2005, \mnras, 364, L56

\bibitem[{Riechers {et~al.}(2008)Riechers, Walter, Carilli, Bertoldi, \&
  Momjian}]{riechers:qso.host.wet.merger.remnant.z4}
Riechers, D.~A., Walter, F., Carilli, C.~L., Bertoldi, F., \& Momjian, E. 2008,
  \apj, in press, arXiv:0808.3774

\bibitem[{{Rigby} {et~al.}(2006){Rigby}, {Rieke}, {Donley}, {Alonso-Herrero},
  \& {P{\'e}rez-Gonz{\'a}lez}}]{rigby:qso.hosts}
{Rigby}, J.~R., {Rieke}, G.~H., {Donley}, J.~L., {Alonso-Herrero}, A., \&
  {P{\'e}rez-Gonz{\'a}lez}, P.~G. 2006, \apj, 645, 115

\bibitem[{{Risaliti} {et~al.}(2005){Risaliti}, {Elvis}, {Fabbiano}, {Baldi}, \&
  {Zezas}}]{risaliti:2005.nh.variability.1365}
{Risaliti}, G., {Elvis}, M., {Fabbiano}, G., {Baldi}, A., \& {Zezas}, A. 2005,
  \apjl, 623, L93

\bibitem[{{Risaliti} {et~al.}(2002){Risaliti}, {Elvis}, \&
  {Nicastro}}]{risaliti:nh.column.variability}
{Risaliti}, G., {Elvis}, M., \& {Nicastro}, F. 2002, \apj, 571, 234

\bibitem[{{Risaliti} {et~al.}(1999){Risaliti}, {Maiolino}, \&
  {Salvati}}]{risaliti:seyfert.2.nh.distrib}
{Risaliti}, G., {Maiolino}, R., \& {Salvati}, M. 1999, \apj, 522, 157

\bibitem[{{Robertson} {et~al.}(2006{\natexlab{a}}){Robertson}, {Bullock},
  {Cox}, {Di Matteo}, {Hernquist}, {Springel}, \&
  {Yoshida}}]{robertson:disk.formation}
{Robertson}, B., {Bullock}, J.~S., {Cox}, T.~J., {Di Matteo}, T., {Hernquist},
  L., {Springel}, V., \& {Yoshida}, N. 2006{\natexlab{a}}, \apj, 645, 986

\bibitem[{{Robertson} {et~al.}(2006{\natexlab{b}}){Robertson}, {Cox},
  {Hernquist}, {Franx}, {Hopkins}, {Martini}, \& {Springel}}]{robertson:fp}
{Robertson}, B., {Cox}, T.~J., {Hernquist}, L., {Franx}, M., {Hopkins}, P.~F.,
  {Martini}, P., \& {Springel}, V. 2006{\natexlab{b}}, \apj, 641, 21

\bibitem[{{Robertson} {et~al.}(2006{\natexlab{c}}){Robertson}, {Hernquist},
  {Cox}, {Di Matteo}, {Hopkins}, {Martini}, \&
  {Springel}}]{robertson:msigma.evolution}
{Robertson}, B., {Hernquist}, L., {Cox}, T.~J., {Di Matteo}, T., {Hopkins},
  P.~F., {Martini}, P., \& {Springel}, V. 2006{\natexlab{c}}, \apj, 641, 90

\bibitem[{{Rosolowsky}(2007)}]{rosolowsky:m31.gmcs}
{Rosolowsky}, E. 2007, \apj, 654, 240

\bibitem[{{Rowan-Robinson} {et~al.}(2009){Rowan-Robinson}, {Valtchanov}, \&
  {Nandra}}]{rowanrobinson:xr.ir.comparison.of.torii}
{Rowan-Robinson}, M., {Valtchanov}, I., \& {Nandra}, K. 2009, \mnras, in press
  [arXiv:0905.4389]

\bibitem[{{Salucci} {et~al.}(1999){Salucci}, {Szuszkiewicz}, {Monaco}, \&
  {Danese}}]{salucci:bhmf}
{Salucci}, P., {Szuszkiewicz}, E., {Monaco}, P., \& {Danese}, L. 1999, \mnras,
  307, 637

\bibitem[{{S{\'a}nchez} {et~al.}(2006){S{\'a}nchez}, {Davies}, {Eisenhauer},
  {Tacconi}, {Genzel}, \& {Sternberg}}]{sanchez:circinus.torus.mass}
{S{\'a}nchez}, F.~M., {Davies}, R.~I., {Eisenhauer}, F., {Tacconi}, L.~J.,
  {Genzel}, R., \& {Sternberg}, A. 2006, \aap, 454, 481

\bibitem[{{Sanders}(1999)}]{sanders:agn.vs.sf.in.ulirgs}
{Sanders}, D.~B. 1999, \apss, 266, 331

\bibitem[{{Sanders} \& {Mirabel}(1996)}]{sanders96:ulirgs.mergers}
{Sanders}, D.~B., \& {Mirabel}, I.~F. 1996, \araa, 34, 749

\bibitem[{{Sanders} {et~al.}(1988{\natexlab{a}}){Sanders}, {Soifer}, {Elias},
  {Madore}, {Matthews}, {Neugebauer}, \& {Scoville}}]{sanders88:quasars}
{Sanders}, D.~B., {Soifer}, B.~T., {Elias}, J.~H., {Madore}, B.~F., {Matthews},
  K., {Neugebauer}, G., \& {Scoville}, N.~Z. 1988{\natexlab{a}}, \apj, 325, 74

\bibitem[{{Sanders} {et~al.}(1988{\natexlab{b}}){Sanders}, {Soifer}, {Elias},
  {Neugebauer}, \& {Matthews}}]{sanders88:warm.ulirgs}
{Sanders}, D.~B., {Soifer}, B.~T., {Elias}, J.~H., {Neugebauer}, G., \&
  {Matthews}, K. 1988{\natexlab{b}}, \apjl, 328, L35

\bibitem[{{Sazonov} \& {Revnivtsev}(2004)}]{sazonov.revnivtsev04:local.hx.qlf}
{Sazonov}, S.~Y., \& {Revnivtsev}, M.~G. 2004, \aap, 423, 469

\bibitem[{{Schartmann} {et~al.}(2005){Schartmann}, {Meisenheimer}, {Camenzind},
  {Wolf}, \& {Henning}}]{schartmann:2005.torus.modeling}
{Schartmann}, M., {Meisenheimer}, K., {Camenzind}, M., {Wolf}, S., \&
  {Henning}, T. 2005, \aap, 437, 861

\bibitem[{{Schartmann} {et~al.}(2009){Schartmann}, {Meisenheimer}, {Klahr},
  {Camenzind}, {Wolf}, \&
  {Henning}}]{schartmann:2009.stellar.fb.effects.on.torus}
{Schartmann}, M., {Meisenheimer}, K., {Klahr}, H., {Camenzind}, M., {Wolf}, S.,
  \& {Henning}, T. 2009, \mnras, 393, 759

\bibitem[{{Schinnerer} {et~al.}(2000){Schinnerer}, {Eckart}, {Tacconi},
  {Genzel}, \& {Downes}}]{schinnerer:interfer.obs.1068}
{Schinnerer}, E., {Eckart}, A., {Tacconi}, L.~J., {Genzel}, R., \& {Downes}, D.
  2000, \apj, 533, 850

\bibitem[{{Schinnerer}
  {et~al.}(2008)}]{schinnerer:submm.merger.w.compact.mol.gas}
{Schinnerer}, E., {et~al.} 2008, \apjl, 689, L5

\bibitem[{{Schmitt} {et~al.}(1997){Schmitt}, {Kinney}, {Storchi-Bergmann}, \&
  {Antonucci}}]{schmitt:1997.radio.alignment.w.host}
{Schmitt}, H.~R., {Kinney}, A.~L., {Storchi-Bergmann}, T., \& {Antonucci}, R.
  1997, \apj, 477, 623

\bibitem[{{Scoville} {et~al.}(1986){Scoville}, {Sanders}, {Sargent}, {Soifer},
  {Scott}, \& {Lo}}]{scoville86}
{Scoville}, N.~Z., {Sanders}, D.~B., {Sargent}, A.~I., {Soifer}, B.~T.,
  {Scott}, S.~L., \& {Lo}, K.~Y. 1986, \apjl, 311, L47

\bibitem[{{Scoville} {et~al.}(1987){Scoville}, {Yun}, {Sanders}, {Clemens}, \&
  {Waller}}]{scoville:gmc.properties}
{Scoville}, N.~Z., {Yun}, M.~S., {Sanders}, D.~B., {Clemens}, D.~P., \&
  {Waller}, W.~H. 1987, \apjs, 63, 821

\bibitem[{{Setti} \& {Woltjer}(1989)}]{setti:1989.xrb.from.agn}
{Setti}, G., \& {Woltjer}, L. 1989, \aap, 224, L21

\bibitem[{{Shankar} {et~al.}(2004){Shankar}, {Salucci}, {Granato}, {De Zotti},
  \& {Danese}}]{shankar:bhmf}
{Shankar}, F., {Salucci}, P., {Granato}, G.~L., {De Zotti}, G., \& {Danese}, L.
  2004, \mnras, 354, 1020

\bibitem[{{Shankar} {et~al.}(2009){Shankar}, {Weinberg}, \&
  {Miralda-Escud{\'e}}}]{shankar:bol.qlf}
{Shankar}, F., {Weinberg}, D.~H., \& {Miralda-Escud{\'e}}, J. 2009, \apj, 690,
  20

\bibitem[{{Shen} {et~al.}(2010){Shen}, {Shao}, \&
  {Gu}}]{shen:2010.torus.alignment}
{Shen}, S., {Shao}, Z., \& {Gu}, M. 2010, \apjl, 725, L210

\bibitem[{Shi {et~al.}(2006)}]{shi:silicate.contraints.on.torii}
Shi, Y., {et~al.} 2006, The Astrophysical Journal, 653, 127

\bibitem[{{Shi} {et~al.}(2007)}]{shi:qso.host.sf.lf}
{Shi}, Y., {et~al.} 2007, \apj, 669, 841

\bibitem[{{Shlosman} \& {Heller}(2002)}]{shlosman:nested.bar.evol}
{Shlosman}, I., \& {Heller}, C.~H. 2002, \apj, 565, 921

\bibitem[{{Silk} \& {Rees}(1998)}]{silkrees:msigma}
{Silk}, J., \& {Rees}, M.~J. 1998, \aap, 331, L1

\bibitem[{{Silverman} {et~al.}(2005)}]{silverman:hx.spacedensity.ldde}
{Silverman}, J.~D., {et~al.} 2005, \apj, 624, 630

\bibitem[{{Silverman} {et~al.}(2008)}]{silverman:qso.hosts}
---. 2008, \apj, 675, 1025

\bibitem[{{Simcoe} {et~al.}(1997){Simcoe}, {McLeod}, {Schachter}, \&
  {Elvis}}]{simcoe:1997.agn.host.alignment}
{Simcoe}, R., {McLeod}, K.~K., {Schachter}, J., \& {Elvis}, M. 1997, \apj, 489,
  615

\bibitem[{{Simpson}(2005)}]{simpson:type1.frac}
{Simpson}, C. 2005, \mnras, 360, 565

\bibitem[{{Simpson} \& {Rawlings}(2000)}]{simpson00:ir.photometry.radio.qsos}
{Simpson}, C., \& {Rawlings}, S. 2000, \mnras, 317, 1023

\bibitem[{{Simpson} {et~al.}(1999){Simpson}, {Rawlings}, \&
  {Lacy}}]{simpson99:thermal.imaging.of.radio.gal}
{Simpson}, C., {Rawlings}, S., \& {Lacy}, M. 1999, \mnras, 306, 828

\bibitem[{{Soifer} {et~al.}(1984)}]{soifer84b}
{Soifer}, B.~T., {et~al.} 1984, \apjl, 283, L1

\bibitem[{{Solomon} {et~al.}(1987){Solomon}, {Rivolo}, {Barrett}, \&
  {Yahil}}]{solomon:gmc.scalings}
{Solomon}, P.~M., {Rivolo}, A.~R., {Barrett}, J., \& {Yahil}, A. 1987, \apj,
  319, 730

\bibitem[{{Soltan}(1982)}]{soltan82}
{Soltan}, A. 1982, \mnras, 200, 115

\bibitem[{{Spaans} \& {Silk}(2005)}]{spaans:2005.gmc.eos}
{Spaans}, M., \& {Silk}, J. 2005, \apj, 626, 644

\bibitem[{{Springel}(2005)}]{springel:gadget}
{Springel}, V. 2005, \mnras, 364, 1105

\bibitem[{{Springel} {et~al.}(2005{\natexlab{a}}){Springel}, {Di Matteo}, \&
  {Hernquist}}]{springel:red.galaxies}
{Springel}, V., {Di Matteo}, T., \& {Hernquist}, L. 2005{\natexlab{a}}, \apjl,
  620, L79

\bibitem[{{Springel} {et~al.}(2005{\natexlab{b}}){Springel}, {Di Matteo}, \&
  {Hernquist}}]{springel:models}
---. 2005{\natexlab{b}}, \mnras, 361, 776

\bibitem[{{Springel} \& {Hernquist}(2002)}]{springel:entropy}
{Springel}, V., \& {Hernquist}, L. 2002, \mnras, 333, 649

\bibitem[{{Springel} \& {Hernquist}(2003)}]{springel:multiphase}
---. 2003, \mnras, 339, 289

\bibitem[{{Steffen} {et~al.}(2003){Steffen}, {Barger}, {Cowie}, {Mushotzky}, \&
  {Yang}}]{steffen03:agn.obsc.vs.l.z}
{Steffen}, A.~T., {Barger}, A.~J., {Cowie}, L.~L., {Mushotzky}, R.~F., \&
  {Yang}, Y. 2003, \apjl, 596, L23

\bibitem[{{Stevens} {et~al.}(2005){Stevens}, {Page}, {Ivison}, {Carrera},
  {Mittaz}, {Smail}, \& {McHardy}}]{stevens:xray.qso.hosts}
{Stevens}, J.~A., {Page}, M.~J., {Ivison}, R.~J., {Carrera}, F.~J., {Mittaz},
  J.~P.~D., {Smail}, I., \& {McHardy}, I.~M. 2005, \mnras, 360, 610

\bibitem[{{Tan} {et~al.}(2006){Tan}, {Krumholz}, \&
  {McKee}}]{tan:mol.cloud.formation.times}
{Tan}, J.~C., {Krumholz}, M.~R., \& {McKee}, C.~F. 2006, \apjl, 641, L121

\bibitem[{{Teyssier} {et~al.}(2010){Teyssier}, {Chapon}, \&
  {Bournaud}}]{teyssier:2010.clumpy.sb.in.mergers}
{Teyssier}, R., {Chapon}, D., \& {Bournaud}, F. 2010, \apjl, 720, L149

\bibitem[{{Thatte} {et~al.}(2000){Thatte}, {Tecza}, \&
  {Genzel}}]{thatte:m83.double.nucleus}
{Thatte}, N., {Tecza}, M., \& {Genzel}, R. 2000, \aap, 364, L47

\bibitem[{Thompson {et~al.}(2009)Thompson, Levenson, Uddin, \&
  Sirocky}]{thompson:dust.em.from.unobscured.agn}
Thompson, G.~D., Levenson, N.~A., Uddin, S.~A., \& Sirocky, M.~M. 2009, eprint
  arXiv, 0903, 2422

\bibitem[{{Thompson} {et~al.}(2005){Thompson}, {Quataert}, \&
  {Murray}}]{thompson:rad.pressure}
{Thompson}, T.~A., {Quataert}, E., \& {Murray}, N. 2005, \apj, 630, 167

\bibitem[{{Tortora} {et~al.}(2009){Tortora}, {Antonuccio-Delogu}, {Kaviraj},
  {Silk}, {Romeo}, \& {Becciani}}]{tortora:2009.agn.jet.fb.and.ell.colors}
{Tortora}, C., {Antonuccio-Delogu}, V., {Kaviraj}, S., {Silk}, J., {Romeo},
  A.~D., \& {Becciani}, U. 2009, \mnras, 396, 61

\bibitem[{{Tran}(2003)}]{tran:hblr.2}
{Tran}, H.~D. 2003, \apj, 583, 632

\bibitem[{{Treister} \& {Urry}(2006)}]{treister:obscured.frac.z.evol}
{Treister}, E., \& {Urry}, C.~M. 2006, \apjl, 652, L79

\bibitem[{Treister {et~al.}(2009)Treister, Urry, \&
  Virani}]{treister:compton.thick.fractions}
Treister, E., Urry, C.~M., \& Virani, S. 2009, The Astrophysical Journal, 696,
  110

\bibitem[{{Tremaine}(1995)}]{tremaine:m31.nuclear.disk.model}
{Tremaine}, S. 1995, \aj, 110, 628

\bibitem[{{Tremaine}(2001)}]{tremaine:slow.keplerian.modes}
---. 2001, \aj, 121, 1776

\bibitem[{{Trump} {et~al.}(2009){Trump}, {Impey}, {Taniguchi}, {Brusa},
  {Civano}, {Elvis}, {Gabor}, {Jahnke}, {Kelly}, {Koekemoer}, {Nagao},
  {Salvato}, {Shioya}, {Capak}, {Huchra}, {Kartaltepe}, {Lanzuisi}, {McCarthy},
  {Maineri}, \& {Scoville}}]{trump:lowl.agn.dilution}
{Trump}, J.~R., {et~al.} 2009, \apj, in press, arXiv:0910.2672

\bibitem[{{Ueda} {et~al.}(2003){Ueda}, {Akiyama}, {Ohta}, \&
  {Miyaji}}]{ueda03:qlf}
{Ueda}, Y., {Akiyama}, M., {Ohta}, K., \& {Miyaji}, T. 2003, \apj, 598, 886

\bibitem[{{Wada} \& {Norman}(2002)}]{wada:starburst.torus.model}
{Wada}, K., \& {Norman}, C.~A. 2002, \apjl, 566, L21

\bibitem[{{Wada} {et~al.}(2009){Wada}, {Papadopoulos}, \&
  {Spaans}}]{wada:torus.mol.gas.hydro.sims}
{Wada}, K., {Papadopoulos}, P., \& {Spaans}, M. 2009, \apj, in press,
  arXiv:0906.5444

\bibitem[{{Wang} {et~al.}(2007){Wang}, {Chen}, {Yan}, {Hu}, \&
  {Bian}}]{wang:seyfert.nuclear.sf}
{Wang}, J.-M., {Chen}, Y.-M., {Yan}, C.-S., {Hu}, C., \& {Bian}, W.-H. 2007,
  \apjl, 661, L143

\bibitem[{{Ward-Thompson} {et~al.}(1994){Ward-Thompson}, {Scott}, {Hills}, \&
  {Andre}}]{wardthompson94:protostellar.core.sizes}
{Ward-Thompson}, D., {Scott}, P.~F., {Hills}, R.~E., \& {Andre}, P. 1994,
  \mnras, 268, 276

\bibitem[{{Willott} {et~al.}(2000){Willott}, {Rawlings}, {Blundell}, \&
  {Lacy}}]{willott00:optical.qso.frac.vs.l}
{Willott}, C.~J., {Rawlings}, S., {Blundell}, K.~M., \& {Lacy}, M. 2000,
  \mnras, 316, 449

\bibitem[{{Yip} {et~al.}(2004)}]{yip:qso.eigenspectra}
{Yip}, C.~W., {et~al.} 2004, \aj, 128, 2603

\bibitem[{{Younger} {et~al.}(2009){Younger}, {Hayward}, {Narayanan}, {Cox},
  {Hernquist}, \& {Jonsson}}]{younger:warm.ulirg.evol}
{Younger}, J.~D., {Hayward}, C.~C., {Narayanan}, D., {Cox}, T.~J., {Hernquist},
  L., \& {Jonsson}, P. 2009, \mnras, 396, L66

\bibitem[{{Younger} {et~al.}(2008){Younger}, {Hopkins}, {Cox}, \&
  {Hernquist}}]{younger:minor.mergers}
{Younger}, J.~D., {Hopkins}, P.~F., {Cox}, T.~J., \& {Hernquist}, L. 2008,
  \apj, 686, 815

\bibitem[{{Yu} \& {Tremaine}(2002)}]{yutremaine:bhmf}
{Yu}, Q., \& {Tremaine}, S. 2002, \mnras, 335, 965

\bibitem[{{Yuan} \& {Narayan}(2004)}]{xbongs}
{Yuan}, F., \& {Narayan}, R. 2004, \apj, 612, 724

\bibitem[{{Yuan} {et~al.}(2010){Yuan}, {Kewley}, \&
  {Sanders}}]{yuan:ulirgs.are.agn.sb.composites.sequence}
{Yuan}, T., {Kewley}, L.~J., \& {Sanders}, D.~B. 2010, \apj, 709, 884

\bibitem[{{Zakamska} {et~al.}(2006)}]{zakamska:qso.hosts}
{Zakamska}, N.~L., {et~al.} 2006, \aj, 132, 1496

\bibitem[{{Zhang} {et~al.}(2009){Zhang}, {Soria}, {Zhang}, {Swartz}, \&
  {Liu}}]{zhang:2009.agn.vs.hubble.type}
{Zhang}, W.~M., {Soria}, R., {Zhang}, S.~N., {Swartz}, D.~A., \& {Liu}, J.~F.
  2009, \apj, 699, 281

\end{thebibliography}

\end{document}